\documentclass{aa}
\usepackage{graphicx}
\usepackage{txfonts}
\usepackage{psfig}
\usepackage{psfrag}
\usepackage{latexsym}
\usepackage{times,lscape,verbatim}
\usepackage{natbib}
\usepackage[mathcal]{euscript}
\usepackage{mathrsfs}
\begin{document}
	\title{Grain growth across protoplanetary discs: 10~$\mu$m silicate feature versus millimetre slope}
   \titlerunning{Correlating $\mu$m and mm observations in YSOs}

   \author{Dave J. P. Lommen\inst{1}
   	\and Ewine F. van Dishoeck\inst{1,2}
	\and Chris M. Wright\inst{3}
	\and Sarah T. Maddison\inst{4}
	\and Michiel Min\inst{5,6}
	\and David J. Wilner\inst{7}
	\and Demerese M. Salter\inst{1}
	\and Huib Jan van Langevelde\inst{8,1}
	\and Tyler L. Bourke\inst{7}
	\and Remco F. J. van der Burg\inst{1}
	\and Geoffrey A. Blake\inst{9}
	     }

   \offprints{Dave Lommen, \\ \email{dave@strw.leidenuniv.nl}}

   \institute{Leiden Observatory, Leiden University, P.O. Box 9513, 2300 RA Leiden, The Netherlands
   	\and Max-Planck-Institut f\"{u}r extraterrestrische Physik, Garching, Germany
   	\and School of Physical, Environmental and Mathematical Sciences, UNSW@ADFA, Canberra ACT 2600, Australia
	\and Centre for Astrophysics and Supercomputing, Swinburne University of Technology, PO Box 218, Hawthorn, VIC 3122, Australia 
	\and Astronomical Institute Utrecht, Princetonplein 5, 3584 CC Utrecht, The Netherlands
	\and Astronomical institute Anton Pannekoek, University of Amsterdam, Kruislaan 403, 1098 SJ Amsterdam, The Netherlands
	\and Harvard-Smithsonian Center for Astrophysics, 60 Garden Street, 02138 Cambridge, MA, USA
	\and Joint Institute for VLBI in Europe, PO Box 2, 7990 AA Dwingeloo, The Netherlands
	\and California Institute of Technology, Pasadena, CA 91125, USA
             }

   \date{Received ...; accepted ...}

 \abstract 
  {Young stars are formed within dusty discs. The grains in the disc are originally of the same size as interstellar dust,
   around 0.1~$\mu$m. Models predict that these grains will grow in size through coagulation. Observations of the 
   silicate features at $\mu$m wavelengths are consistent with growth from submicron to micron sizes in selected sources whereas the slope 
   of the SED at longer wavelengths traces growth up to mm sizes and larger.}
 {We here look for a correlation between these two grain growth indicators.}
  {A large sample of T-Tauri and Herbig-Ae/Be stars, spread over five different star-forming regions, was observed with the Spitzer Space Telescope
  at 5--13~$\mu$m; a subsample was observed at mm wavelengths. We complement this subsample with data from the literature to maximise the overlap 
  between $\mu$m and mm observations and search for correlations. Synthetic spectra are produced to determine which 
  processes may produce the dust evolution.}
{Dust disc masses in the range $<1$ to $7 \times 10^{-4}$~M$_\odot$ are obtained. Most sources have a mm spectral slope consistent with
  grain growth. There is a tentative correlation between the 10-$\mu$m silicate feature and the slope of the SED between 1 and 3~mm.
  The observed sources seem to be grouped per star-forming region in the $\mu$m-vs-mm diagram. 
  The modelling results show that first the 10-$\mu$m feature becomes flatter and subsequently
  the mm slope becomes shallower. Grain size distributions shallower than that of the ISM and/or bright central stars are required to explain
  specific features. Settling of larger grains towards the disc midplane affects 10-$\mu$m feature, but hardly the mm slope.}
{The tentative correlation between the strength of the 10-$\mu$m feature and the mm slope suggests that the inner and 
  outer disc evolve simultaneously. Dust with a mass dominated by $\sim$mm-sized
  grains is required to explain the shallowest mm slopes. Other processes besides grain growth may also be responsible for the removal of small 
  grains. Observations with future, more sensitive telescopes are required to provide the necessary statistics to study these processes of disc and
  dust evolution.}


   \maketitle
%

\section{Introduction}\label{sect: introduction}

	A long-standing problem in planet formation is how tiny interstellar dust particles of less than a micron in size coagulate
	and grow to eventually form planets, thousands of kilometres in size. It is in the very nature of this field that it has to be studied
	at various levels, since different physical processes dominate during the various phases. The first steps, which lead to dust grains
	of about a decimetre in size, are studied both in the laboratory and with computer simulations 
	\citep[see][for detailed reviews]{dominik:2007, blum:2008}. Local concentrations of boulders and subsequent gravitational collapse may then lead
	to the formation of planetesimals several hundreds of kilometres in size \citep[e.g.,][]{johansen:2007}. This paper focuses on the
	observational signatures of (sub)micron-sized grains up to centimetre-sized pebbles.

	The InfraRed Spectrograph (IRS) on-board the Spitzer Space Telescope has provided a wealth of mid-infrared (5--40 $\mu$m) spectra from 
	discs around pre-main-sequence stars \citep[e.g.][]{kessler-silacci:2006,furlan:2006}. The spectra of these objects are often dominated by 
	silicate emission features at 10 and 20~$\mu$m. In young stellar objects, these features are formed in the upper atmosphere of the hot inner disc.	
	The varying strength and the shape of these features can be naturally explained by different grain sizes in the upper layers 
	of the inner disc, with strong, pointed features being representative of $\sim$0.1~$\mu$m-sized grains and flatter features coming from
	dust grains of several $\mu$m in size \citep{kessler-silacci:2006}. These results confirm earlier
	results from the Infrared Space Observatory \citep[][]{bouwman:2001, meeus:2001} and from ground-based observations
	\citep[e.g.,][]{przygodda:2003}. It has been suggested that crystallisation has a similar effect on the 10-$\mu$m feature as grain growth
	\citep[e.g.,][]{honda:2003, meeus:2003}. However, this effect is minimal and the dominating factor
	for the strength and shape of the 10-$\mu$m feature is the characteristic grain size \citep{olofsson:2009}.
	
	Because the 10-$\mu$m feature only probes the surface layers of the inner disc, a stronger, more peaked feature could 
	also be due to the settling of larger, micron-sized grains towards the mid-plane. As the larger grains settle and the small ones remain
	suspended in the upper layers, the surface becomes dominated by small grains, creating a strong silicate band. \citet{dullemond:2008} 
	investigated this idea through theoretical models. They find that settling can in principle explain the 
	different shapes of the	10-$\mu$m feature, but only in quite specific cases, so that overall grain growth is still the most likely
	explanation for the flattening of these features. 
	Recent interferometric observations of the 10-$\mu$m spectral region 
	in discs around $\sim$1 and 2-3-M$_\odot$ objects show that the grains closer to the central star are both larger and more crystalline than 
	those further out in the disc \citep[see, e.g., the recent review by][]{vanboekel:2008}. Hence, the evolution of the 10-$\mu$m feature
	may be caused by a combination of grain growth and crystallisation and appears to progress from the inner disc outwards. On the other hand,
	analysis of the longer wavelenth mid-infrared crystalline features indicates significant growth and crystallisation in the outer disc as well
	\citep{olofsson:2009}.
	
	Whereas the mid-infrared region potentially provides information on the growth of grains from interstellar, submicron
	sizes to sizes of several microns, the growth to larger sizes can only be probed by submillimetre (submm),
	millimetre (mm), and occasionally centimetre (cm) observations. 
	Ground-breaking work was done by \citet{beckwith:1990} and \citet{beckwith:1991}, both analytically studying the
	emission of dust grains and obtaining the first submm slopes by observing a large sample of young stellar objects 
	at mm wavelengths. More recently,
	\citet{andrews:2005} performed a sensitive single-dish submm continuum survey of 153 young stellar objects in the Taurus-Auriga 
	star-formation region, including a large amount of archival and literature data. They found that the submm slope
	between 350~$\mu$m and 1.3~mm could be well described by $\alpha = 2.0 \pm 0.5$, where
	$F_\nu \propto \nu^{\alpha}$, while the value for the interstellar medium is $\alpha \approx 3.7$ \citep[cf.][]{draine:2006}.
	\citet{andrews:2005} interpreted this shallow slope as a combined effect of a contribution from optically thick regions in the disc and
	grain growth. It should be noted, however, that the sources in this study were spatially unresolved, and the (sub)mm 
	emission may have a significant contribution from surrounding (envelope) material. More recently, interferometric
	studies of several dozen T-Tauri stars gave values of $\alpha \lesssim 3.0$ \citep[][]{rodmann:2006,andrews:2007,lommen:2007}. Similar results
	were found for a number of more massive Herbig-Ae/Be stars \citep[e.g.,][]{natta:2004}. From this mm slope one can estimate the opacity index 
	$\beta \approx (1 + \Delta) \times (\alpha - 2)$, where $\Delta$ is found to be $\sim$0.20 \citep{rodmann:2006, lommen:2007}, and values of 
	$\beta \approx 1.0$ for $\lambda \gtrsim 1$~mm	were found. Such a slope can be naturally explained by a significant fraction of grains at least 
	several mm in size present in the discs \citep{draine:2006}. 
	
	A subsample of the sources observed by \citet{lommen:2007} overlapped with the 
	{\em Spitzer} Infrared Spectrograph (IRS) observations published by \citet{kessler-silacci:2006} and \citet{lommen:2007} found a 
	tentative correlation between the mm slope of the spectral energy distribution (SED) and the strength and shape of the 10-$\mu$m
	silicate feature for these sources. Note that the 10-$\mu$m feature primarily probes the hot surface layers of the inner disc, whereas the
	(sub)mm observations provide information of the cold mid-plane of the outer disc. A correlation between the two is therefore not obvious at all
	and a confirmation of this correlation would give very valuable information on the processes of dust growth in
	protoplanetary discs, as it would imply that grain growth from submicron to mm sizes is both fast and occurs simultaneously throughout the
	whole disc. 

	\citet{acke:2004a} calculated the (sub)mm spectral indices of 26 Herbig-Ae/Be stars, for which the infrared
	SED could also be determined. They found a correlation between the strength of the ratio
	of the near- to mid-infrared excess and the slope of the (sub)mm energy distribution for these sources, which
	they attributed to a correlation between the disc geometry (flared versus self-shadowed) and the size of the 
	grains in the disc. However, the authors did not find a correlation between the strength and the shape of the 
	10-$\mu$m silicate feature and the (sub)mm spectral index \citep[see also][]{acke:2004b}.

	The aim of this paper is to investigate the tentative correlation between the strength and shape of the 10-$\mu$m silicate feature and the 
	spectral slope in the (sub)mm regime, found by \citet{lommen:2007}, for a larger sample. A subsample
	of sources studied with the Spitzer IRS were observed with mm and cm interferometers (Section 2). 
	Interferometers were used to ascertain that the emission is dominated by disc emission, since extended emission from
	surrounding material will be filtered out. Also, spatially resolving the disc ensures that the emission is not optically thick
	\citep[e.g.,][]{natta:2004}. The results of the observations, including dust disc masses and mm slopes, are shown in Section 3, and in Section 4 we 
	present model results for discs. The observations and models are compared and discussed in Section 5; conclusions are formulated in Section 6.


\section{Observations}\label{sect: observations}

	For this study, we compared Spitzer IRS observations covering the 10-$\mu$m silicate feature with mm observations from the Very Large
	Array (VLA, operated by NRAO\footnote{The National Radio Astronomy Observatory is a facility of the National Science
	Foundation operated under cooperative agreement by Associated Universities, Inc.}), the Combined Array for Research in 
	Millimeter-wave Astronomy (CARMA\footnote{Support for CARMA construction was derived from the Gordon and Betty Moore
	Foundation, the Kenneth T. and Eileen L. Norris Foundation, the Associates of the California Institute of Technology, the
	states of California, Maryland, and Illinois, and the National Science Foundation. Ongoing CARMA development and operations
	are supported by the National Science Foundation under a cooperative agreement, and by the CARMA partner universities.}), the Submillimeter
	Array (SMA\footnote{The Submillimeter Array is a joint project between the Smithsonian Astrophysical 
	Observatory and the Academia Sinica Institute of Astronomy and Astrophysics and is funded by the Smithsonian 
   	Institution and the Academia Sinica.}), and the Australia Telescope Compact Array (ATCA\footnote{The 
	Australia Telescope Compact Array is part of the Australia Telescope which is funded by the Commonwealth of Australia 
	for operation as a National Facility managed by CSIRO.}). 
	The sources for which new observations are obtained for this work are listed in Table~\ref{tab: source list}. A full log of the newly obtained mm 
	and cm observations is listed in Appendix~\ref{app: observations}. A full log of the newly obtained mm and cm results is listed in 
	Appendix~\ref{app: results}.

\subsection{Source selection and Spitzer data}\label{sect: source selection}

	To look for possible environmental effects, sources in a total of five star-forming regions were observed, spread over the 
	constellations Lupus, Chamaeleon, Corona Australis, Serpens, and the Gum nebula in Vela at distances of about 150--200, 160, 130, 260,
	and 400	pc, respectively. Furthermore, data from the literature for the Taurus-Auriga star-forming region at about 140 pc were 
	included to improve the	statistics further, see Table~\ref{tab: star-forming regions}. The sources were pre-selected to have a large 
	spread in the strengths and shapes of the 10-$\mu$m features from Spitzer IRS data, mainly the ``From Molecular Cores to 
	Planet-forming Discs'' programme \citep[c2d,][Program IDs 139 and 172--179]{evans:2003}, the ``The evolution of dust mineralogy in 
	southern star forming clouds'' programme (C.M.~Wright PI, Project ID 20611), and ``A complete IRS survey of the evolution of 
	circumstellar disks within 3 Myr: New clusters of sequential star formation in Serpens'' (K.M.~Pontoppidan PI, Project ID 30223). The 
	spectra from the c2d project were previously 
	published in \citet{kessler-silacci:2006} and \citet{olofsson:2009}. Program P20611 includes Spitzer IRS observations 
	from embedded YSOs, T-Tauri stars, and Herbig/Vela-type stars. The results for the T-Tauri stars are presented in this 
	work. 
	
	The data from Project ID 20611 are presented here for the first time. The data from the other programmes are re-reduced for 
	this work using the updated c2d IRS reduction pipeline \citep{lahuis:2006} for uniformity of the comparisons.
	Spectra were obtained both integrated over the full aperture of the instrument as well as convolved with the point spread function
	(PSF) at each wavelength. The spectra obtained using the Full-Aperture extraction method were used in here, unless the final 
	spectrum quality of the PSF extraction method was considerably better. Furthermore, only data from the short-low module (SL, 
	5.2--14.5~$\mu$m) were included, unless data from the short-high module (SH, 9.9--19.6~$\mu$m) were present and of significantly 
	higher quality.
	\begin{table}
		\caption[]{Distances to and ages of star-forming clouds.}
		\label{tab: star-forming regions}
		\centering
		\begin{tabular}{lcc}
			\hline
			\hline
			Cloud			& Age$^\mathrm{a}$		& $D^\mathrm{b}$	\\
						& (Myr)				& (pc)			\\
			\hline
			Lupus 1 and 2		& $\lesssim 1$			& $150 \pm 20$		\\
			Lupus 3			& 1--1.5			& $200 \pm 20$		\\
			Lupus 4			& 1--1.5			& $165 \pm 15$		\\
			Cha I			& 3--4 (southern subcluster)	& $160 \pm 15$		\\
						& 5--6 (northern subcluster)	& 			\\
			Corona Australis	& 5--13				& $\sim$130		\\
			Serpens			& 1--15				& $259 \pm 37$		\\
			Gum nebula		& 2--6				& $400 \pm 60$		\\
			Taurus-Auriga		& 1--10				& $140 \pm 15$		\\
			\hline
		\end{tabular}
		\begin{list}{}{}
			\item[$^\mathrm{a}$]	Ages adopted from \citet{heiles:1998}, \citet{comeron:2003}, 
				\citet{james:2006}, \citet{luhman:2007}, \citet{comeron:2008}, \citet{kenyon:2008}, \citet{neuhauser:2008}, 
				\citet{oliveira:2009}, and references therein.
			\item[$^\mathrm{b}$]	Distances adopted from \citet{brandt:1971}, \citet{kenyon:1994}, 
				\citet{straizys:1996}, \citet{whittet:1997}, \citet{bertout:1999}, \citet{dezeeuw:1999}, 
				\citet{comeron:2008}, \citet{neuhauser:2008}, and references therein.
		\end{list}
	\end{table}
	
	In binary systems, it is possible that circumstellar discs get truncated due to binary interaction, affecting grain 
	growth in the discs. To check for such effects, a number of binaries were included in the sample. Furthermore, the 
	sources were selected to include so-called ``cold'' or ``transitional'' discs \citep[e.g.][]{brown:2007}. The cold discs show a lower flux 
	in the mid-infrared, which can be naturally explained by a lack of small warm dust close to the star. Several of the cold 
	discs were recently found to be circumbinary discs, with a large hole or gap in the centre, e.g., CS~Cha \citep{espaillat:2007a} and
	HH~30 \citep{guilloteau:2008}. However, some cold discs are supposedly single stars, requiring a different mechanism to clear the 
	inner discs of small, hot grains \citep[e.g.,][]{pontoppidan:2008}. One such mechanism could be grain growth into larger particles. 
	Another possibility would be that a planet has cleared the inner disc from most of the large grains, leaving behind a protoplanetary
	disc dominated by small, micron-sized grains.
	A number of cold discs of \citet{brown:2007} and \citet{merin:2008} were 
	included in the sample with the aim to explore this possibility. A full list of the sources (35 single sources and five binaries) is given in Table~\ref{tab: 
	source list}. As will be shown in the next Section, 33 of these turn out to have a detected 10-$\mu$m feature and 13 yield a mm slope,
	more than doubling the sample of sources studied in \citet{lommen:2007}.
	\begin{table*}
	 \caption[]{List of sources observed with the SMA, ATCA, CARMA, and VLA. }
	 \label{tab: source list}
	 \centering
	  \begin{tabular}{lccclccccl}
	   	\hline
	   	\hline
	   	Source				& RA		& Dec		& Sp.~T.	& Cloud		& Spitzer	& 1~mm 			& 3~mm 			& 7~mm 			& Comments		\\
						& (J2000)	& (J2000)	&		&		&		&			&			&			&			\\
	   	\hline
		\multicolumn{10}{c}{Vela}																							\\
		\hline
	   	HBC~553				& 08 08 22.2	& -36 03 47.0	& M1.5		& Vela		& Wright	& ---			& ---			& ---			&			\\
	   	HBC~556				& 08 10 30.9	& -36 01 46.5	& M4		& Vela		& Wright	& ---			& ATCA			& ATCA			&			\\
	   	HBC~557				& 08 12 47.0	& -36 19 18.0	& K3:		& Vela		& Wright	& ---			& ATCA			& ATCA			&			\\
	   	HBC~559				& 08 13 56.1	& -36 08 02.1	& 		& Vela		& Wright	& ---			& ATCA			& ATCA			&			\\
	   	HBC~560				& 08 14 21.9	& -36 10 03.4	& K8		& Vela		& Wright	& ---			& ---			& ---			&			\\
	   	HBC~561				& 08 15 55.3	& -35 57 58.1	& K8		& Vela		& Wright	& ---			& ---			& ---			& Binary$^\mathrm{a}$	\\
	   	\hline
		\multicolumn{10}{c}{Chamaeleon}																							\\
		\hline
	   	SZ~Cha				& 10 58 16.9	& -77 17 17.6	& K0e		& Cha~I		& GTO		& ---			& ATCA			& 			&			\\
	   	Sz~32				& 11 09 53.4	& -76 34 25.5	& K4.7		& Cha~I		& c2d		& ---			& ATCA			& ATCA			&			\\
	   	\hline
		\multicolumn{10}{c}{Lupus}																							\\
		\hline
	   	IK~Lup				& 15 39 27.8	& -34 46 17.2	& K7		& Lupus~1	& Wright	& SMA			& ATCA			& ---			& Binary$^\mathrm{b}$	\\
	   	Sz~66				& 15 39 28.3	& -34 46 18.0	& M2		& Lupus~1	& Wright	& SMA			& ATCA			& ---			& Binary$^\mathrm{b}$	\\
		HM~Lup				& 15 47 50.6	& -35 28 35.3	& M4		& Lupus~1	& c2d		& SMA			& --- 			& ---			& 			\\
		Sz~73				& 15 47 56.9	& -35 14 34.7	& M0		& Lupus~1	& c2d		& SMA			& ---			& ---			& 			\\
	   	HN~Lup				& 15 48 05.2	& -35 15 52.8	& M1.5		& Lupus~1	& Wright	& SMA			& ---			& ---			&			\\
		Sz~76				& 15 49 30.7	& -35 49 51.4	& M1		& Lupus~1	& c2d		& SMA			& ---			& ---			& 			\\
	   	Sz~77				& 15 51 47.0	& -35 56 42.8	& M0		& Lupus~1	& Wright	& SMA			& ---			& ---			&			\\
	   	IM~Lup				& 15 56 09.2	& -37 56 05.9	& M0		& Lupus~2	& c2d		& SMA$^\mathrm{c}$	& ATCA$^\mathrm{c}$	& ATCA			&			\\
	   	RY~Lup				& 15 59 28.4	& -40 21 51.2 	& G0V:		& Lupus~3	& c2d		& SMA			& ATCA			& ATCA			&			\\
	   	MY~Lup				& 16 00 44.6	& -41 55 29.6	&		& Lupus~4	& Wright	& SMA			& ATCA			& ATCA			&			\\
	   	EX~Lup				& 16 03 05.5	& -40 18 25.3	& M0		& Lupus~3	& c2d		& SMA			& ---			& ---			&			\\
	   	Sz~91				& 16 07 11.6	& -39 03 47.1	& M0.5		& Lupus~3	& ---		& SMA			& --- 			& ---			& Cold disc		\\
	   	Sz~96				& 16 08 12.6	& -39 08 33.3	& M1.5		& Lupus~3	& c2d		& SMA			& ---			& ---			&			\\
		Sz~102				& 16 08 29.7	& -39 03 11.0	& K:e		& Lupus~3	& c2d		& SMA			& ---			& ---			&			\\
	   	Sz~111				& 16 08 54.7	& -39 37 43.1	& M1.5		& Lupus~3	& ---		& SMA			& ATCA			& ATCA			& Cold disc		\\
	   	SSTc2d~J161029.57-392214.7	& 16 10 29.6	& -39 22 14.7	&		& Lupus~3	& c2d		& SMA			& ---			& ---			& Cold disc		\\
	   	SSTc2d~J161159.81-382338.5	& 16 11 59.8	& -38 23 38.5	&		& Lupus~3	& c2d		& SMA			& ---			& ---			&			\\
	   	RX~J1615.3-3255			& 16 15 20.2	& -32 55 05.0	& K5		& Isolated	& c2d		& SMA			& ATCA			& ATCA			&			\\
	   	\hline
		\multicolumn{10}{c}{Serpens}																							\\
		\hline
	   	VV~Ser				& 18 28 47.9	& +00 08 40.0	& A2e		& Serpens	& c2d		& CARMA			& CARMA			& VLA			& Herbig Ae		\\
	   	SSTc2d~J182850.20+000949.7	& 18 28 50.2	& +00 09 49.7	& M5$\pm$4	& Serpens	& c2d		& CARMA			& CARMA			& VLA			& 			\\
	   	SSTc2d~J182858.08+001724.4	& 18 28 58.1	& +00 17 24.4	& G3$\pm$5	& Serpens	& c2d		& CARMA			& CARMA			& ---			& Cold disc		\\
	   	SSTc2d~J182900.88+002931.5	& 18 29 00.9	& +00 29 31.5	& K7$\pm$2	& Serpens	& c2d		& CARMA			& CARMA			& VLA			&			\\
	   	CoKu~Ser-G3			& 18 29 01.8	& +00 29 54.6	& K0$\pm$7	& Serpens	& c2d		& CARMA			& CARMA			& VLA			&			\\
	   	IRAS 18268-0025			& 18 29 28.1	& -00 22 58	& 		& Serpens	& c2d		& CARMA			& CARMA			& VLA			&			\\
	   	SSTc2d~J182936.19+004216.7	& 18 29 36.2	& +00 42 16.7	& F9$\pm$5	& Serpens	& c2d		& CARMA			& CARMA			& ---			&			\\
	   	SSTc2d~J182944.10+003356.1	& 18 29 44.1	& +00 33 56.1	& M0$\pm$1.5	& Serpens	& Pontoppidan	& CARMA			& CARMA			& ---			& Cold disc		\\
	   	EC~82 	     	       		& 18 29 56.8	& +01 14 46.0	& M0		& Serpens 	& c2d	      	& CARMA		      	& CARMA			& VLA		      	& 		      	\\
	   	EC~90 	     	       		& 18 29 57.5	& +01 14 07	& M4$\pm$2	& Serpens 	& c2d	     	& CARMA		      	& CARMA			& VLA		      	& Binary$^\mathrm{d}$	\\
	   	EC~97      			& 18 29 58.2	& +01 15 22	& 		& Serpens 	& c2d	      	& CARMA		      	& --- 		      	& VLA		      	& 		      	\\
	   	GSC~00446-00153			& 18 30 06.9	& +00 42 34	& F3V		& Serpens	& ---		& CARMA			& CARMA			& ---			&			\\
	   	\hline
		\multicolumn{10}{c}{Corona Australis}																						\\
		\hline
	   	S~CrA				& 19 01 08.6	& -36 57 20.0	& K3		& CrA		& Wright	& SMA			& ATCA			& ATCA			& Binary$^\mathrm{e}$	\\
	   	DG~CrA				& 19 01 55.2	& -37 23 40.5	& 		& CrA		& Wright	& SMA			& ATCA			& ATCA			&			\\
	   	VV~CrA				& 19 03 06.7	& -37 12 49.7	& K7		& CrA		& Wright	& SMA			& ATCA			& ATCA			& Binary$^\mathrm{f}$	\\
	   	\hline
	  \end{tabular}
	 \begin{list}{}{}
		\item[$^\mathrm{a}$]	Separation $0\farcs63$ \citep{correia:2006}.
	  	\item[$^\mathrm{b}$]	IK~Lup (Sz~65) and Sz~66 form a binary with a separation of $6\farcs4$ (this work).
	  	\item[$^\mathrm{c}$]	Published in \citet{lommen:2007}. 
	  	\item[$^\mathrm{d}$]	Separation $1\farcs5$ \citep[e.g.,][and references therein]{ciardi:2005}.
	  	\item[$^\mathrm{e}$]	Separation $1\farcs3$ \citep[e.g.,][and references therein]{forbrich:2007}.
	  	\item[$^\mathrm{f}$]	Separation $2\farcs0$ (this work).
	 \end{list} 
	\end{table*}

\subsection{SMA observations}\label{sect: sma observations}

	15 single sources and one binary were observed with the SMA for the project	2007B-S033. The 
	observations were carried out on 14 March and 19 April 2008. The data of 14 March were unusable due to phase 
	instabilities and the track was reobserved on 7 May 2009.
	On 19 April 2008, the phases were stable 
	and the zenith optical depth at 225~GHz was around $\tau_{225} = 0.13$ all through the night. The synthesised beam
	was about $4.8 \times 2.8$~arcsec (natural 
	weighting). On 7 May 2009, the phases were stable and $\tau_{225}$ was low with values ranging from 0.05 to 0.08.
	The synthesised beam was about $4.1 \times 2.2$~arcsec.
	The two sidebands were combined into one continuum channel to improve the signal-to-noise ratio, resulting in an 
	effective wavelength of 1.33~mm.

	The sources VV~CrA (binary), S~CrA (binary), and DG~CrA (single source) were observed as part of the SMA ``filler'' project 2008A-S111 on
	1 October 2008. Only six of the eight antennas were available for this track. However, $\tau_{225} \approx 0.1$
	and the phases were stable, resulting in extremely good data. The synthesised beam of the resulting maps was
	about $5.0 \times 2.1$~arcsec (natural weighting). The correlator was tuned to 218 and 228~GHz. Combination of 
	the two sidebands resulted in an effective wavelength of 1.35~mm. 
	
	The absolute flux calibration of the first track (19 April 2008) was carried out on Mars and the resulting fluxes are estimated to be 
	accurate to about 20\%. The second and third tracks (1 October 2008 and 7 May 2009) were flux calibrated on Callisto. The uncertainty in the
	absolute fluxes for those tracks is estimated to be 15\% or better.
	
	Hence, a total of 16 single sources and three binaries located in the Lupus star-forming region were observed with the SMA for this
	project. The sources are listed in Table~\ref{tab: source list}, a detailed log of the observations is given in Table~\ref{tab:
	observation log SMA}, and detailed results are presented in Table~\ref{tab: results log SMA} and Fig.~\ref{fig: SMA UVdist}.
	
\subsection{ATCA observations}\label{sect: atca observations}

	The data for the ATCA project C1794 were taken over the period July to August 2008 when the array was in the H214
	configuration. A total of 15 sources were observed: 14 sources (including the binary IK~Lup+Sz~66) were measured
	at 3~mm and 11 sources at 7~mm. The sources are listed in Table~\ref{tab: source list}, a detailed log of the observations is given in
	Table~\ref{tab: observation log ATCA}, and detailed results are presented in Table~\ref{tab: results log ATCA} and Figs.~\ref{fig: ATCA
	UVdist 3mm} and \ref{fig: ATCA UVdist 7mm}. The weather changed considerably over the course of the observations. A short
	indication of the circumstances for each day is included in Appendix~\ref{app: observations}.
	Physical baselines ranged from 82 to 247 metres, resulting in synthesised beam sizes of about 2~arcsec at 3~mm
	and about 4~arcsec at 7~mm. Combining the two sidebands in the 3 mm band resulted in an effective wavelength of 
	3.17~mm, those taken in the 7~mm band in an effective wavelength of 6.82~mm. 
	
	The absolute flux calibration for the first track was carried out on Mars, whereas the flux calibration for the other tracks was carried 
	out on Uranus. Only the shortest baselines were taken into account when determining the absolute gain offset so as to minimise the possible 
	effect of the planets' being resolved. Furthermore, the planets were observed at elevations close to those at which the gain calibrators were 
	observed. Overall, the uncertainty in the absolute fluxes is estimated to range from 15 to about 25\%.

\subsection{CARMA observations}\label{sect: carma observations}

	For this work, eleven single sources and one binary located in Serpens were observed with CARMA at 1 and 3~mm in the period April to June 
	2008 for project c0165. The sources are listed in Table~\ref{tab: source list}, a detailed log of the observations is 
	given in Table~\ref{tab: observation log CARMA}, and detailed results are presented in Table~\ref{tab: results log 
	CARMA} and Figs.~\ref{fig: CARMA UVdist 1mm} and \ref{fig: CARMA UVdist 3mm}. Weather conditions varied over the 
	course of the observations, with a typical water path
	length of 3--6 mm. 
	
	The gain calibrator originally selected for the observations at 1~mm, QSO J1743-038, turned out to 
	be too weak to perform a decent gain calibration, rendering most of the C-configuration observations unusable. 
	For the second part of the observations the telescope was in the D configuration (baselines 11-148 metres),
	yielding a synthesised beam of about $3 \times 2$~arcsec at 1~mm and about $6 \times 4$~arcsec at 3~mm. The 
	effective wavelength of the 1~mm-band observations was 1.33~mm, that of the 3~mm-band observations 3.15~mm.
	
	The absolute fluxes were calibrated on the quasars QSO J2253+161 (3c454.3), QSO J1229+020 (3c273), and QSO J1256-057 (3c279),
	whose fluxes were bootstrapped from planet observations on short baselines on dates as close as possible to the observation
	dates. The
	fluxes of these quasars vary considerably over the course of weeks to months at 1 and 3~mm, but day-to-day variations are usually less than 
	10\%. Taking this into account, the effective uncertainty in the absolute fluxes for our target sources is estimated to be less than 30\%.

\subsection{VLA observations}\label{sect: vla observations}

	Of the sources in the Serpens star-forming region observed with CARMA, seven single sources and the binary EC~90 were observed with the 
	VLA at 7~mm and at 1.3, 3.6, and 6.3~cm under programme AL720. The sources are listed in Table~\ref{tab: source list}, a detailed log of 
	the observations is given in Table~\ref{tab: observation log VLA}, and detailed results are presented in Table~\ref{tab: results log 
	VLA}. The observations were carried out from 10--15 March 2008, 
	when the array was in the C configuration, with baselines of up to 3.6~km and a synthesised beam of about 0.5 arcsec 
	at 7~mm. All observations were performed in the default continuum mode in which, at each frequency, the full 100-MHz 
	bandwidth was used in two adjacent 50 MHz bands. Although weather conditions were good in general, a few hours 
	of observing time were lost at the end of the last two tracks due to high winds. 
	
	The VLA data were flux calibrated on the quasar QSO J1331+305 (3c286). The flux as a function of wavelength is modelled by the AIPS
	reduction package. The resulting uncertainty in the absolute flux calibration is estimated to be about 20\% at 7~mm and 1.3~cm and better
	than 10\% at 3.6 and 6.3~cm.

\section{Results}\label{sect: results}

\subsection{Mm and cm source fluxes and dust disc masses}\label{sect: source fluxes}

	A full log of the results is listed in Appendix~B. The results of the interferometric observations at 1, 3, and 7~mm are listed in 
	Table~\ref{tab: results}. 
	\begin{table*}
	 \caption[]{Fluxes from point-source fits in the ($u, v$) plane obtained from interferometric data and single-dish 1.20-1.27~mm SEST fluxes.}
	 \label{tab: results}
	 \centering
	  \begin{tabular}{lcccccccc}
	   \hline
	   \hline
	   Source					& \multicolumn{2}{c}{1.3~mm$^\mathrm{a}$}	& \multicolumn{2}{c}{3.2~mm$^\mathrm{b}$}	& \multicolumn{2}{c}{6.8~mm}		& \multicolumn{2}{c}{SEST 1.20-1.27~mm$^\mathrm{c}$}	\\
	   						& Flux			& rms			& Flux			& rms			& Flux			& rms		& Flux			& rms			\\
							& (mJy)			& (mJy/bm)		& (mJy)			& (mJy/bm)		& (mJy)			& (mJy/bm)	& (mJy)			& (mJy)			\\
	   \hline
	   HBC~556					& ---      		& ---	  		& $<3.7^\mathrm{d}$  	& 1.2			& $<0.7^\mathrm{d}$	& 0.22		& ---			& ---			\\
	   HBC~557					& ---      		& ---	  		& $<3.2^\mathrm{d}$  	& 1.1			& $<0.6^\mathrm{d}$	& 0.18		& ---			& ---			\\
	   HBC~559					& ---      		& ---	  		& $<2.9^\mathrm{d}$  	& 1.0			& $<0.3^\mathrm{d}$	& 0.09		& ---			& ---			\\
	   \hline
	   SZ~Cha					& ---      		& ---	  		& 2.3	  		& 0.4			& ---			& ---		& 77.5			& 20.3			\\
	   Sz~32					& ---      		& ---	  		& $<2.9^\mathrm{d}$	& 1.0			& 0.77			& 0.14		& 93.1			& 20.8			\\
	   \hline
	   IK~Lup					& 28			& 2.8			& 3.4			& 0.4			& ---			& ---		& 56			& 10			\\
	   Sz~66					& $<8^\mathrm{d}$	& 2.8			& 2.2			& 0.4			& ---			& ---		& 47			& 12			\\
	   HM~Lup					& $<10^\mathrm{d}$	& 3.4			& ---			& ---			& ---			& ---		& $<45^\mathrm{d}$	& 15			\\
	   Sz~73~a					& 16			& 2.9			& ---			& ---			& ---			& ---		& 26			& 8			\\
	   Sz~73~b					& 16			& 2.9			& ---			& ---			& ---			& ---		& ---			& ---			\\
	   HN~Lup					& 15			& 3.0			& ---			& ---			& ---			& ---		& $<51^\mathrm{d}$	& 17			\\
	   Sz~76					& 12			& 3.3			& ---			& ---			& ---			& ---		& $<45^\mathrm{d}$	& 15			\\
	   Sz~77					& $<10^\mathrm{d}$	& 3.2			& ---			& ---			& ---			& ---		& $<45^\mathrm{d}$	& 15			\\
	   IM~Lup					& 188		  	& 4.3			& 8.9	  		& 1.3			& 2.2			& 0.16		& 260			& 9			\\
	   RY~Lup					& 78			& 4.9	  		& 2.8	  		& 0.7			& $<0.6^\mathrm{d}$	& 0.21		& ---			& ---			\\
	   MY~Lup					& 56      		& 3.4	  		& 8.7	  		& 0.4			& $<0.6^\mathrm{d}$	& 0.20		& ---			& ---			\\
	   EX~Lup					& 19			& 3.9			& ---			& ---			& ---			& ---		& ---			& ---			\\
	   Sz~91					& $<13^\mathrm{d}$	& 4.3			& ---			& ---			& ---			& ---		& $<27^\mathrm{d}$	& 9			\\
	   Sz~96					& $<13^\mathrm{d}$	& 4.2			& ---			& ---			& ---			& ---		& $<45^\mathrm{d}$	& 15			\\
	   Sz~102					& $<11^\mathrm{d}$	& 3.8			& ---			& ---			& ---			& ---		& $<30^\mathrm{d}$	& 10			\\
	   Sz~111					& 49			& 4.8	  		& 5.7	  		& 0.7			& $<0.6^\mathrm{d}$	& 0.19		& ---			& ---			\\
	   SSTc2d~J161029.57-392214.7			& $<13^\mathrm{d}$	& 4.4			& ---			& ---			& ---			& ---		& ---			& ---			\\
	   SSTc2d~J161159.81-382338.5			& $<13^\mathrm{d}$	& 4.2			& ---			& ---			& ---			& ---		& ---			& ---			\\
	   RX~J1615.3-3255				& 132			& 3.9	  		& 6.7	  		& 0.6			& $<0.5^\mathrm{d}$	& 0.17		& ---			& ---			\\
	   \hline
	   EC~82					& $<15.7^\mathrm{d}$	& 5.2	 		& $<2.9^\mathrm{d}$	& 1.0			& $<0.5^\mathrm{d}$	& 0.2		& ---			& ---			\\
	   EC~90					& 95.7	 		& 10.4	 		& $11.5$		& 1.2			& $<1.0^\mathrm{d}$	& 0.3		& ---			& ---			\\
	   EC~97					& $<23.3^\mathrm{d}$	& 7.8			& ---			& ---			& $<0.6^\mathrm{d}$	& 0.2		& ---			& ---			\\
	   SSTc2d~J182900.88+002931.5			& 26.4  		& 4.9	 		& 3.4			& 0.5			& $<0.7^\mathrm{d}$	& 0.2		& ---			& ---			\\
	   IRAS 18268-0025				& $<15.6^\mathrm{d}$  	& 5.2	 		& $<1.9^\mathrm{d}$	& 0.6			& ---			& ---		& ---			& ---			\\
	   CoKu~Ser-G3					& $<17.1^\mathrm{d}$  	& 5.7	 		& $<1.8^\mathrm{d}$	& 0.6			& $<1.2^\mathrm{d}$	& 0.4		& ---			& ---			\\
	   SSTc2d~J182858.08+001724.4			& $<24.6^\mathrm{d}$	& 8.2	 		& $<1.9^\mathrm{d}$	& 0.6			& ---			& ---		& ---			& ---			\\
	   VV~Ser					& $<14.8^\mathrm{d}$	& 4.9	 		& $<1.8^\mathrm{d}$	& 0.6			& $<0.7^\mathrm{d}$	& 0.2		& ---			& ---	 		\\
	   SSTc2d~J182850.20+000949.7			& $<22.8^\mathrm{d}$	& 7.6	 		& $<1.9^\mathrm{d}$	& 0.6			& $<0.6^\mathrm{d}$	& 0.2		& ---			& ---			\\
	   SSTc2d~J182944.10+003356.1			& $<15.0^\mathrm{d}$	& 5.0	 		& $<1.7^\mathrm{d}$	& 0.6			& ---			& ---		& ---			& ---			\\
	   SSTc2d~J182936.19+004216.7			& $<8.7^\mathrm{d}$  	& 2.9	 		& $<2.9^\mathrm{d}$	& 1.0			& ---			& ---		& ---			& ---			\\
	   GSC~00446-00153				& 90.8  		& 3.5	 		& 6.8			& 1.0			& ---			& ---		& ---			& ---			\\
	   \hline
	   VV~CrA					& 376			& 4.5	  		& 26.8	 		& 1.1			& 8.3			& 0.25		& 469, {\it 584}	& 21			\\
	   S~CrA					& 303			& 3.2	  		& 24.9  		& 1.0			& 3.7			& 0.20		& {\it 290}		& ---			\\
	   DG~CrA					& $<6.6$		& 2.2	  		& $<2.5^\mathrm{d}$	& 0.8			& $<0.5^\mathrm{d}$	& 0.16		& ---			& ---			\\
	   \hline
	  \end{tabular}
	 \begin{list}{}{}
	  \item[$^\mathrm{a}$]	1~mm band observations are at 1.33~mm (SMA, Lupus and CARMA, Serpens) and at 1.35~mm (SMA, Corona Australis).
	  \item[$^\mathrm{b}$]	3~mm band observations are at 3.17~mm (ATCA, Lupus and Corona Australis) and at 3.15~mm (CARMA, Serpens).
	  \item[$^\mathrm{c}$]	SEST fluxes are from \citet[][Lupus]{nurnberger:1997}, \citet[][Chamaeleon]{henning:1993}, and 
	  	\citet[][Corona Australis]{henning:1994}, with an adopted centre frequency of 236~GHz (1.27~mm). The values in italic are from
		\citet{chini:2003}, with an adopted centre frequency of 250~GHz (1.20~mm).
	  \item[$^\mathrm{d}$]	Quoted value is 3$\sigma$ upper limit.
	 \end{list} 
	\end{table*}
	
	A total of 16 single sources and three binaries in Lupus are observed with the SMA. Nine of the single sources are detected and one of those, Sz~73, turned
	out to harbour two sources, with a projected separation of about 4~arcsec. It is possible that the detection of Sz~73 with SEST \citep{nurnberger:1997} 
	included both sources. The binaries VV~CrA and S~CrA are detected and unresolved. Of the binary system IK~Lup (Sz~65) and 
	Sz~66, only IK~Lup is detected, although a second peak is detected at 2~arcsec from the 2MASS position of Sz~66. Sz~66 was previously
	detected with a $S/N$ of almost four using the SEST bolometer. 
	All sources in Lupus observed with the ATCA at 3.2~mm are detected; the binary system IK~Lup and Sz~66 remained unresolved.
	Only one Lupus source, IM~Lup, is detected at 6.8~mm. MY~Lup would
	have been detected at 6.8~mm with a signal-to-noise ratio of about ten if it had a similar mm slope as IM~Lup. 
	
	None of the three sources in the Gum nebula observed with the ATCA at 3.2 and at 6.8~mm are detected at either wavelength down to
	3$\sigma$ upper limits of $\sim$3~mJy at 3.2~mm and of $\sim$0.5~mJy at 6.8~mm. This can be attributed to the large distance between us and this
	star-forming region. If the sources in the Gum nebula had similar luminosities as those in the Lupus clouds, they would have had a flux of 
	$\sim$0.7~mJy at 3.2~mm, which is below the noise level. Note that, although the Vela molecular ridge has been observed at mm wavelengths 
	\citep{massi:1999,massi:2007}, no published mm continuum data of the Gum nebula exist in the literature.
	
	The source SZ~Cha is detected at 2.3~mJy at 3.2~mm. Sz~32 is not detected down to a 3$\sigma$ upper limit of 
	2.9~mJy at 3.2~mm. It is, however, detected with a flux of 0.77~mJy at 6.8~mm.
	
	VV~CrA and S~CrA are clearly detected at 1.3~mm with the SMA, with fluxes of 376 and 303~mJy. DG~CrA, however, is not detected, down to a
	3$\sigma$ upper limit of only 6.6~mJy. VV~CrA and S~CrA are also easily detected with the ATCA at 3 and 7~mm. 
	
	Of the sources in the Serpens star-forming region that were observed with the CARMA, only three are detected: the single sources SSTc2d~J182900.88+002931.5
	and GSC~00446-00153 and the binary system EC~90, which remained unresolved. This can in part be
	explained by the distance to the star-forming region in Serpens, which is larger than those in Chamaeleon, Lupus, and Corona Australis.
	Furthermore, some of the sources, of which six are new {\em Spitzer} sources, may have an intrinsically lower luminosity. None of the 
	sources are detected at 6.8~mm using the VLA.
	
	Four cold discs are observed at 1.3 and 3.2~mm for this work. Only one of those, Sz~111, is detected. Unfortunately, Sz~111
	was not observed with the {\em Spitzer} IRS.
	
	All four binaries that are observed at 1.3 and 3.2~mm are detected at both wavelengths. However, in the case of the binary consisting 
	of the stars IK~Lup and Sz~66, only the former is detected at 1.3~mm. EC~90 and S~CrA remain unresolved. The binary IK~Lup+Sz~66 is
	resolved with the ATCA at 3.2~mm. VV~CrA is resolved with a binary separation of 2.0~arcsec with the ATCA at 3.2~mm if the source is 
	imaged using uniform weighting (optimised for resolution). However, this binary remains unresolved with the SMA at 1.3~mm (beam size
	$4.7 \times 1.9$~arcsec) and with the ATCA at 6.8~mm (beam size $5.3 \times 3.0$~arcsec).
	
	The detection rate of the sources observed in this study is rather low. This can in part be understood by the distance to the star-forming 
	regions, with the Serpens star-forming region being almost twice as far away as the Taurus-Auriga star-forming region and the Gum nebula in Vela 
	almost three times as far away. This reduces the observed flux for similar sources by a factor of about four to nine. The low detection rate for
	Lupus is largely a selection effect: most of the brightest sources had been observed before \citep{lommen:2007}. These previously detected 
	sources were not reobserved for this work, but their published values will be included in the analysis below.
	
	Dust disc masses are obtained from the fluxes at 3.2~mm under the rather crude assumptions of an isothermal disc and a fixed opacity.
	Assuming also optically thin mm emission, the dust disc mass is given by $M_{\rm disc} = F_\nu D^2 / \kappa_\nu B_\nu$($T_{\rm dust}$), where
	$D$ is the distance to the source, $\kappa_\nu$ the dust opacity \citep[taken to be 0.9~cm$^2$~g$^{-1}$, cf.][]{beckwith:1990},
	and $B_\nu$($T_{\rm dust}$) the brightness at the frequency $\nu = 94$~GHz for a dust temperature $T_{\rm dust}$, as given by the Planck function.
	We assume a dust temperature $T_{\rm dust} = 25$~K and find dust disc masses ranging from $\sim$0.4 to $\sim 7 \times 10^{-4}$~M$_\odot$. The dust disc masses
	are presented in Table~\ref{tab: spectral slopes}.

\subsection{Millimetre slopes}\label{sect: millimetre slopes}

	The fluxes at 1, 3, and 7~mm can be combined to obtain the spectral index $\alpha$, where $F_\nu \propto \nu^{\alpha}$. We are 
	interested in the emission coming from the dusty disc. However, at 7~mm, other emission mechanisms may contribute significantly to the
	flux. Sources may include an ionised wind or chromospheric magnetic activity. \citet{rodmann:2006} compare their fluxes at 7~mm to 
	those at 3 and 6~cm and claim that about 20\% of the emission at 7~mm is due to free-free emission. On the other hand, 
	\citet{lommen:2009} find that the emission at 7~mm can be entirely attributed to dust emission for a small sample of three sources. It
	is possible that the emission due to, e.g., an ionised wind, is independent of the disc mass and thus the relative contribution 
	from such a wind will be larger for young stellar objects that are weaker at mm wavelengths. This could explain the findings of 
	\citet{lommen:2009}, who monitored some of the strongest pre-main-sequence mm emitters in the southern sky. However, a larger and more
	sensitive survey at mm to cm wavelengths is required before more quantitative statements on this subject can be made. Since we do not 
	have fluxes at all three wavelengths for most sources, separate indices will be obtained between 1 and 3~mm and between 3 and 7~mm. 
	The results are given in Table~\ref{tab: spectral slopes}.
   \begin{table*}
   	\caption{Spectral slopes at mm wavelengths, dust disc masses, and properties of the 10-$\mu$m silicate feature. }
	\label{tab: spectral slopes}
	\centering
	  \begin{tabular}{lcccccc}
	   \hline
	   \hline
	   Source					& $\alpha_{1-3}$			& $\alpha_{3-7}$			& Dust disc mass$^\mathrm{a}$	& ($F_{10}-F_{\rm cont}$)/$F_{\rm cont}$	& $F^{10\mu m}_{peak}$	& $F_{11.3}$/$F_{9.8}$	\\
	   						&					&					& ($10^{-4}$M$_\odot$)		&						&			&			\\
	   \hline
	   HBC~553					& ...					& ...					& ...				& 0.22						& 1.43			& 0.86			\\
	   HBC~556					& ...      				& ...	  				& $<5.1$			& 0.40						& 1.64			& 0.99			\\
	   HBC~557					& ...      				& ...	  				& $<4.4$			& 0.21						& 1.37			& 0.95			\\
	   HBC~559					& ...      				& ...	  				& $<4.0$			& 0.39						& 1.60			& 0.85			\\
	   HBC~560					& ...					& ...					& ...				& 0.29						& 1.44			& 1.03			\\
	   HBC~561					& ...					& ...					& ...				& 0.43						& 1.67			& 0.90			\\
	   \hline	
	   SZ~Cha					& $3.8 \pm 0.4^\mathrm{b}$		& ...	  				& 0.6				& 0.71						& 2.30			& 0.85			\\
	   Sz~32					& $>3.7^\mathrm{b}$			& $<1.8$  				& $<0.8$			& 0.14						& 1.24			& 1.11			\\
	   \hline
	   IK~Lup					& $2.7 \pm 0.3^\mathrm{b}$		& ...					& 0.7				& 0.36						& 1.51			& 0.93			\\
	   Sz~66					& $1.9 \pm 0.2^\mathrm{b}$		& ...					& 0.4				& 0.40						& 1.62			& 0.92			\\
	   HM~Lup					& ...					& ...					& ...				& 0.50						& 1.81			& 0.85			\\
	   Sz~73					& ...					& ...					& ...				& 0.26						& 1.38			& 0.95			\\
	   HN~Lup					& ...					& ...					& ...				& 0.18						& 1.29			& 1.03			\\
	   Sz~76					& ...					& ...					& ...				& 0.45						& 1.77			& 0.77			\\
	   Sz~77					& ...					& ...					& ...				& 0.43						& 1.63			& 0.94			\\
	   IM~Lup					& $3.6 \pm 0.4^\mathrm{b}$		& $1.8 \pm 0.3$				& 1.7				& 0.52						& 1.81			& 0.85			\\
	   RY~Lup					& $3.8 \pm 0.5$				& $>2.0$	  			& 0.5				& 1.10						& 3.16			& 0.66			\\
	   MY~Lup					& $2.1 \pm 0.5$				& $>3.5$  				& 1.7				& 0.54						& 1.87			& 0.79			\\
	   EX~Lup					& ...					& ...					& ...				& 0.54						& 2.01			& 0.73			\\
	   Sz~96					& ...					& ...					& ...				& 0.78						& 2.30			& 0.82			\\
	   Sz~102					& ...					& ...					& ...				& 0.80						& 2.36			& 0.82			\\
	   Sz~111					& $2.5 \pm 0.4$				& $>2.9$  				& 1.1				& ...						& ...			& ...			\\
	   SSTc2d~J161029.57-392214.7$^\mathrm{h}$	& ...					& ...					& $<0.3$			& 0.29						& 1.52			& 0.89			\\
	   SSTc2d~J161159.81-382338.5			& ...					& ...					& $<0.3$			& 0.92						& 2.48			& 0.84			\\
	   RX~J1615.3-3255				& $3.4 \pm 0.5$				& $>3.4$  				& 1.3				& 1.25						& 3.04			& 0.96			\\
	   \hline	
	   VV~Ser					& ...					& ...	       				& $<1.0$			& 0.33			        		& 1.45			& 0.97			\\
	   SSTc2d~J182850.20+000949.7			& ...					& ...	       				& $<1.1$			& 0.36						& 1.57			& 0.88			\\
	   SSTc2d~J182900.88+002931.5			& $2.4 \pm 0.4$				& ...	 				& 2.0				& 0.15						& 1.27			& 0.96			\\
	   CoKu~Ser-G3					& ...					& ...	       				& $<1.0$			& -0.09			        		& 0.82			& 1.17			\\
	   IRAS~18268-0025				& ...					& ...					& $<0.7$			& 0.48						& 1.67			& 1.14			\\
	   SSTc2d~J182944.10+003356.1$^\mathrm{i}$	& ...					& ...					& $<0.7$			& 0.30						& 1.53			& 0.96			\\
	   EC~82 aka CK~3				& ...					& ...	 				& $<1.7$			& 1.41						& 3.55			& 0.67			\\
	   EC~90 aka CK~1				& $2.5 \pm 0.5$				& ...	 				& 6.7				& 0.26						& 1.40			& 1.12			\\
	   EC~97 aka CK~4				& ...					& ...					& ...				& 0.34						& 1.56			& 0.90			\\
	   GSC~00446-00153				& $3.0 \pm 0.5$				& ...	 				& 4.0				& ...						& ...			& ...			\\
	   \hline
	   S~CrA					& $2.9 \pm 0.7^\mathrm{e}$		& $2.5 \pm 0.4$				& 3.6				& 0.34						& 1.51			& 0.97			\\
	   DG~CrA					& ...					& ...	  				& $<0.4$			& 1.04						& 2.94			& 0.67			\\
	   VV~CrA					& $2.5 \pm 0.5^\mathrm{d,e,f,g}$	& $2.4 \pm 0.5^\mathrm{g}$		& 3.9				& ...						& ...			& ...			\\
	   \hline
	  \end{tabular}
	 \begin{list}{}{}
	  \item[$^\mathrm{a}$]	Dust disc masses estimated from ATCA and CARMA fluxes, assuming a dust opacity
	  	$\kappa_\nu = 0.9$~cm$^2$~g$^{-1}$ \cite[cf.][]{beckwith:1990}, and a dust temperature $T_{\rm dust} = 25$~K.
	  \item[$^\mathrm{b}$]	Using the SEST 1.27~mm (single-dish) flux from \citet{nurnberger:1997}.
	  \item[$^\mathrm{c}$]	Using the SEST 1.27~mm (single-dish) flux from \citet{henning:1993}.
	  \item[$^\mathrm{d}$]	Using the SEST 1.20~mm (single-dish) flux from \citet{chini:2003}.
	  \item[$^\mathrm{e}$]	Using the SEST 1.27~mm (single-dish) flux from \citet{henning:1994}.
	  \item[$^\mathrm{f}$]	Taking a combined flux of 50.2~mJy at 3.17~mm.
	  \item[$^\mathrm{g}$]	Using the PSF extraction method for the Spitzer IRS spectrum.
	  \item[$^\mathrm{h}$]	Observed with Spitzer IRS for Program ID 30843 (B.~Mer{\'{\i}}n PI). The full IRS spectrum will be presented
	  	in Mer{\'{\i}}n et~al. (2010, in prep.).
	  \item[$^\mathrm{i}$]	Observed with Spitzer IRS for Program ID 30223 (K.M.~Pontoppidan PI). The full IRS spectrum will be presented
	  	in Oliveira et~al. (2010, in prep.).		
	 \end{list} 
   \end{table*}	
   
	The slopes between 1 and 3~mm lie between $2.38 \pm 0.36$ and $3.83 \pm 0.46$. The opacity index $\beta$ can be calculated from the
	mm slope $\alpha$ through $\beta \approx (1 + \Delta) \times (\alpha - 2)$, where $\Delta$ is the ratio of optically thick to optically
	thin emission \citep{beckwith:1990, rodmann:2006}. \citet{rodmann:2006} and \citet{lommen:2007} found values of $\Delta \approx 0.2$
	for the sources in their samples. Adopting this value, opacity indices $\beta$ of about $0.5$ to $2.2$ are found here. The 
	Kolmogorov-Smirnov test gives a probability of 50\% that the values from this sample and that of \citet{lommen:2007} are drawn from 
	the same distribution. This rather low value is due to the steep slopes for the sources RY~Lup ($3.83 \pm 0.46$) and SZ~Cha 
	($3.78 \pm 0.43$). Note that the corresponding values for $\beta$ are $\gtrsim 2$, whereas the value for the interstellar medium is 
	$\beta_{\rm ISM} \approx 2.0$ \citep[e.g.,][]{draine:1984}. 
	
	A mm slope between 3 and 7~mm could only be determined for three sources, whereas lower limits are found for four more sources and a
	strict upper limit of $\alpha_{3-7} < 1.8$ for Sz~32. Interestingly, a lower limit of $\alpha_{1-3} > 3.7$ is found for Sz~32 
	between 1 and 3~mm. Other emission mechanisms (due to, e.g., a wind or chromospheric activity) may contribute at 7~mm. Although
	it is found that for most sources this contribution is only of the order of 20\% \citep{rodmann:2006}, it is possible that it is
	higher for Sz~32, causing the very shallow slope between 3 and 7~mm. The slopes of $\alpha_{3-7} = 2.4 \pm 0.5$ 
	and $2.5 \pm 0.4$ for VV~CrA and S~CrA are consistent with those of $\alpha_{1-3} = 2.5 \pm 0.5$ and $2.9 \pm 0.7$ and also the
	slopes between 3 and 7~mm found for RY~Lup ($> 2.0$), Sz~111 ($> 2.9$), RX~J1615.3-3255 ($> 3.4$), and MY~Lup ($> 3.5$) are consistent
	with the values between 1 and 3~mm. The slope between 3 and 7~mm for IM~Lup, however, is very shallow compared to that between 1 and
	3~mm: $\alpha_{3-7} = 1.8 \pm 0.3$ vs $\alpha_{1-3} = 3.6 \pm 0.4$. \citet{pinte:2008} found a mm spectral index of $2.80 \pm 0.25$
	and their modelling results suggested that IM~Lup has grains of at least mm sizes in the disc. A shallowing of the slope beyond 3~mm 
	may indicate the presence of at least cm-sized grains. A similar effect on the cm SED was found for TW~Hya \citep{wilner:2005}.

\subsection{Results from Spitzer infrared observations}\label{sect: spitzer results}

\begin{figure*}
		\centering
		\includegraphics[width=\textwidth]{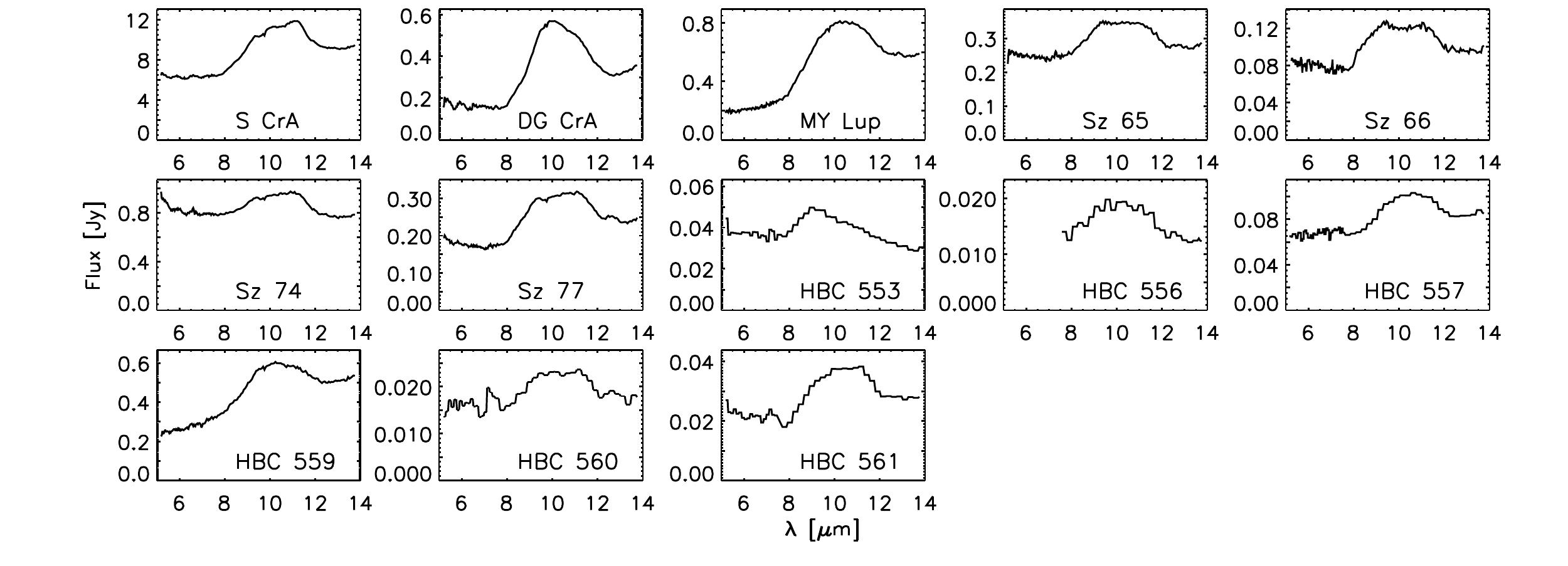}
		\caption[]{Spitzer IRS spectra from the T-Tauri stars observed for Spitzer 
				project P20611 (C.M.~Wright PI). Spectra with a maximum flux below 0.1~Jy were binned
				four times to improve the signal-to-noise ratio.}
		\label{fig: spitzer spectra}
	\end{figure*}

	The spectra of the T-Tauri stars observed for Spitzer project P20611, including sources in Lupus, Corona Australis, and the Gum 
	nebula, are published for the first time here and shown in Fig.~\ref{fig: spitzer spectra}. The spectrum of VV~CrA is saturated below 
	10~$\mu$m and excluded from the sample. The spectrum of SSTc2d~J161029.57-392214.7 (P30843, B.~Mer{\'{\i}}n PI) will be published in 
	Mer{\'{\i}}n et~al., (2010, in prep.). The spectrum of SSTc2d~J182944.11+003356.1 (P30223, 
	K.M.~Pontoppidan PI) will be published in Oliveira et~al. (2010, in prep.). 
	
	The 10-$\mu$m silicate features were analysed in the
	ways of both \citet{furlan:2006} and \citet{kessler-silacci:2006}. \citet{furlan:2006} fitted a third-order polynomial to the
	continuum around the 10-$\mu$m feature and determined the integrated flux above and below the continuum. The strength of the 
	10-$\mu$m feature was then defined as the ratio of the integrated flux due to the feature divided by the integrated flux due to the
	continuum, ($F_{10} - F_{\rm cont}$)/$F_{\rm cont}$, resulting in a strength larger than 0 for a feature in emission. 
	\citet{kessler-silacci:2006} determined the continuum in three different ways, depending on the full mid-infrared SED and which data
	were available for each source, and subsequently determined the normalised spectra $S_\nu$ according to
	\begin{equation}
		S_\nu = 1 + \frac{(F_\nu - F_{\nu, c})}{<F_{\nu, c}>},
	\end{equation}
	where $F_\nu$ is the observed spectrum, $F_{\nu, c}$ is the fitted continuum, and $<F_{\nu, c}>$ is the frequency-averaged continuum
	flux (see their paper for details). They defined the strength of the 10-$\mu$m feature as the maximum value of $S_\nu$ between 5 and
	13~$\mu$m, $S^{10\mu m}_{peak}$, resulting in a value larger than 1 for a feature in emission. Furthermore, 
	\citet{kessler-silacci:2006} defined the shape of the 10-$\mu$m feature as the ratio $S_{11.3}$/$S_{9.8}$.
	
	For this work, the continuum was consistently chosen for all sources by fitting a third-order polynomial to data between 5.0 and 
	7.5~$\mu$m and between 13.0 and 16.0~$\mu$m \citep[cf.][]{furlan:2006}. The regular continuum was used rather than the
	frequency-averaged continuum, resulting in the peak strength $F^{10\mu m}_{peak}$ and the shape $F_{11.3}$/$F_{9.8}$. This does not
	change the results significantly \citep[see][]{kessler-silacci:2006}.
	
\begin{figure}
		\centering
		\includegraphics[width=\columnwidth]{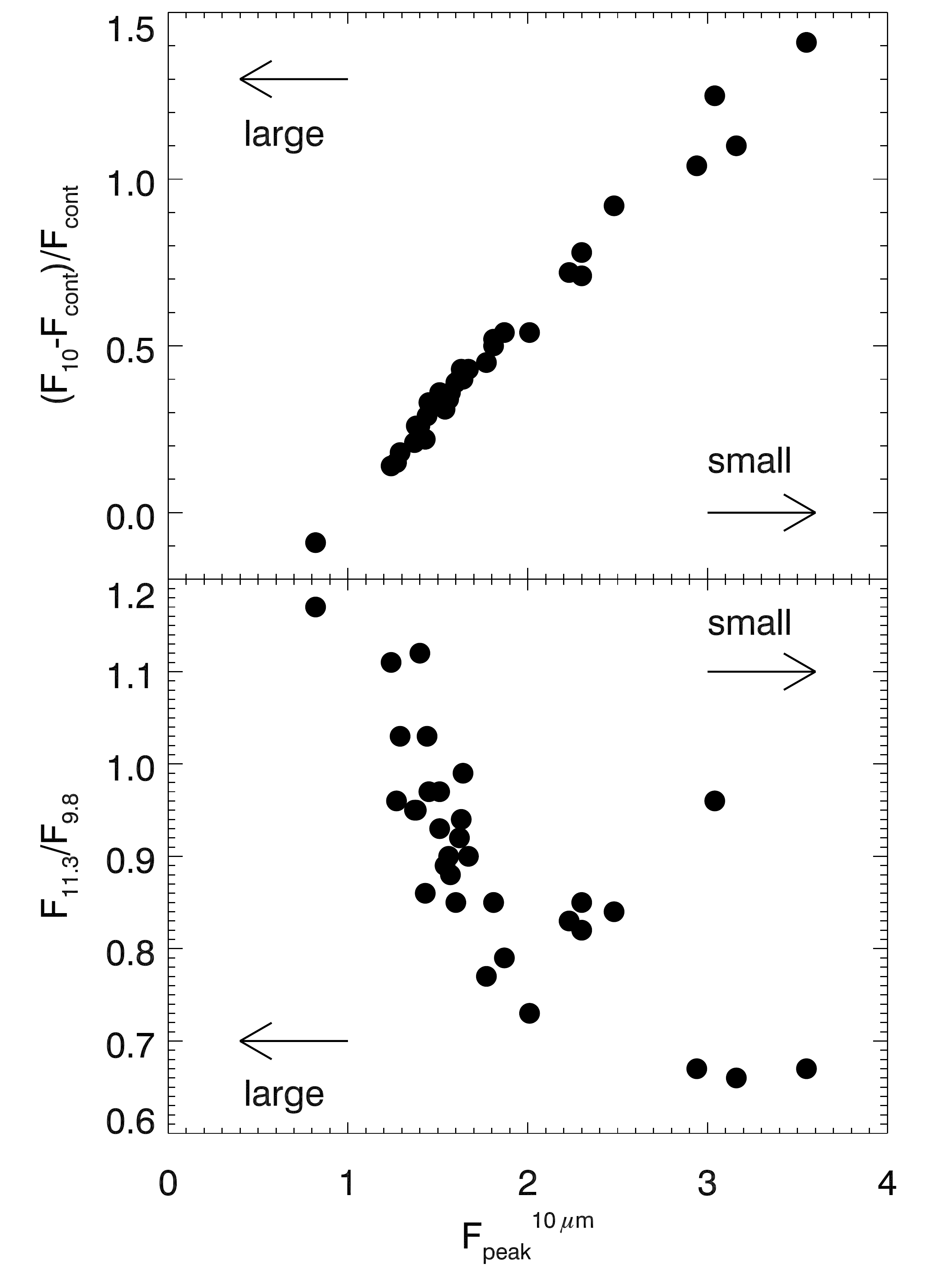}
		\caption[]{{\em Upper panel}: the peak strength of the 10-$\mu$m
				feature as defined in \citet{kessler-silacci:2006} vs the strength as defined in
				\citet{furlan:2006}. {\em Lower panel}: the peak strength of the 10-$\mu$m 
				feature vs the shape as in \citet{kessler-silacci:2006}. Included are the
				sources observed in this work for which the 10-$\mu$m feature could be obtained.
				The object towards the top-right off the general trend in the lower panel
				is the Lupus source RX~J1615.3-3255, which is isolated from the Lupus clouds.}
		\label{fig: peak vs stuff}
	\end{figure}

	The results are listed in Table~\ref{tab: spectral slopes} and shown in Fig.~\ref{fig: peak vs stuff}. The upper panel of 
	Fig.~\ref{fig: peak vs stuff} gives ($F_{10} - F_{\rm cont}$)/$F_{\rm cont}$ vs $F^{10\mu m}_{peak}$, showing a clear correlation 
	between the two definitions of the strength of the 10-$\mu$m feature. The lower panel gives $F^{10\mu m}_{peak}$ vs 
	$F_{11.3}$/$F_{9.8}$, also showing a correlation, confirming the results of \citet{kessler-silacci:2006}. This figure also confirms
	that our sample covers a large range in silicate-feature characteristics. It follows that the three
	methods to quantify the strength or shape of the 10-$\mu$m feature give completely consistent results. When comparing the 10-$\mu$m 
	feature with the mm slope in Sect.~\ref{sect: discussion} below, the strength defined as ($F_{10} - F_{\rm cont}$)/$F_{\rm cont}$ will
	be used. The source that lies towards the top and to the right of the general trend in the lower panel is RX J1615.3-3255, an isolated
	source slightly to the north of the Lupus star-forming clouds.
	
	
\subsection{10-$\mu$m feature vs mm slope}\label{sect: observation results}

\begin{figure}
		\centering
		\includegraphics[width=\columnwidth]{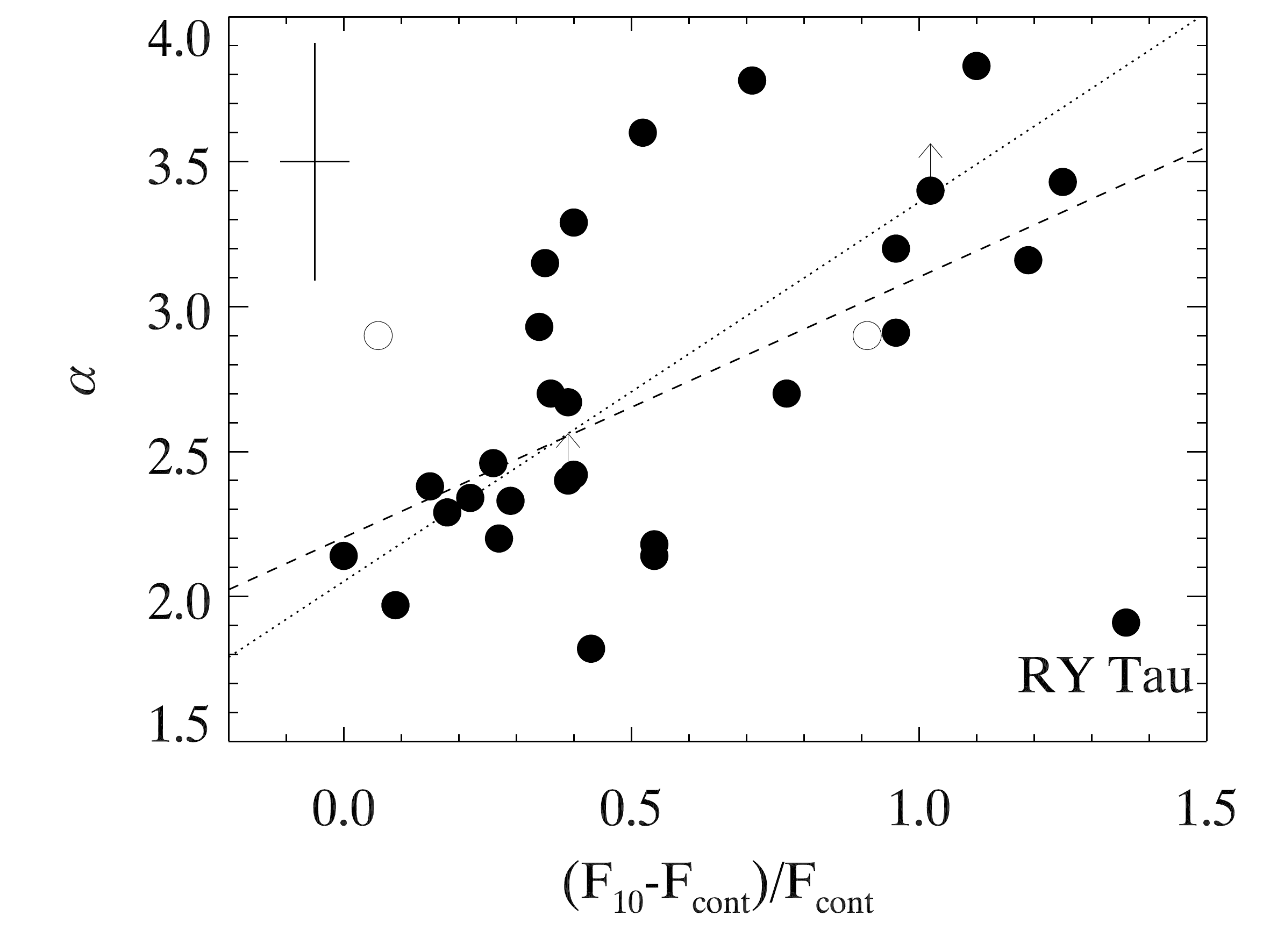}
		\caption[]{The mm slope as measured between 1 and 3~mm as a function of the strength of
				the 10-$\mu$m feature. The open symbols are for T~Cha (to the left), which does not show any silicate 
				emission and is not used in the analysis, and for CS~Cha, a circumbinary disc. The dashed line shows 
				a linear fit to all the data. The dotted line shows a
				linear fit to the data with RY~Tau excluded. Included are the
				sources studied in this work, as well as those from \citet{rodmann:2006}, \citet{andrews:2007}, and
				\citet{lommen:2007}. The cross in the upper left shows typical uncertainties.}
		\label{fig: strength vs slope}
	\end{figure}

\begin{figure}
		\centering
		\includegraphics[width=\columnwidth]{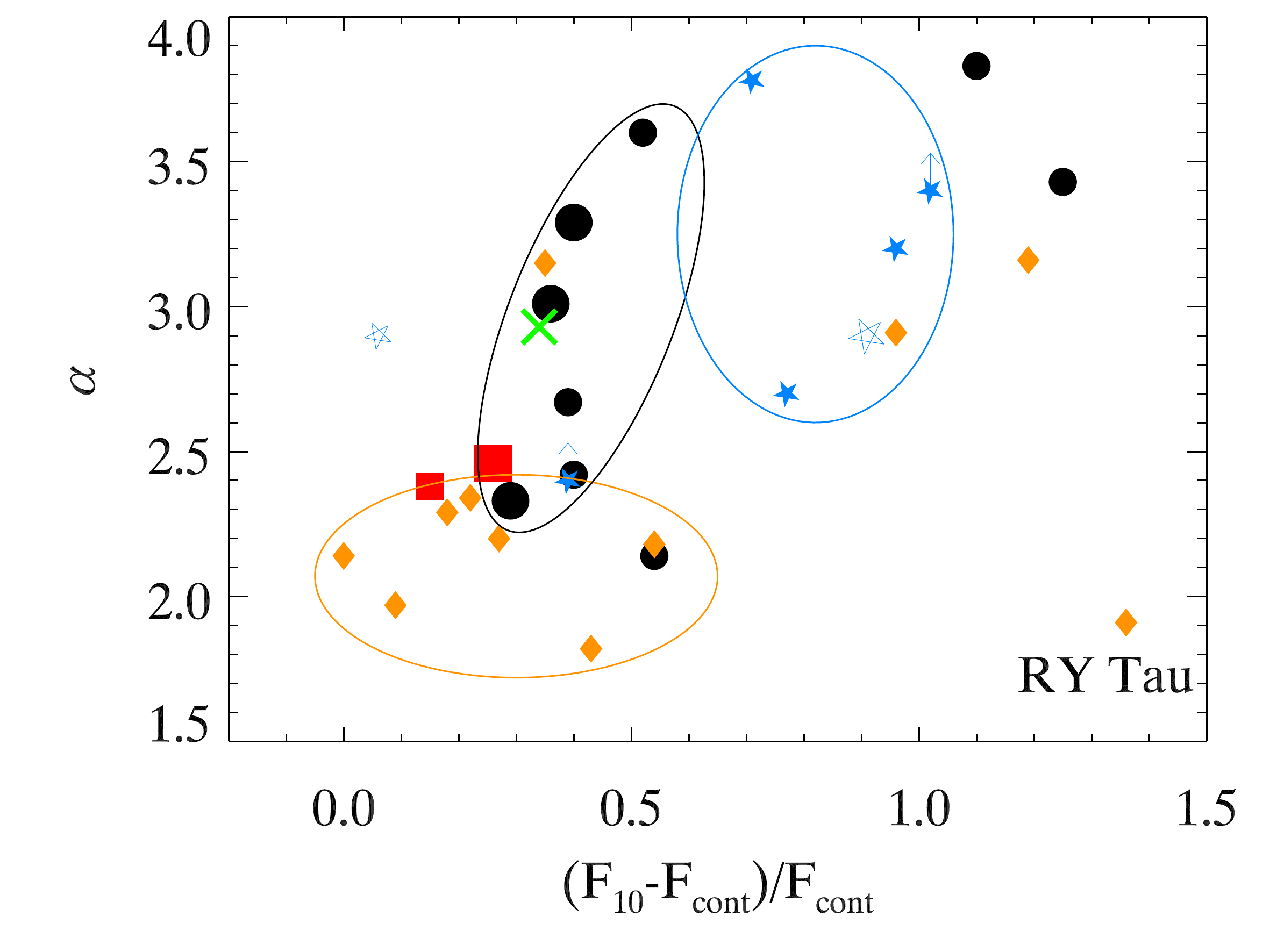}
		\caption[]{Same as Fig.~\ref{fig: strength vs slope}, with the different sources sorted
				by star-forming region: {\em filled circles}: Lupus, {\em five-pointed stars}: Chamaeleon, {\em cross}: 
				Corona Australis, {\em diamonds}: Taurus-Auriga, and {\em squares}: Serpens. The
				ellipses show the concentrations of sources located in the Taurus-Auriga star-forming region
				(lower left), the Chamaeleon~I cloud (top centre), and the Lupus~1 and Lupus~2 clouds (upper left).
				The remaining two Lupus sources in the upper right are an isolated source (RX~J1615-.3-3255, right-most dot)
				and a source from the Lupus~3 cloud (RY~Lup, upper-most dot). The small symbols designate the single
				stars and the large symbols designate multiple systems. The open five-pointed star to the left is for
				T~Cha, an evolved cold disc which shows no silicate emission around 10 $\mu$m. The open five-pointed star
				in the centre is CS~Cha, a circumbinary disc.}
		\label{fig: correlation stuff}
	\end{figure}

	Figs.~\ref{fig: strength vs slope} and \ref{fig: correlation stuff} show the mm slope $\alpha$, measured between 1 and 3~mm, as 
	a function of the strength of the 10-$\mu$m feature $(F_{10}-F_{\rm cont})/F_{\rm cont}$. Only the slope between 1 and 3~mm
	is used, to make the sample as consistent as possible. However, as noted before, the slope between 3 and 7~mm is consistent with
	that between 1 and 3~mm for most sources. Included are the sources from this
	study, as well as eleven sources located in the Taurus-Auriga star-forming region discussed in \citet{rodmann:2006} and 
	\citet{andrews:2007} for which the spectral slope between 1 and 3~mm could be determined, and the sources located in Lupus and 
	Chamaeleon discussed in \citet{lommen:2007}. The total number of sources used is 31; the complete list is given in 
	Table~\ref{tab: sources used in analysis}. In 
	Fig.~\ref{fig: correlation stuff}, the sources are sorted by their star-forming region. 
	The smaller symbols designate single stars and the larger symbols binaries (or stars that are members of a multiple system). The
	open symbol to the left is T~Cha, an evolved system that does not show any silicate emission and is not used in the analysis, and the 
	open symbol in the centre designates the ``cold disc'' CS~Cha. 
	\begin{table*}
		\caption[]{List of sources used in the analysis. }
		\label{tab: sources used in analysis}
		\centering
		\begin{tabular}{lccl}
	   		\hline
	   		\hline
	   		Source				& ($F_{10}-F_{\rm cont}$)/$F_{\rm cont}$	& $\alpha_{1-3}$			& Notes				\\
			\hline
			\multicolumn{4}{c}{This work}																\\
			\hline
			RY~Lup				& 1.10						& 3.83					&				\\
			RX~J1615.3-3255			& 1.25						& 3.43					&				\\
			IK~Lup				& 0.36						& 3.01					& Binary with Sz~66		\\
			Sz~66				& 0.40						& 3.29					& Binary with IK~Lup		\\
			IM~Lup				& 0.52						& 3.60					&				\\
			SZ~Cha				& 0.71						& 3.78					&				\\
			S~CrA				& 0.34						& 2.93					& Binary			\\
			EC~90				& 0.26						& 2.46					& Binary			\\
			SSTc2d~J182900.88+002931.5	& 0.15						& 2.38					&				\\
			\hline
			\multicolumn{4}{c}{Values taken from \citet{lommen:2007}}												\\
			\hline
			CR~Cha				& 0.96						& 3.20					&				\\
			WW~Cha				& 0.77						& 2.70					&				\\
			HT~Lup				& 0.29						& 2.33					& Binary			\\
			GW~Lup				& 0.40						& 2.42					&				\\
			RU~Lup				& 0.39						& 2.67					&				\\
			T~Cha				& 0.06						& 2.90					& Cold disc			\\
			CS~Cha				& 0.91						& 2.90					& Cold disc			\\
			DI~Cha				& 0.39						& $>2.40$				&				\\
			Glass~I				& 1.02						& $>3.40$				&				\\
			\hline
			\multicolumn{4}{c}{Sources from \citet{rodmann:2006}}											\\
			\hline
			DG~Tau				& 0.00						& 2.14					&				\\
			DO~Tau				& 0.18						& 2.29					&				\\
			\hline
			\multicolumn{4}{c}{Sources from \citet{andrews:2007}}											\\
			\hline
			AA~Tau				& 0.35						& 3.15					&				\\
			CI~Tau				& 0.54						& 2.18					&				\\
			DL~Tau				& 0.09						& 1.97					&				\\
			DM~Tau				& 0.96						& 2.91					&				\\
			DN~Tau				& 0.22						& 2.34					&				\\
			DR~Tau				& 0.27						& 2.20					&				\\
			FT~Tau				& 0.43						& 1.82					&				\\
			GM~Aur				& 1.19						& 3.16					&				\\
			RY~Tau				& 1.36						& 1.91					&				\\
			AS~205				& 0.27						& 0.67					&				\\
	   		\hline
		\end{tabular}
	\end{table*}
	
	The sources in the sample shown in Fig.~\ref{fig: strength vs slope} lie in a broad band roughly running from the lower left (shallow
	mm slope and weak 10-$\mu$m feature) to the upper right (steep mm slope and strong 10-$\mu$m feature). The sole exception is the 
	source RY~Tau, which lies in the lower right corner. The mm slope and the strength of the 10-$\mu$m feature correlate weakly for the
	full sample: the Spearman rank correlation coefficient is 0.50, with a 99.5\% confidence level. However, if the point for RY~Tau is
	excluded, the Spearman rank coefficient becomes 0.66, with a 99.99\% confidence level. Note that RY~Tau is a peculiar source: it is
	found to be a rapidly rotating UX Or-type star powering a microjet \citep[e.g.,][]{petrov:1999, agra-amboage:2009}. 
	A possible explanation for its location in the 10-$\mu$m-feature vs mm-slope diagram is a rather evolved disc in which a recent 
	collision event produced small grains. This may be similar to the effect recently observed in EX~Lup, in which a significantly more
	crystalline 10-$\mu$m feature was observed after an outburst \citep{abraham:2009}.
	RY~Tau will not be included in the further discussion.

	Fig.~\ref{fig: correlation stuff} suggests a grouping in the $\mu$m-vs-mm diagram according to parental cloud,
	with the sources from the Taurus-Auriga star-forming region more concentrated in the lower left, the Lupus
	sources more to the upper left, and the Chamaeleon sources more to the centre right. Note that the six Lupus
	sources that are on the left part of the diagram (from top to bottom: IM~Lup, Sz~66, Sz~65, RU~Lup, GW~Lup, and
	HT~Lup) are all located in the Lupus~1 and Lupus~2 clouds, whereas the remaining two Lupus sources are located in
	Lupus~3 (top-most source, RY~Lup) and off-cloud (RX~J1615.3-3255). Larger-number statistics are needed to confirm
	this grouping by star-forming region in the $\mu$m-vs-mm diagram.

\begin{figure}
		\centering
		\includegraphics[width=\columnwidth]{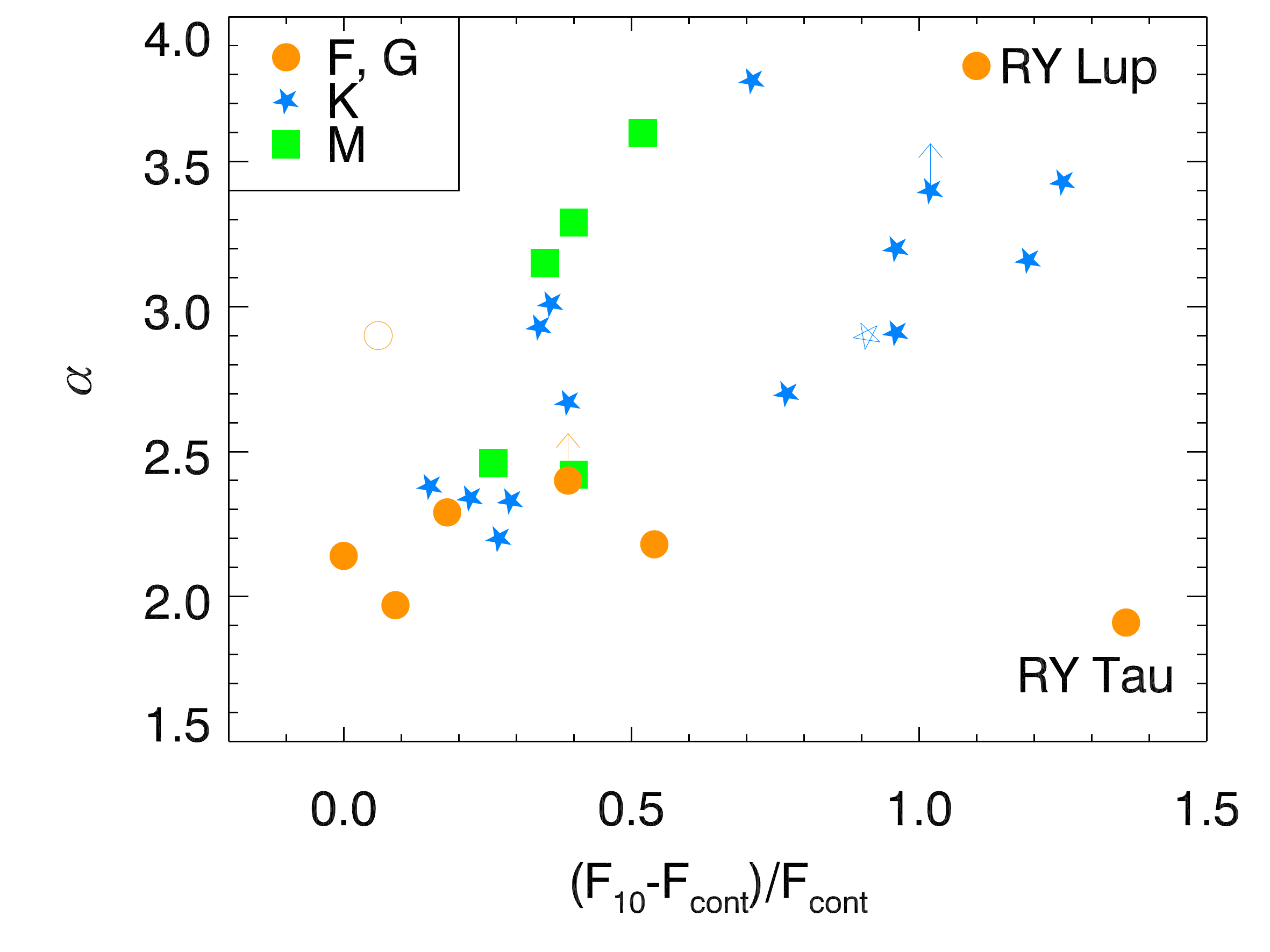}
		\caption[]{Same as Fig.~\ref{fig: strength vs slope}, with the different sources sorted by spectral type: {\em circles}: F and
			G, {\em five-pointed stars}: K, and {\em squares}: M.}
		\label{fig: correlation SpT}
	\end{figure}

	\citet{kessler-silacci:2006} found a correlation between the spectral type of a source and the strength and shape of the 10-$\mu$m
	silicate feature, brown dwarfs having predominantly flatter and Herbig-Ae/Be stars having more peaked features. It was found that this
	is most likely due to the location of the silicate emission region: \citet{kessler-silacci:2007} showed
	that the radius of the 10-$\mu$m silicate emission zone in the disc goes roughly as $(L_*/L_\odot)^{0.56}$. Hence, the 10-$\mu$m 
	feature probes a radius further from the star for early-type stars than for late-type stars. In this context it is interesting to see
	whether a correlation with spectral type is found in the 10-$\mu$m-feature vs mm-slope diagram (Fig.~\ref{fig: correlation SpT}). The 
	M stars in the sample presented here are concentrated to the left, the F and G stars to the lower left, and the K stars are found both
	in the lower left and in the upper right. Hence, no clear correlation with spectral type is found here. It is interesting to note, 
	though, that the F and G sources RY~Tau and RY~Lup show up isolated from the other F and G sources. This may indicate that these 
	sources are indeed different from the other sources in the sample, justifying the choice not to include RY~Tau in the analysis.

\section{Modelling}\label{sect: modelling}

\subsection{Disk model parameters and SEDs}\label{sect: disk model params}

	Variations in the strength and shape of the 10-$\mu$m feature \citep[e.g.,][]{kessler-silacci:2006} as well as in the 
	(sub)mm slope \citep[e.g.,][]{beckwith:1990} can be explained by variations in the dominating grain size in the 
	circumstellar discs, so that one may expect a correlation between properties of the 10-$\mu$m feature and the mm slope. Such
	a correlation is found for the sample as a whole (see previous Section) and this may imply that grain growth occurs 
	in the whole disc simultaneously, or that grains grow in the inner disc and the new grain size distribution is very 
	efficiently spread to the outer disc through radial mixing. Both processes will have the effect of a shift of dust 
	mass from small particles to larger grains. To study this more quantitatively, we ran a number of models with varying 
	grain size distributions.
	
	We use the axisymmetric radiative-transfer code RADMC, developed by \citet{dullemond:2002} and 
	\citet{dullemond:2004}. The model consists of a flaring disc, heated passively by radiation from the central star, and 
	includes a hot inner wall, which is directly irradiated by the central star (Natta et~al. 2001, see also Dullemond et~al. 2001).
	\nocite{natta:2001} \nocite{dullemond:2001}
	The surface density of the disc as a function of radius $\Sigma(r)$ is defined to be:
	\begin{equation}\label{eq: surface density}
		\Sigma(r) = \Sigma_{\rm out} \left(r/R_{\rm out}\right)^n,
	\end{equation}
	with $n = -1$. The total gas+dust disc mass was fixed to
	$M_{\rm disc} = 5 \times 10^{-3}$~M$_\odot$. The gas-to-dust mass ratio $\Psi$ was set to 100, which implies that the total mass
	in dust grains is also fixed, to a value of $5 \times 10^{-5}$~M$_\odot$. The inner radius is fixed to $R_{\rm in} = 0.1$~AU and the
	outer radius $R_{\rm out}$ is varied as outlined below. The photons, originating from the central star, are allowed to move in three 
	dimensions within the axisymmetric grid. In the models, the geometry of the disc is determined by vertical thermal equilibrium. The
	temperature and luminosity of the central source are varied, as are the power-law slope of the grain size distribution, the disc outer
	radius, and the maximum grain size.
	
	For the dust opacities, we use a mixture of 80\% amorphous olivine and 20\% armorphous carbon (percentages by mass). The
	opacities are calculated using a Distribution of Hollow Spheres \citep[DHS, see][]{min:2003}. The total volume of the spheres
	occupied by the inclusion $f$ is taken in the range $f = [0, 0.8]$.
	It was found \citep[e.g.][]{chiang:2001,dalessio:2001,dullemond:2004} that the mm slope changes if one goes from a disc with only
	``small'' particles to a disc that also contains some ``large'' grains. \citeauthor{dullemond:2004} started with a disc in which the 
	dust is made up of only 0.1-$\mu$m-sized particles and subsequently replaced 90\%, 99\%, 99.9\%, 99.99\%, and 99.999\% of the dust by 
	large, 2-mm-sized grains. The mm slope changes considerably when the mass fraction in large grains is changed from 0 to 90\%, but it
	does not change further if a larger fraction of the dust mass is put in large grains \citep[see Fig.~7 in][]{dullemond:2004}. This is
	a result of the fact that at 1~mm the opacity is dominated by the large grains, virtually independent of the mass fraction 
	(K.~Pontoppidan, priv. comm.).
	Although it is possible that a more gradual change in the mm slope is seen when smaller mass fractions are put in large grains, it 
	does seem to be more important what the largest grain size is, rather than which fraction of the dust is contained in such large 
	grains. We therefore chose not to use a bimodal dust distribution, but a distribution in which the size of the grains ranges from a 
	minimum value $a_{\rm min}$ to a maximum value $a_{\rm max}$ according to
	\begin{equation}
		n(a) \propto \left(\frac{a}{a_{\rm min}}\right)^{-m}.
	\end{equation}
	This power-law distribution is expected on theoretical grounds whenever grain-grain collisions lead to 
	shattering \citep{dohnanyi:1969}. It should be noted that models which include grain growth may lead to different 
	grain size distributions \citep[e.g.,][]{dullemond:2005,tanaka:2005}. The value $a_{\rm min}$ was fixed to 0.003~$\mu$m. Note that 
	$a_{\rm min}$ can still have a considerable effect on the 10-$\mu$m feature, with large values for $a_{\rm min}$ giving significantly 
	weaker features	\citet{olofsson:2009}. This effect is strongest for steep grain size distributions, e.g., $m = 4.0$. The 
	maximum grain size $a_{\rm max}$ was varied in steps of ten from 0.1~$\mu$m to 1.0~cm. A value of $m = 3.5$ is representative of 
	interstellar grains \citep{mathis:1977}. A shallower slope of $m = 2.5$ is expected when grains coagulate to larger sizes 
	\citep{natta:2004, natta:2007}, whereas a slope of $m = 4.0$ is expected when also fragmentation is taken into account 
	\citep{brauer:2008, dominik:2008}.
	The different model parameters are summarised in Table~\ref{tab: model parameters}.
	\begin{table}
	 \caption[]{Model parameters.}
	 \label{tab: model parameters}
	 \centering
	  \begin{tabular}{lcc}
	   \hline
	   Parameter					& standard			& range				\\
	   \hline
	   Mass $M_{\rm star}$				& 1.0~M$_\odot$			&				\\
	   $T_{\rm eff}$				& 3000~K			& 3000~K, 4000~K		\\
	   Luminosity $L_{\rm star}$			& 1.0~L$_\odot$			& 1.0~L$_\odot$, 6.0~L$_\odot$	\\	
	   Radius $R_{\rm star}$			& 3.7~R$_\odot$			& 3.7~R$_\odot$, 5.1~L$_\odot$	\\
	   \hline
	   Mass $M_{\rm disc}$ (M$_\odot$)		& $5 \times 10^{-3}$		&				\\
	   Disc inclination angle $i$ ($^\circ$)	& 45				& 5.7, 15, 30, 45, 60, 75	\\
	   Surface mass density gradient $n$		& --1.0				&				\\
	   Gas-to-dust ratio $\Psi$			& 100				&				\\
	   Inner radius $R_{\rm in}$ (AU)		& 0.1				&				\\
	   Outer radius $R_{\rm out}$ (AU)		& 300				& 100, 200, 300			\\
	   \hline
	   Minimum grain size $a_{\rm min}$ ($\mu$m)	& 0.003				&				\\
	   Maximum grain size $a_{\rm max}$ ($\mu$m)	& 				& 0.1, 1, 10, 100,		\\
	   						&				& 1000, 10000			\\
	   Power-law slope $m$				& 3.5				& 2.5, 3.0, 3.5, 4.0		\\
	   \hline
	  \end{tabular}
	\end{table}
	
\begin{figure}
		\centering
		\includegraphics[width=\columnwidth]{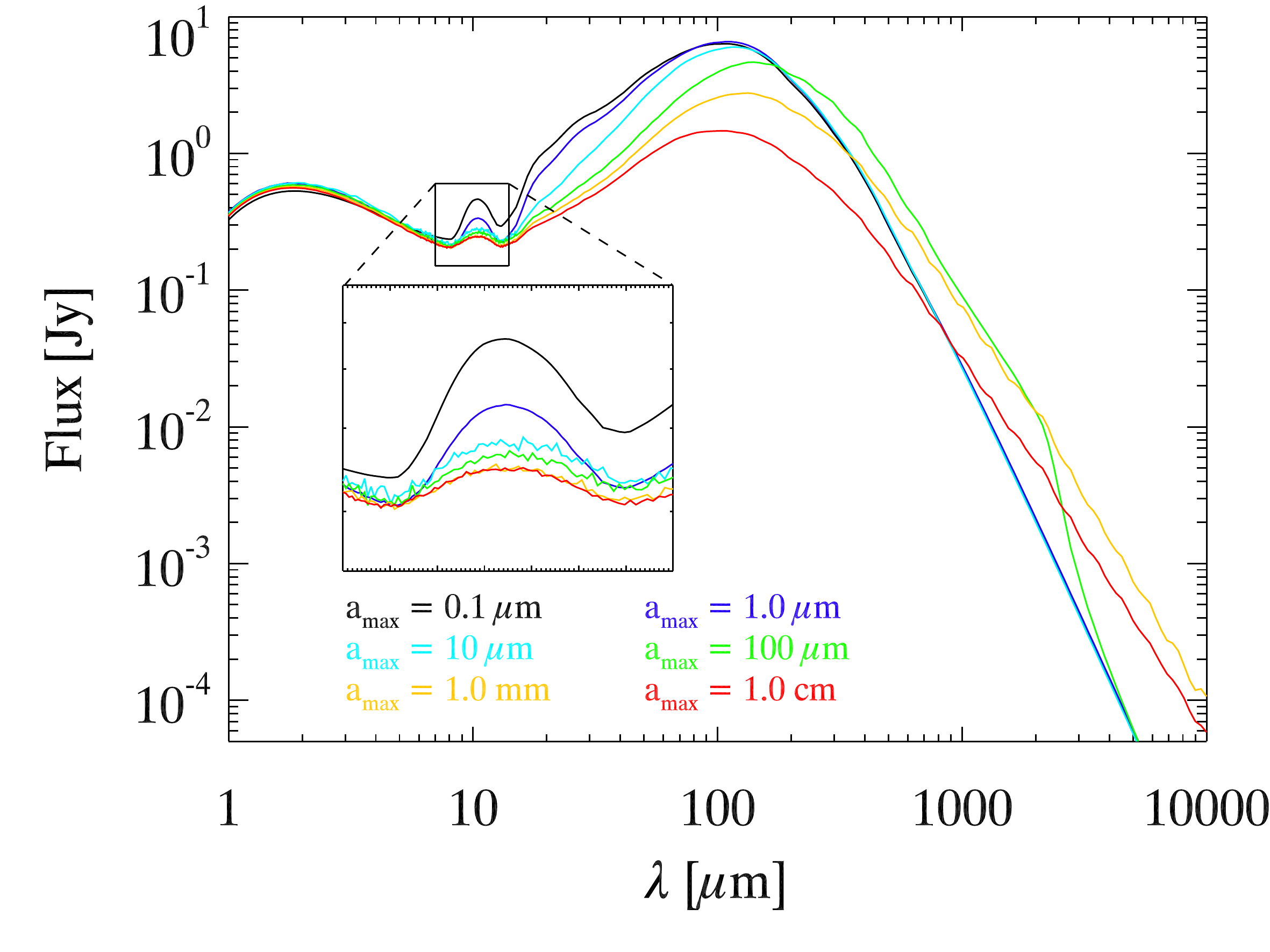}
		\caption[]{Spectral energy distributions (SEDs) for models of a $5 \times 10^{-3}$~M$_\odot$ disc with a varying
			grain size distribution, seen under an inclination $i = 45^\circ$. The minimum grain size is 
			0.003~$\mu$m for all models, and the maximum grain size varies from 0.1~$\mu$m (black curve) to 1.0~cm 
			in steps of ten (dark blue, light blue, green, yellow, red). The inset shows a blow-up of the 
			10-$\mu$m features. Note that the SEDs for $a_{\rm max} = 0.1$~$\mu$m,	1.0~$\mu$m, and 
			10~$\mu$m completely overlap in the mm part. The sharp drop in flux for $a_{\rm max} = 100$~$\mu$m around
			2000~$\mu$m is due to a drop in opacity at about $\lambda = 2 \times \pi \times a_{\rm max}$.}
		\label{fig: seds}
	\end{figure}

	The resulting SEDs from six models, with $a_{\rm max}$ varying and the other parameters kept fixed, is shown in 
	Fig.~\ref{fig: seds}.
	In these models, $R_{\rm out}$ was fixed to $300$~AU and the scale height was kept fixed at the same value
	in all models to show only the effect of varying $a_{\rm max}$. Strong variations are seen in all wavelength regimes, from the near-infrared through the mm. At wavelengths 
	$\lambda \lesssim 2$~$\mu$m, grain size distributions without grains larger than 1~$\mu$m give such a high opacity
	that the central star is significantly reddened. 
	In the mid- and far infrared, the flux drops with increasing maximum grain size.
	The (sub)mm part of the SED does not
	change appreciably unless grains with sizes of $\sim$100~$\mu$m or larger are included. After that, the (sub)mm slope 
	becomes shallower quite rapidly with increasing $a_{\rm max}$. This figure also demonstrates that care must be
	taken when estimating the disc mass from the (sub)mm emission alone: 
	even when the dust composition is kept the same, assuming a different grain size distribution may already change the opacity at
	1~mm by an order of magnitude, which will give an equally large uncertainty in the mass estimate from an observed flux at
	that wavelength.

\subsection{10-$\mu$m feature vs mm slope}\label{sect: model results}
	
\begin{figure*}
		\centering
		\includegraphics[width=\textwidth]{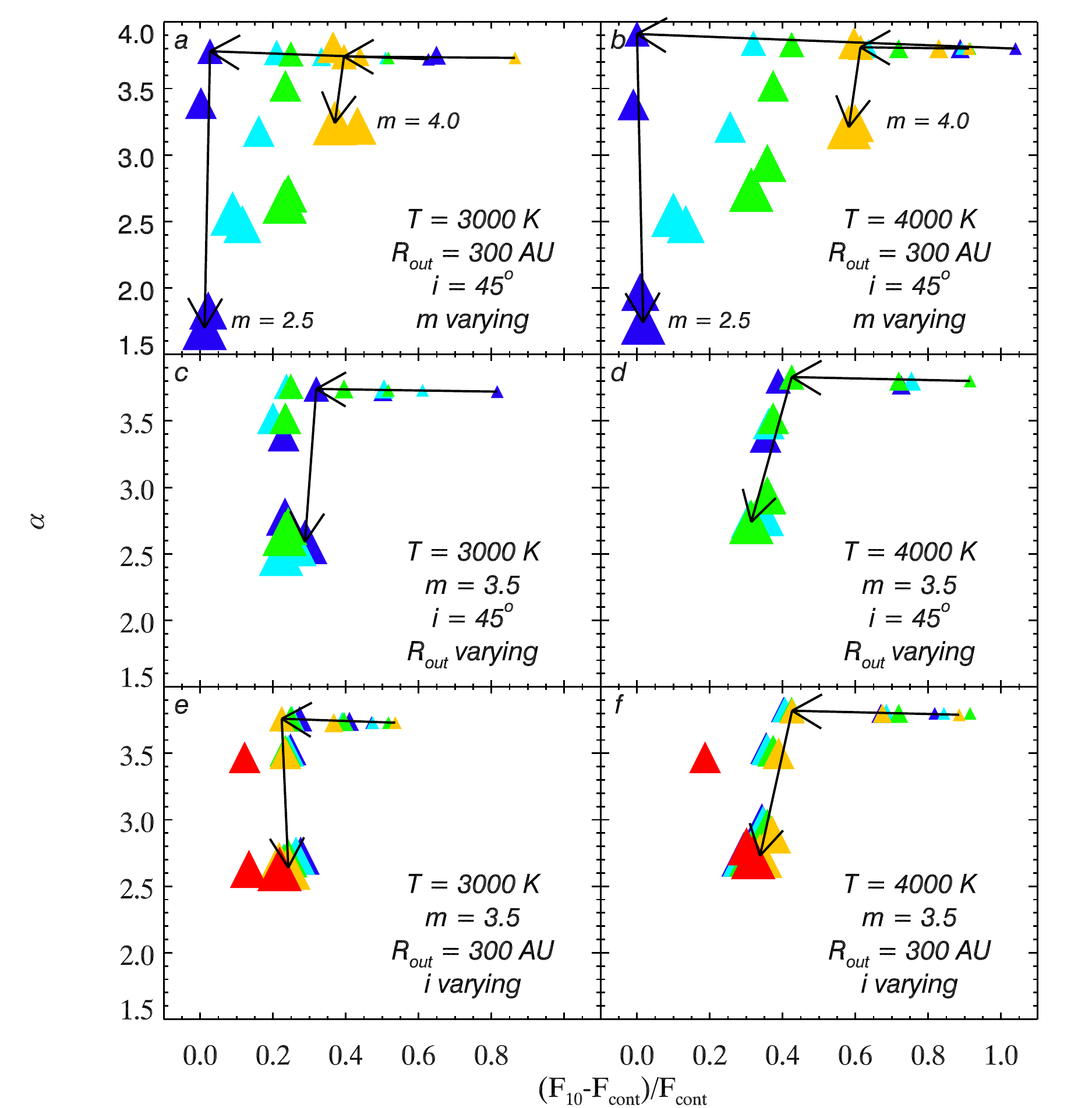}
		\caption[]{The mm slope $\alpha$ between 1 and 3~mm vs the strength of the 10-$\mu$m feature $(F_{10}-F_{\rm cont})/F_{\rm 
				cont}$. See text for disc parameters.
				The size of the triangles indicates the size of the largest dust grains $a_{\rm max}$ and the colour indicates
				the variation of the other parameters. The arrows indicate a possible evolution, that of increasing grain size.
				{\em a}: $T_{\rm eff} = 3000$~K and $L = 1$~L$_\odot$. The power-law slope of the grain size 
				distribution is varied from 2.5 (dark blue) to 3.0 (light blue), 3.5 (green), and 4.0 (yellow).
				{\em b}: $T_{\rm eff} = 4000$~K and $L = 6$~L$_\odot$. The power-law slope of the grain size
				distribution is varied from 2.5 (dark blue) to 3.0 (light blue), 3.5 (green), and 4.0 (yellow).
				{\em c}: $T_{\rm eff} = 3000$~K and $L = 1$~L$_\odot$. The disc radius is varied from
				100~AU (dark blue) to 200~AU (light blue) to 300~AU (green).
				{\em d}: $T_{\rm eff} = 4000$~K and $L = 6$~L$_\odot$. The disc radius is varied from
				100~AU (dark blue) to 200~AU (light blue) to 300~AU (green).
				{\em e}: $T_{\rm eff} = 3000$~K and $L = 1$~L$_\odot$. The inclination under which the disc is
				viewed is varied from 15 (dark blue) to 30 (light blue), 45 (green), 60 (yellow), and 75 (red) degrees.
				{\em f}: $T_{\rm eff} = 4000$~K and $L = 6$~L$_\odot$. The inclination under which the disc is
				viewed is varied from 15 (dark blue) to 30 (light blue), 45 (green), 60 (yellow), and 75 (red) degrees.}
		\label{fig: shape vs slope models set 2}
	\end{figure*}

	In Fig.~\ref{fig: shape vs slope models set 2}, we plot the strength of the 10-$\mu$m
	feature vs the mm slope for different models. The strength of the 10-$\mu$m feature 
	$(F_{10}-F_{\rm cont})/F_{\rm cont}$ is defined as in \citet{furlan:2006} and the mm
	slope $\alpha$ is determined between 1.0 and 3.0~mm. The main aim of this figure is to 
	show the variation of the 10-$\mu$m-feature strength and mm slope with various parameters.
	While the quantitative details will depend on the specific dust and disc parameters used,
	the qualitative trends found in these figures should be robust.
	
	In each of the panels, the results for different maximum grain sizes are shown. The size of the triangles is an indication for the maximum grain size under
	consideration. A general trend is observed, in the sense that the models with only small grains end up in the upper right corner of the micron-vs-mm diagram (strong
	10-$\mu$m feature and steep mm slope), the models which include grains of mm sizes or larger end up more to the lower left of the diagram (weak 10-$\mu$m feature and
	shallower mm slope), and those with grain sizes of up to 10 or 100~$\mu$m end up towards the upper left corner of the diagram (weak 10-$\mu$m feature and steep mm
	slope). A possible evolutionary sequence, in which the maximum grain size in the disc gradually increases, is indicated by the arrows: first, the 10-$\mu$m feature
	becomes weaker and later, the mm slope becomes shallower. A test to check whether
	radial variation of $a_{\rm max}$ -- larger grains closer to the star, where the
	densities are higher -- did not show any significant difference.
	
	The models show the effect of the temperature and luminosity of the central star.
	on the strength of the 10-$\mu$m feature and the steepness of the mm slope. The 
	left column shows the results for a central star with $T_{\rm eff} = 3000$~K and $L = 
	1$~L$_\odot$ and the right column for $T_{\rm eff} = 4000$~K and $L = 6$~L$_\odot$. In
	Figs.~\ref{fig: shape vs slope models set 2}a and b, the power-law slope of the grain 
	size distribution is varied from $m = 2.5$ to 3.0, 3.5, and 4.0. 
	It appears that only grain size distributions with $m = 2.5$ produce completely flat 
	10-$\mu$m silicate features as well as mm slopes with $\alpha < 2.0$, whereas grain 
	size distributions with $m = 4.0$ never produce a mm slope with $\alpha < 3.0$. 
	Furthermore, the strongest 10-$\mu$m features are only obtained with a central star of
	4000~K and $L = 6$~L$_\odot$.
	
	In Figs.~\ref{fig: shape vs slope models set 2}c and d, the power-law slope of the grain size distribution is fixed to $m = 3.5$. The
	disc radius $R_{\rm out}$ is varied between 100, 200, and 300~AU. This has a small effect on the strength of the 10-$\mu$m feature,
	particularly for $T_{\rm eff} = 3000$~K and $a_{\rm max} = 0.1$~$\mu$m. This can be understood in the sense that for a smaller disc
	with the same dust mass, a larger amount of mass is closer to the star and thus radiates in the infrared. The mm slope of the SED is
	practically unaffected.
	
	Figs.~\ref{fig: shape vs slope models set 2}e and f show the results for models in 
	which the power-law slope of the grain size distribution was fixed to $m = 3.5$, the 
	disc outer radius to $R_{\rm out} = 300$~AU, and for which the inclination $i$ under 
	which the disc is observed is varied. In most cases, the inclination has a limited 
	effect on both the strength of the 10-$\mu$m feature and the mm slope of the SED. Only
	under very high inclination (e.g., 75$^\circ$, where 90$^\circ$ is edge-on) does the 
	10-$\mu$m feature appear in absorption (not shown). A similar effect is seen
	if the discs are more flaring than found in vertical hydrostatic equilibrium: the
	10-$\mu$m feature is primarily weakened, because of the enhanced extinction under most
	inclinations.
	
	A second set of models is run to investigate the effects of dust settling, i.e., the process in which larger grains fall to the
	disc midplane under the influence of gravity, while the smaller grains stay suspended in the disc atmosphere. As mentioned before, 
	\citet{dullemond:2008} found that a bimodal grain size distribution can explain variations in the strength of the 10-$\mu$m feature, 
	but only under specific circumstances. They looked at grains that are mainly responsible for the 10-$\mu$m feature, in particular 
	grains of 3~$\mu$m and of $\ll$ 1~$\mu$m. To study the effect of the settling of larger grains, we ran a number of models with up to
	six different grain size distributions: grains with sizes between 0.003 and 0.1~$\mu$m, between 0.1 and 1~$\mu$m, 1 and 10, 10 and 
	100, 100 and 1000, and finally between 1000 and 10,000~$\mu$m. The degree of settling is given by a parameter $s$, varying between 0.25, 0.50,
	0.75, and 1.00, and is chosen to be different for each of the grain size distributions: the larger the grains, the larger the degree
	of settling. For example, if $H$ denotes the self-consistent scale height, a settling parameter $s = 0.75$ indicates that:
	\begin{list}{\leftmargin=1em}
		\item grains between 0.003 and 0.1~$\mu$m are at $H$;
		\item grains between 0.1 and 1~$\mu$m are at $0.75 \times H$;
		\item grains between 1 and 10~$\mu$m are at $0.75^2 \times H$;
		\item grains between 10 and 100~$\mu$m are at $0.75^3 \times H$;
		\item grains between 100 and 1000~$\mu$m are at $0.75^4 \times H$;
		\item grains between 1000 and 10,000~$\mu$m are at $0.75^5 \times H$.
	\end{list}
	Hence, a larger number for $s$ indicates a smaller degree of settling and $s = 1.00$ corresponds to no settling (all grains 
	are at the self-consistent scale height).
	
\begin{figure}
		\centering
		\includegraphics[width=\columnwidth]{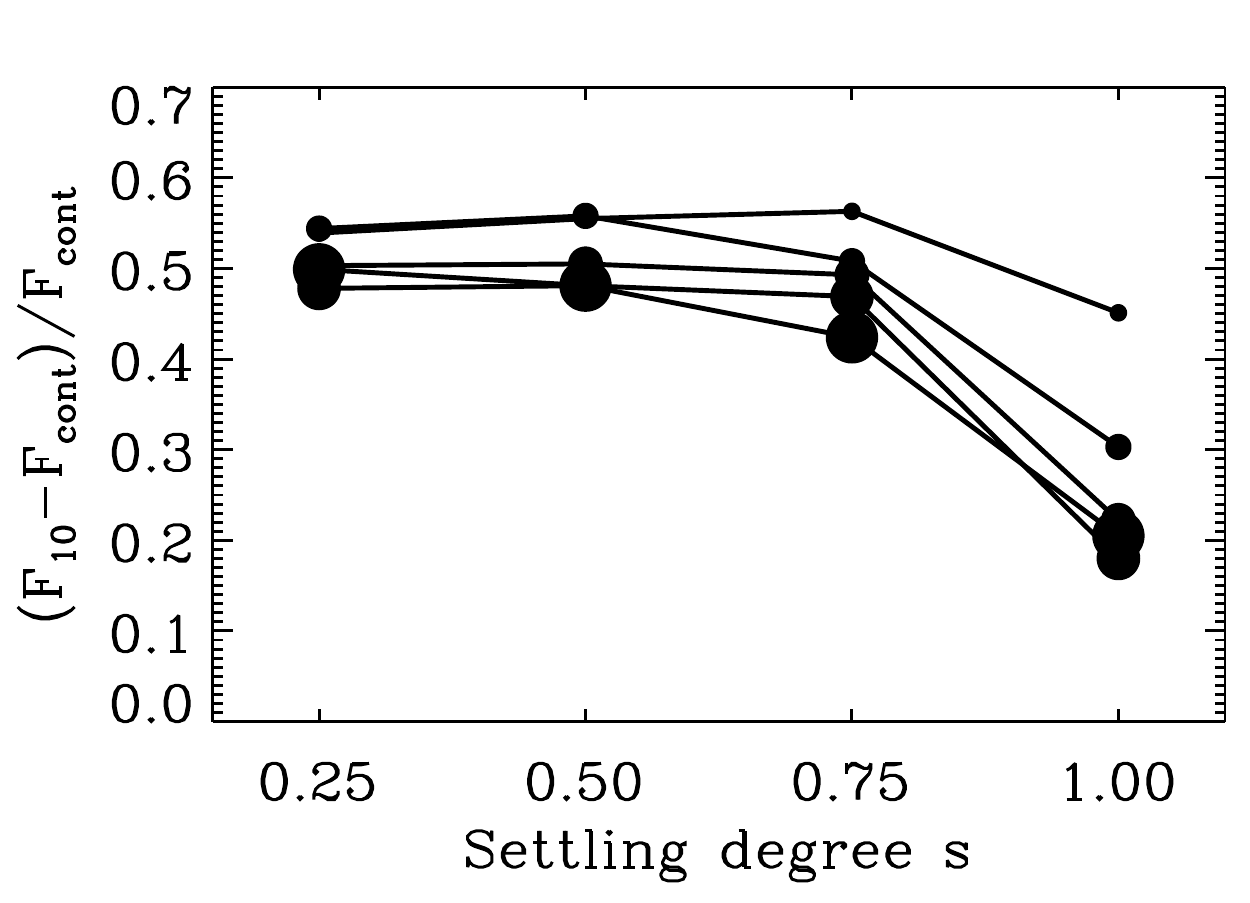}
		\caption[]{Strength of the 10-$\mu$m feature as a function of the settling parameter $s$ and the maximum grain size 
			$a_{\rm max}$. The size of the points indicates $a_{\rm max}$, which varies from 1 to 10, 100, 1000, and 
			10,000~$\mu$m. Note that a small value for $s$ corresponds to a large degree of settling; $s = 1.00$ corresponds to no
			settling (all dust grains at the self-consistent scale height). All 
			other disc parameters are as for the standard model in 
			Table~\ref{tab: model parameters}.}
		\label{fig: settling}
	\end{figure}

	These models are run using the radiative transfer code MCMax \citep{min:2009}. MCMax and RADMC were benchmarked against the 
	results of \citet{pascucci:2004} and the differences in the resulting SEDs are minimal, with in particular the 10-$\mu$m features 
	being practically indistinguishable \citep[see the Appendix in][]{min:2009}. The results for the settling are shown in Fig.~\ref{fig: settling}.
	Note that the strength of the 10-$\mu$m feature increases when settling is switched on, but does not increase significantly 
	more when the value of $s$ is decreased more (i.e., when the degree of settling is increased). The slope in the mm part of the SED is
	practically unaffected by the degree of settling.
	
	It can be concluded that a variation of the maximum grain size $a_{\rm max}$ 
	affects both the strength of the 10-$\mu$m feature and the steepness of the mm slope 
	$\alpha$: a larger maximum grain size yields a weaker 10-$\mu$m feature and a 
	shallower mm slope. This effect is robust against variations in
	the degree of settling $s$, which only affects the strength of the 10-$\mu$m 
	feature. Only varying the maximum grain size is, however, not enough to explain the 
	range in 10-$\mu$m features and mm slopes that is observed. Variations in the 
	power-law index of the grain size distribution and the temperature and luminosity of 
	the central source are required as well to cover the full observed range. For example,
	only models with a central-star temperature of at least 4000~K yield a 10-$\mu$m 
	feature with a strength of $(F_{10}-F_{\rm cont})/F_{\rm cont} \approx 1.0$; models
	with relatively flat or shallow grain size distributions are required to get mm slopes of 
	$\alpha \lesssim 2.0$ \citep[see also][]{natta:2007, ricci:2010}.
	
\section{Discussion}\label{sect: discussion}

	A correlation between the strength of the 10-$\mu$m silicate feature and the slope between 1 and 3~mm is observed in a sample of in
	total 31 T-Tauri stars. 
	This seems to imply that, while in the upper layers of the hot inner disc small grains are still coagulating to a few microns in size,
	in the cold mid-plane of the outer disc large grains of at least several millimetres in size are already being formed. One should keep
	in mind, though, that the 10-$\mu$m silicate feature only probes the hot surface layers of the inner disc. It is therefore possible,
	and even likely, that large grains or boulders exist in the mid-plane of the inner disc. The tentative correlation seen in 
	our observations show, however, that the inner and outer discs of young stellar objects do not evolve independently from each other. 
	Furthermore, the gradual decline of the 10-$\mu$m feature as the mm slope becomes shallower implies that micron-sized grains remain 
	present in the disc atmosphere as cm-sized pebbles are already forming in the mid-plane.

	The radiative-transfer programmes RADMC and MCMax were used to run a number of models with varying disc parameters, stellar 
	parameters, and power-law grain size distributions. The only parameter that affects the strength of the 10-$\mu$m feature and
	the mm slope of the SED simultaneously is the maximum grain size $a_{\rm max}$: a larger $a_{\rm max}$ gives both a weaker 10-$\mu$m 
	feature and a shallower mm slope. This result is very robust against variations of the other parameters. There are, however, a few 
	striking results when the models are compared to the observations. Firstly, the strongest 10-$\mu$m silicate features, with 
	$(F_{10}-F_{\rm cont})/F_{\rm cont} \approx 1.0$, can only be reproduced for a star with $T_{\rm eff} = 4000$~K and $L = 6$~L$_\odot$,
	whereas a tempearature of 3000~K and a luminosity of 1~L$_\odot$ seem to be more common for observed T-Tauri stars
	\citep[see][]{evans:2009}. This may in part be a selection effect: the sources with the hottest central stars are the most likely
	to have strong mm fluxes, making it more likely for them to provide a good mm slope. A subsample of the sources of \citet{evans:2009}
	which had enough data points in the SED to provide a decent model fit also gave on average a higher $T_{\rm eff}$ than would be
	expected from the near-infrared colours alone (B.~Mer{\'{\i}}n, priv. comm.). The physical reason for the increase in strength
	of the 10-$\mu$m feature with temperature lies in the fact that the photosphere is still an important continuum source. When the
	temperature of the star is increased, the relative contribution from the photosphere becomes smaller and the peak/continuum ratio of
	the feature goes up. 
	
	Secondly, the sources with $(F_{10}-F_{\rm cont})/F_{\rm cont} \approx 0.0$ as well as those with a mm slope $\alpha \lesssim 2.0$ can
	only be reproduced if the power-law slope of the grain size distribution is as low as $m = 2.5$. A slope $m = 4.0$ does not allow the
	mm slope to get below 3.0. The strength of the 10-$\mu$m feature remains at least as large as 1.4, and even 1.6 in the case 
	$T_{\rm eff} = 4000$~K. This appears to indicate that, as grains are growing to mm and cm sizes, the power-law slope of the grain size
	distribution becomes shallower. It is, however, more likely that a power-law grain size distribution with a mimimum grain size of 
	$a_{\rm min} = 0.003$~$\mu$m no longer applies. Either the effective minimum grain size increases, as suggested by the work of
	\citet{olofsson:2009}, or a different grain size distribution is more applicable, such as naturally obtained from models including 
	fragmentation \citep[e.g.,][]{dullemond:2005}.
	
	It is important at this point to note once more that the 10-$\mu$m feature originates entirely from the disc atmosphere; 
	(sub)micron-sized dust grains that are closer to the midplane are not observable due to the optical thickness of the infrared emission.
	The emission at (sub)mm wavelengths, on the other hand, is predominantly optically thin; hence, it may originate from the whole disc. 
	In fact, the bulk of the (sub)mm emission comes from the midplane, since the larger grains will settle to the midplane. To study the effect of the settling of larger (mm and cm-sized) grains, we ran a 
	number of models with up to six different grain size distributions. The smallest grains ($0.003 < a < 0.1$~$\mu$m) followed the gas,
	while the increasingly larger grains were located increasingly closer to the midplane. It turns out that the strength of the 10-$\mu$m
	feature always increases 
	when settling is switched on, but does not increase significantly more when the value of the settling parameter $s$ is decreased more
	(i.e., when the degree of settling is increased). This can be understood from the fact that a small degree of settling already removes
	the larger grains from the disc atmosphere. Since the 10-$\mu$m feature originates in the disc atmosphere, it does not matter where
	the larger grains reside, as long as they are not too close to the surface. The slope in the mm part of the SED is practically
	unaffected by the degree of settling. This is because the disc is optically thin at these long wavelengths and hence it does
	not matter where in the disc the large grains, which determine the brightness at long 
	wavelengths, are located. While the absolute strength and slope depend on the detailed dust
	and disc model parameters, the trends found here should be robust. Further modelling and
	exploration of the large parameter space are left to a future paper.

	The observations indicate a clustering of the sources per star-forming region. The T-Tauri stars in Chamaeleon show the strongest
	10-$\mu$m features, as well as the steepest mm slopes. The T-Tauri stars located in the Lupus 1 and Lupus 2 clouds have somewhat 
	weaker 10-$\mu$m features and mm slopes that are a bit shallower than the Chamaelon sources. The sources located in the Taurus-Auriga
	star-forming region, finally, have the weakest 10-$\mu$m features and the shallowest mm slope. Although the number statistics are
	too low to draw strong conclusions from this grouping per star-forming region, it is intersesting to hypothesise as to what might be the
	origin of this grouping. It could in principle be due to a selection
	effect. To check for this, the strengths of the 10-$\mu$m features of the eleven sources from \citet{furlan:2006} included in this 
	work were compared with	those of the 72 Taurus sources not included in this work and the two samples were found to be 
	statistically indistinguishable. 
	Furthermore, after this paper was submitted, \citet{ricci:2010} published mm slopes for a total of 21 sources in the 
	Taurus-Auriga star-forming region. \citeauthor{ricci:2010} found the slopes between 1 and 3~mm all to be $\alpha \leq 2.7$; the 
	corresponding strengths of the 10-$\mu$m features are $(F_{10}-F_{\rm cont})/F_{\rm cont} < 0.75$ for 17 of those 21 sources
	\citep{furlan:2006}. Thus, these sources mainly populate the lower left corner of the $\mu$m-vs-mm diagram.
	It is then tempting to attribute the clustering to an evolutionary sequence, with the Chamaeleon 
	sources being the least evolved and the Taurus-Auriga sources the most evolved. If the evolution proceeds equally fast in each 
	star-forming region, Taurus-Auriga would then be the oldest star-forming region and Chamaeleon the youngest. However, ages of 
	pre-main-sequence stars are notoriously difficult to determine and there is a large spread of stellar ages within each region (see 
	Table~\ref{tab: star-forming regions}). Note that a chemical differentiation, with, e.g., Chamaeleon a region with more silicates than
	Taurus, would mainly have an effect on the strength of the 10-$\mu$m feature, whereas the mm slope of the SED is largely determined by
	the sizes of the grains in the disc alone \citep{draine:2006}. Also, sources are
	likely to go through their evolutionary stages at different rates, which may be determined
	by other parameters such as initial conditions of the core.

	Our final sample only contained two cold discs for which the strength of the 10-$\mu$m feature and the slope of the SED 
	between 1 and 3~mm could be obtained. These are T~Cha, an evolved disc with no silicates in the spectrum that is not included in the 
	analysis, and CS~Cha, a	circumbinary disc. Several processes have been proposed which may cause the deficiency of hot dust in the 
	inner disc, such as photo-evaporation, the clearing out of the inner disc by a planet, or grain growth up to mm sizes and larger. If 
	grain growth were the dominating reason for the decrease in infrared flux, one would expect the cold disc to show up in the lower part
	of the 10-$\mu$m-feature vs mm-slope diagram. However, CS~Cha is located rather in the centre of the diagram and it would appear that 
	grain growth is not the main reason for the lack of infrared emission. Indeed, since CS~Cha harbours a binary, it may be the case that
	the inner disc is cleared out due to binary interactions, with a ``normal'' outer disc remaining.

\section{Conclusions}\label{sect: conclusions}

	We observed five binary systems and 35 single T-Tauri stars (of which one turned out to harbour two sources as well) at infrared and mm 
	wavelengths. New {\em Spitzer} IRS spectra of the wavelength region containing the
	10-$\mu$m silicate feature are presented for 13 sources and the slopes in the mm regime of the SED are obtained, also for 13 sources.
	Furthermore, dust disc masses are obtained for 21 new sources, including strict 3$\sigma$ upper limits for nine of the sources. Combining 
	our new observations with data from the literature, a tentative correlation between the strength of the 10-$\mu$m feature and the mm slope is 
	found for a sample of in total 31 T-Tauri stars. This seems to suggest that the inner and outer discs do not evolve 
	independently.
	
	Modelling protoplanetary discs with varying geometries and grain size distributions indicates that grain growth alone cannot explain
	the observed evolution in the strength of the 10-$\mu$m feature and the mm slope of the SED. It would rather seem that as the maximum
	grain size increases, also the power-law slope of the grain size distribution becomes shallower. \citet{ricci:2010} also found
	that for the majority of the sources in their sample a grain size distribution flatter than that of the interstellar medium is
	required \citep[see also][]{natta:2007}. This is an indication that a power-law
	grain size distribution with a fixed minimum grain size is too simple a representation of the dust in protoplanetary discs.
	
	The sample contains only one cold or transitional disc for which the mm slope in the SED and the strength of the 10-$\mu$m silicate feature could 
	be obtained properly. This source, CS~Cha, ends up rather in the centre of the 10-$\mu$m vs mm-slope diagram, indicating that grain
	growth is probably not the source of the removal of dust from the inner disc. Transitional discs are often relatively weak in the
	(sub)mm regime and a new generation of telescopes is required to determine the mm slope for more of these sources.
	
	Although all 10-$\mu$m features used in this work were obtained using the {\em Spitzer} Space Telescope, the mm fluxes and 
	hence the mm slopes were determined with a large number of different telescopes. As this potentially leads to systematic effects, it 
	will be necessary to observe a statistically large enough sample with only one or maybe two (sub)mm telescopes. Ideally, about a dozen 
	sources per star-forming regions for at least four regions should be observed with the same telescope. However, due to the relative 
	weakness of the sources under consideration, the scarcity of available telescope time, and the spread of young star-forming regions 
	over the sky, this is currently hard to achieve. The Atacama Large Millimeter/submillimeter Array (ALMA) will, upon completion, be 
	able to study an order of magnitude more sources than presented here, providing us with the statistics to study interesting 
	relations such as the grouping of YSOs by star-forming region in the $\mu$m-vs-mm diagram. ALMA will also be able to resolve most of
	the sources studied here.
	
	This work has mainly focused on observations at 1 and 3~mm, at which grains with sizes of the order of 1~cm can be studied 
	\citep{draine:2006}. The detection of larger grains, or pebbles, in protoplanetary discs, requires observations at cm wavelengths. 
	However, there may be other sources of emission at cm wavelengths, such as magnetic fields or stellar winds, and it is necessary to 
	monitor sources over extended periods of time to disentangle the different emission mechanisms \citep{wilner:2005, lommen:2009}. The 
	maximum bandwidth of the ATCA was recently improved by a factor of 16 with the implementation of the Compact Array Broadband Backend 
	(CABB). This will for the first time allow the survey of large samples of young stellar objects in the southern hemisphere at cm wavelengths.
	
	Thus, in five to ten years from now we will not only be studying how grain growth occurs in protoplanetary discs, but we will actually be able 
	to pinpoint the locations in the discs where the formation of pebbles and boulders, the precursors to planetesimals and planets, is taking place.
	Furthermore, on-going observations at cm wavelengths will give us a stronger handle on the processes taking place in and around the discs, affecting
	the growth of grains and the formation of planets.

\begin{acknowledgements}
	We are indebted to Kees Dullemond for the use of his RADMC and RAYTRACE codes and to Stephen Bourke for help with 
	AIPS. A special thanks to the ATNF and SMA staff for assistance with the observations. Ruud Visser's help on all
	things computer is greatly appreciated and Carsten Dominik is thanked for his useful comments. Partial support for this work was provided by a Netherlands Research 
	School For Astronomy network 2 grant, and by a Netherlands Organisation for Scientific Research Spinoza grant. 
	C.M.W. acknowledges financial support from an ARC Australian Research Fellowship, Discovery Project DP0345227. This research has made use of 
	the SIMBAD database, operated at CDS, Strasbourg, France.
\end{acknowledgements}

\appendix

\section{Observations}\label{app: observations}


	In Lupus, 15 single sources and the binary IK Lup \& Sz 66 were observed with the Submillimeter Array (SMA) at about 1~mm. Furthermore, the 
	binaries VV CrA and S CrA and the single source DG CrA in Corona Australis were observed with the SMA at about 1~mm. A full log of the SMA 
	observations is given in Table~\ref{tab: observation log SMA}. The results of these observations are shown in Table~\ref{tab: results log SMA} and
	Fig.~\ref{fig: SMA UVdist}.
	\begin{table*}
		\caption[]{Overview of the SMA observations.}
		\centering
		\begin{tabular}{lcccccl}
			\hline
			\hline
			Obs. date		& Wavelengths	& Config.	& Target sources$^\mathrm{a}$	& Gain calibrator(s) 	& Flux cal.	& Notes				\\
						& (mm)		&		&				&		     	&  		& 				\\
			\hline
			20080419		& 1.30, 1.36	& compact	& EX Lup, RX J1615.3-3255	& 1517-243, 1626-298 	& Mars		& PWV $<$ 4.0~mm		\\
						&		&		& RY Lup, 161029.57-392214.7	&		     	&  		&  				\\
						&		&		& 161159.81-382338.5, Sz 111	&		     	&  		&  				\\
						&		&		& Sz 91, Sz 96			&		     	&  		&  				\\
			20081001		& 1.31, 1.37	& compact	& VV CrA, S CrA, DG CrA		& 1924-292	     	& Callisto 	& PWV $<$ 2.5~mm		\\
			20090507		& 1.30, 1.36	& compact	& HM Lup, MY Lup, 		& 1626-298 		& Callisto 	&				    \\
						&		&		& Sz 102, IK Lup \& Sz 66	&		     	&  		&  				\\
						&		&		& Sz 73, Sz 74			&		     	&  		&  				\\
						&		&		& Sz 76, Sz 77			&		     	&  		&  				\\
			\hline
		\end{tabular}
		\label{tab: observation log SMA}
	 \begin{list}{}{}
	  \item[$^\mathrm{a}$]	In the case of SSTc2d names, only the coordinates (in J2000) are shown.
	 \end{list} 
	\end{table*}

	With the Australia Telescope Compact Array (ATCA), 15 sources spread over the constellations Lupus, Vela, Corona 
	Australis, and Chamaeleon were observed at 3 and 7~mm. A log of the ATCA observations is given in Table~\ref{tab:
	observation log ATCA}. 
	\begin{table*}
		\caption[]{Overview of the ATCA observations.}
		\centering
		\begin{tabular}{lcccccl}
			\hline
			\hline
			Obs. date		& Wavelengths	& Config.	& Target sources		& Gain calibrator(s) 	& Flux cal.	      	& Notes				\\
						& (mm)		&		&				&		     	&  		    	& 				\\
			\hline
			20080712		& 3.14, 3.21	& H214		& HBC 556, HBC 557, HBC 559	& 0745-330	     	& Mars		     	& Mostly nice weather		\\
			20080713		& 6.67, 6.99	& H214		& HBC 556, HBC 557, HBC 559	& 0826-373	     	& Uranus		& Weather worsening    		\\
			20080728		& 6.67, 6.99	& H214		& HBC 559			& 0826-373	     	& Uranus		& Mostly nice weather		\\
			20080729		& 3.14, 3.21	& H214		& HBC 559			& 0745-330	     	& Uranus		& Mostly nice weather		\\
						&		&		& SZ Cha, Sz 32			& 1057-797	     	&  		     	&  				\\
			20080730		& 3.14, 3.21	& H214		& SZ Cha			& 1057-797	     	& Uranus		& Weather less than perfect	\\
			20080801		& 6.67, 6.99	& H214		& Sz 111, RY Lup		& 1600-44	     	& Uranus		& Weather improving       	\\
						&		&		& RX J1615.3-3255		& 1622-297	     	&  		     	&  				\\
						&		&		& VV CrA, S CrA, DG CrA		& 1921-293	     	&  		     	&  				\\
			20080802		& 3.14, 3.21	& H214		& SZ Cha			& 1057-797	     	& Uranus		& Weather improving       	\\
						&		&		& Sz 111, RY Lup		& 1600-44	     	&  		     	&  				\\
						&		&		& RX J1615.3-3255		& 1622-297	     	&  		     	&  				\\
						&		&		& VV CrA, S CrA, DG CrA		& 1933-400	     	&  		     	&  				\\
			20080803		& 3.14, 3.21	& H214		& MY Lup			& 1600-44	     	& Uranus		& Mostly nice weather		\\
						&		&		& VV CrA, S CrA, DG CrA		& 1933-400	     	&  		     	&  				\\
			20080804		& 3.14, 3.21	& H214		& Sz 65 \& Sz 66		& 1622-297	     	& Uranus		& Weather less than perfect	\\
			20080805		& 6.48, 6.78	& H214		& MY Lup, IM Lup		& 1600-44	     	& Uranus		& Weather improving       	\\
			\hline
		\end{tabular}
		\label{tab: observation log ATCA}
	\end{table*}

	The Combined Array for Research in Millimeter-wave Astronomy (CARMA) was used to observe 
	eleven single sources and one binary in Serpens at 1 and 3~mm. The log of these observations is presented in Table~\ref{tab: observation 
	log CARMA}; the full results are given in Table~\ref{tab: results log CARMA}.
	\begin{table*}
		\caption[]{Overview of the CARMA observations.}
		\centering
		\begin{tabular}{lcccccl}
			\hline
			\hline
			Wavelength	& Target source			& Config.		& Obs. date(s)		& Gain calibrator(s)	& Flux cal.	& Notes				\\
			(mm)		&				&			&			&			&		&				\\
			\hline
			1.33		& EC 82				& C			& 20080424		& 1743-038		& 3c454.3	& Gain calibrator too weak	\\
					& EC 90				& C			& 20080424		& 1743-038		& 3c454.3	& Gain calibrator too weak	\\
					& SSTc2d J182900.88+002931.5	& C			& 20080424, 20080426	& 1743-038		& 3c454.3	& Gain calibrator too weak	\\
					& IRAS 18268-0025		& C			& 20080426		& 1743-038		& 3c454.3	& Gain calibrator too weak	\\
					& CoKu Ser-G3			& C			& 20080426		& 1743-038		& 3c454.3	& Gain calibrator too weak	\\
					& SSTc2d J182858.08+001724.4	& C			& 20080426		& 1743-038		& 3c454.3	& Gain calibrator too weak	\\
					& SSTc2d J182850.20+000949.7	& C			& 20080426		& 1743-038		& 3c454.3	& Gain calibrator too weak	\\
					& SSTc2d J182944.10+003356.1	& C			& 20080426		& 1743-038		& 3c454.3	& Gain calibrator too weak	\\
					& VV Ser			& C			& 20080426		& 1743-038		& 3c454.3	& Gain calibrator too weak	\\
					& SSTc2d J182936.19+004216.7	& C			& 20080501		& 1743-038		& 3c454.3	& Gain calibrator too weak	\\
					& GSC 00446-00153		& C			& 20080501, 20080518	& 1743-038, 1751+096	& 3c454.3	&				\\
			\hline
			1.33		& EC 82				& D			& 20080620		& 1751+096		& 3c273		&				\\
					& EC 90				& D			& 20080620		& 1751+096		& 3c273		&				\\
					& SSTc2d J182900.88+002931.5	& D			& 20080620		& 1751+096		& 3c273		&				\\
					& IRAS 18268-0025		& D			& 20080620		& 1751+096		& 3c273		&				\\
					& CoKu Ser-G3			& D			& 20080620		& 1751+096		& 3c273		&				\\
					& VV Ser			& D			& 20080620		& 1751+096		& 3c273		&				\\
					& SSTc2d J182858.08+001724.4	& D			& 20080620		& 1751+096		& 3c273		&				\\
					& SSTc2d J182850.20+000949.7	& D			& 20080620, 20080704	& 1751+096		& 3c273		&				\\
					& EC 97				& D			& 20080704		& 1751+096		& 3c279		&				\\
					& SSTc2d J182944.10+003356.1	& D			& 20080704		& 1751+096		& 3c279		&				\\
					& SSTc2d J182936.19+004216.7	& D			& 20080704		& 1751+096		& 3c279		&				\\
					& GSC 00446-00153		& D			& 20080704		& 1751+096		& 3c279		&				\\
			\hline
			3.15		& EC 82				& C			& 20080521		& 1743-038		& 3c273		& Gain calibrator too weak	\\
					& EC 90				& C			& 20080521		& 1743-038		& 3c273		& Gain calibrator too weak	\\
					& SSTc2d J182900.88+002931.5	& C			& 20080521		& 1743-038		& 3c273		& Gain calibrator too weak	\\
					& IRAS 18268-0025		& C			& 20080521		& 1743-038		& 3c273		& Gain calibrator too weak	\\
					& SSTc2d J182858.08+001724.4	& C			& 20080521		& 1743-038		& 3c273		& Gain calibrator too weak	\\
					& SSTc2d J182850.20+000949.7	& C			& 20080521		& 1743-038		& 3c273		& Gain calibrator too weak	\\
					& VV Ser			& C			& 20080521		& 1743-038		& 3c273		& Gain calibrator too weak	\\
			\hline
			3.15		& EC 82				& D			& 20080618		& 1751+096		& 3c279		&				\\
					& EC 90				& D			& 20080618		& 1751+096		& 3c279		&				\\
					& SSTc2d J182900.88+002931.5	& D			& 20080618		& 1751+096		& 3c279		&				\\
					& IRAS 18268-0025		& D			& 20080618		& 1751+096		& 3c279		&				\\
					& CoKu Ser-G3			& D			& 20080618		& 1751+096		& 3c279		&				\\
					& VV Ser			& D			& 20080618		& 1751+096		& 3c279		&				\\
					& SSTc2d J182858.08+001724.4	& D			& 20080619		& 1751+096		& 3c273		&				\\
					& SSTc2d J182850.20+000949.7	& D			& 20080619		& 1751+096		& 3c273		&				\\
					& SSTc2d J182944.10+003356.1	& D			& 20080619		& 1751+096		& 3c273		&				\\
					& SSTc2d J182936.19+004216.7	& D			& 20080619, 20080622	& 1751+096		& 3c273		&				\\
					& GSC 00446-00153		& D			& 20080622		& 1751+096		& 3c273		&				\\
			\hline
		\end{tabular}
		\label{tab: observation log CARMA}
	\end{table*}

	Finally, eight sources in Serpens were observed with the (Very Large Array) VLA at 7~mm and at 1.3, 3.6, and 
	6~cm. A full log of those observations is given in Table~\ref{tab: observation log VLA}.
	\begin{table*}
		\caption[]{Overview of the VLA observations.}
		\centering
		\begin{tabular}{lcccccl}
			\hline
			\hline
			Obs. date	& Wavelengths	& Config.	& Target sources$^\mathrm{a}$			& Gain calibrator(s) 	& Flux cal.	& Notes				\\
					& (mm)		&		&						&		     	&  		& 				\\
			\hline
			20080310	& 6.92, 6.93	& C		& CoKu Ser G3, EC 82				& 1824+013		& 3c286		& Clouds forming		\\
					&		&		& 182900.88+002931.5, EC 90, VV Ser, 		&			&		&				\\
					&		&		& EC 97, 182850.20+000949.7			&			&		&				\\	
			20080311	& 13.4		& C		& CoKu Ser G3, 182900.88+002931.5,		& 1851+005		& 3c286		& Clear sky			\\
					&		&		& EC 90, VV Ser, EC 97				&			&		&				\\
			20080313	& 13.4		& C		& 182850.20+000949.7, 			 	& 1851+005		& 3c286		& Clouds forming		\\
					&		&		& 182909.80+003445.9, EC 82			&			&		&				\\	
			20080314	& 35.5		& C		& EC 82, EC 90, 182950.20+000949.7,		& 1804+010		& 3c286		& Clear sky; high winds		\\
					&		&		& EC 97, VV Ser,				&			&		&				\\
					&		&		& CoKu Ser G3, 182900.88+002931.5,		&			&		&				\\
			20080315	& 61.4, 62.0	& C		& EC 82, EC 90, 182950.20+000949.7,		& 1804+010		& 3c286		& Clear sky; high winds		\\
					&		&		& EC 97, 					&			&		&				\\
					&		&		& CoKu Ser G3, 18290088+0029315,		&			&		&				\\
					&		&		& VV Ser, 18285020+0009497			&			&		&				\\
			\hline
		\end{tabular}
		\label{tab: observation log VLA}
	 \begin{list}{}{}
	  \item[$^\mathrm{a}$]	In the case of SSTc2d names, only the coordinates (in J2000) are shown.
	 \end{list} 
	\end{table*}

\section{Results}\label{app: results}

\begin{figure*}
		\begin{center}
			\includegraphics[width=5cm]{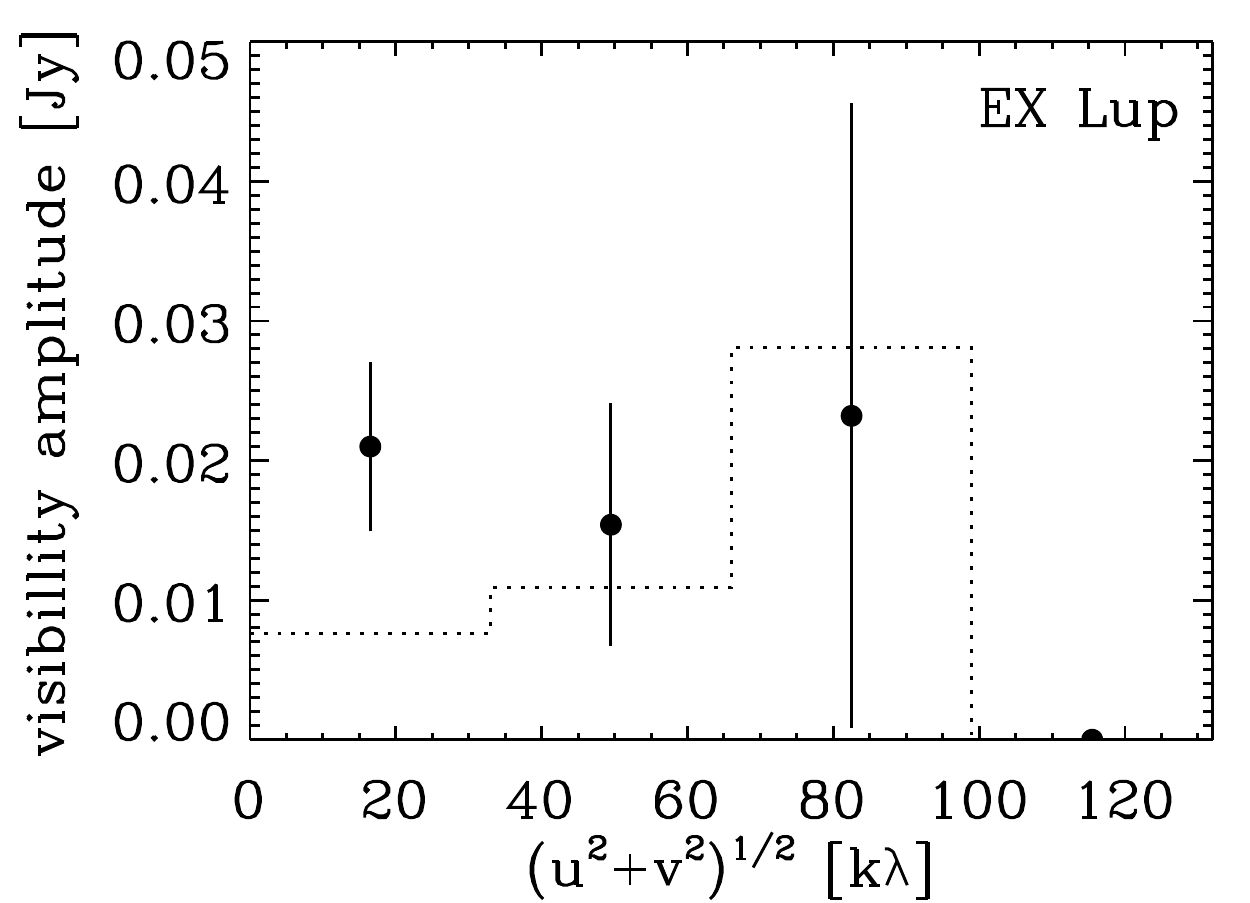}
			\includegraphics[width=5cm]{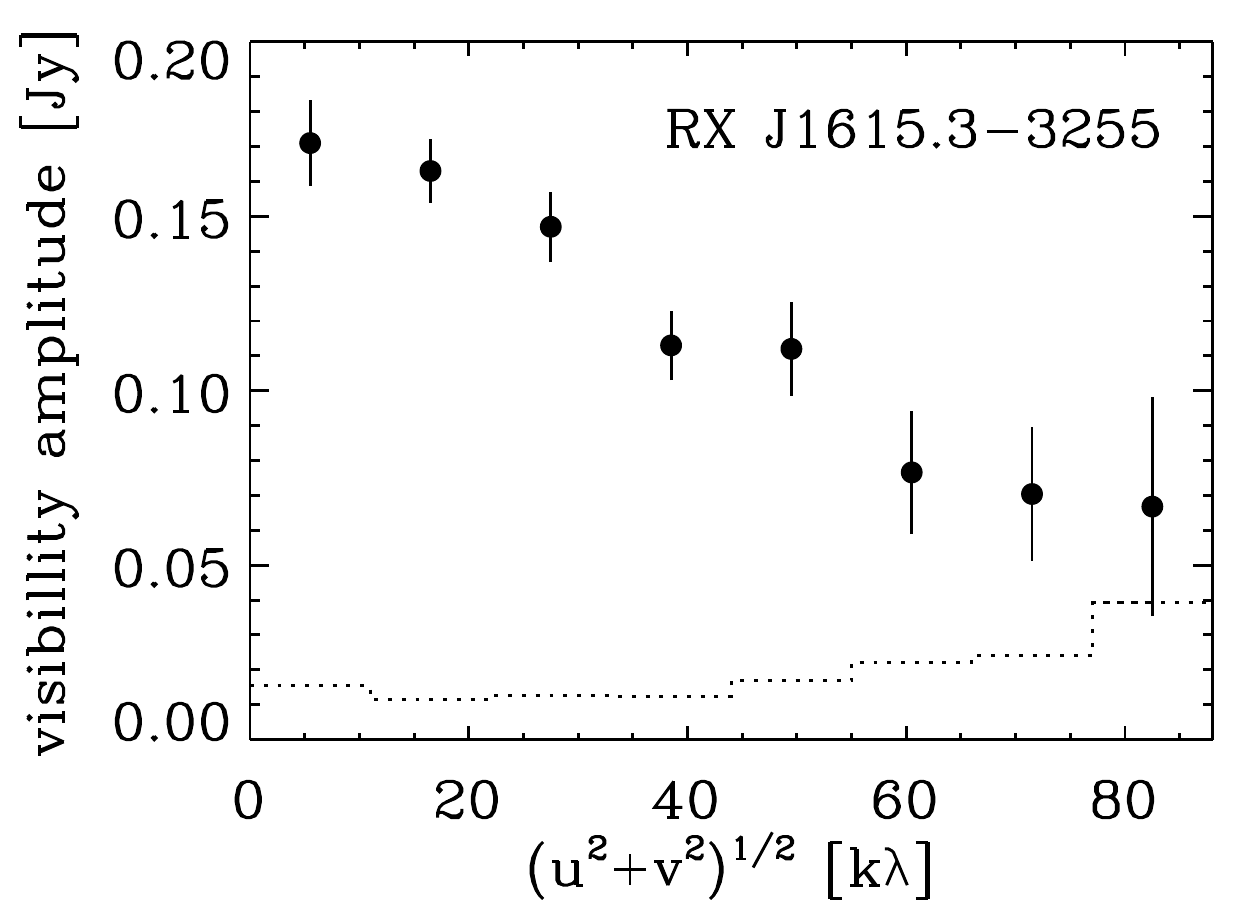}
			\includegraphics[width=5cm]{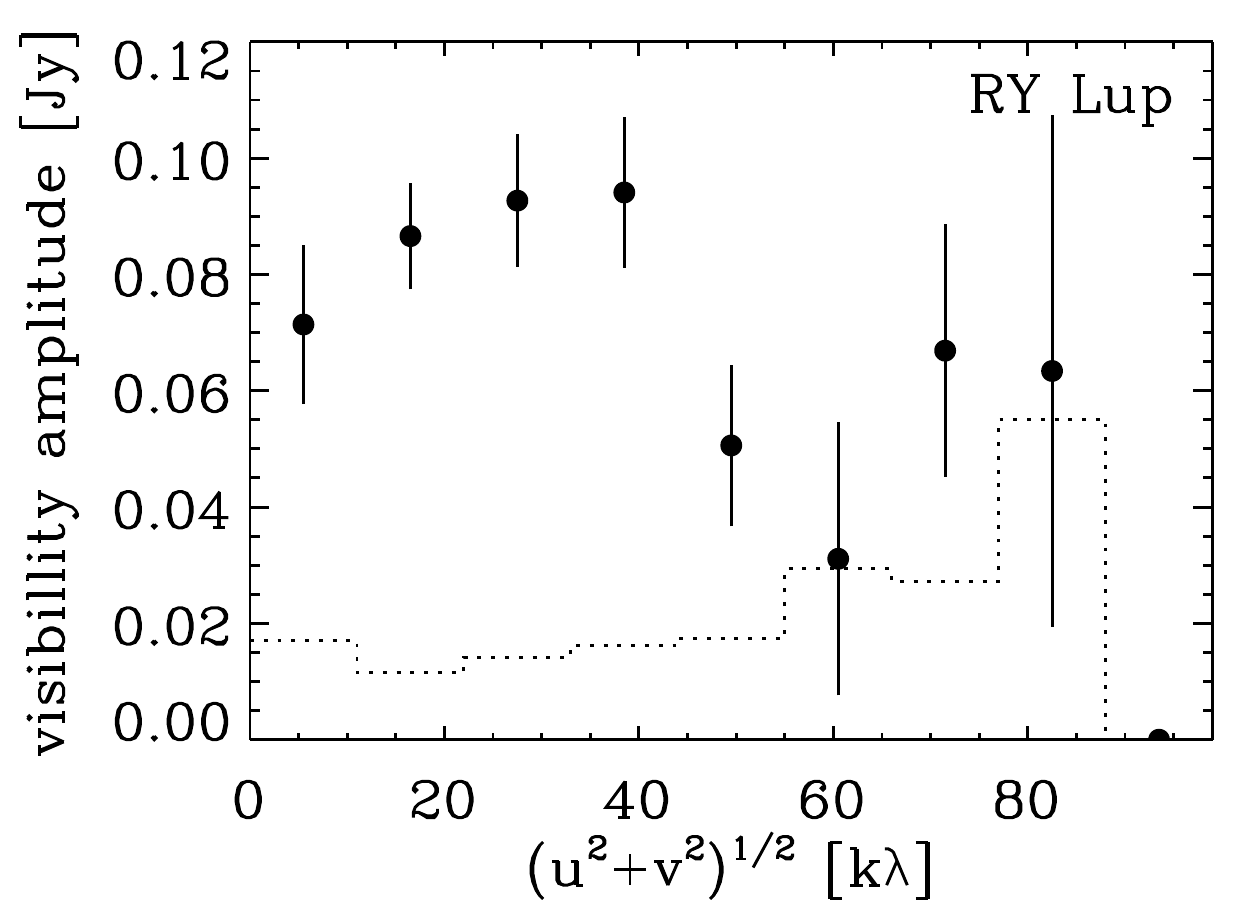}\\
			\includegraphics[width=5cm]{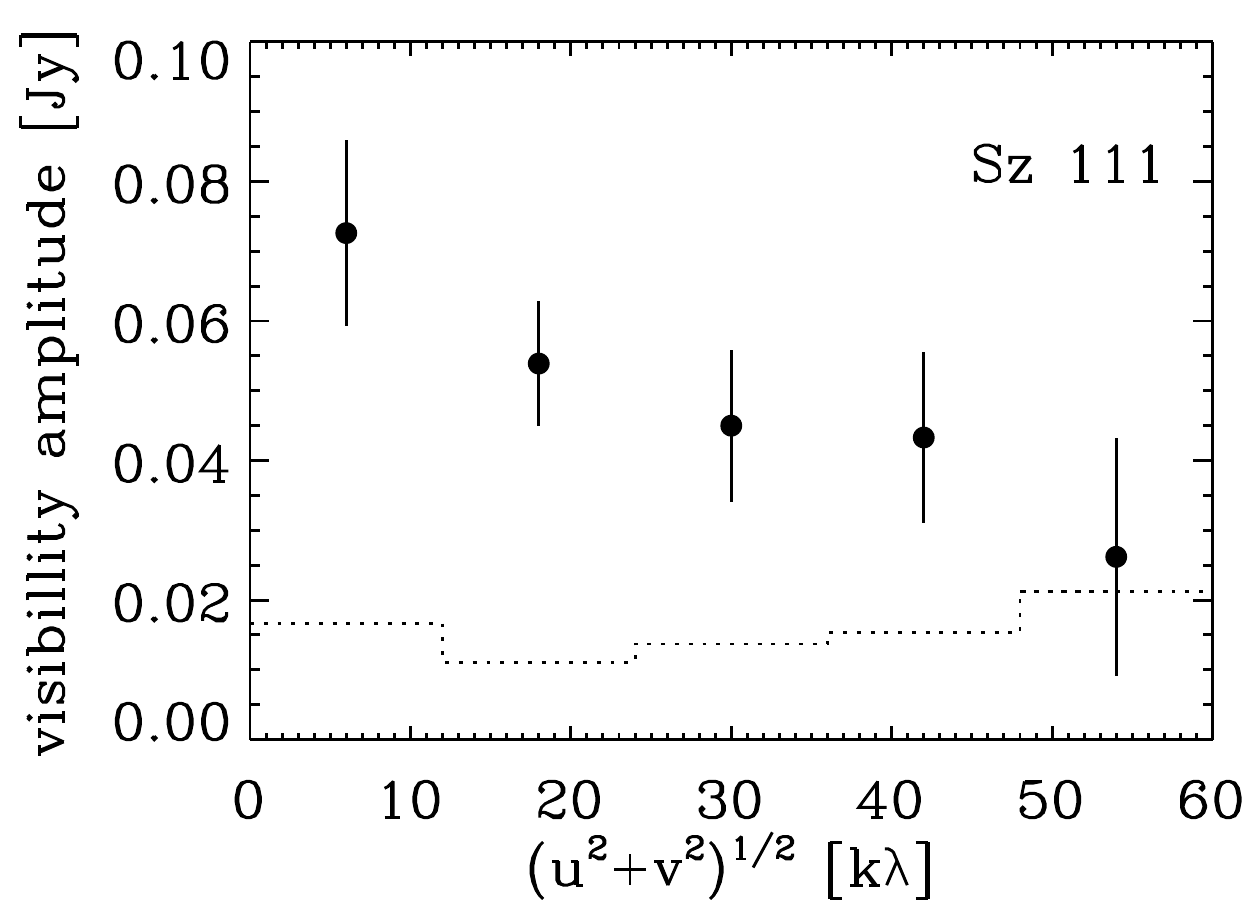}
			\includegraphics[width=5cm]{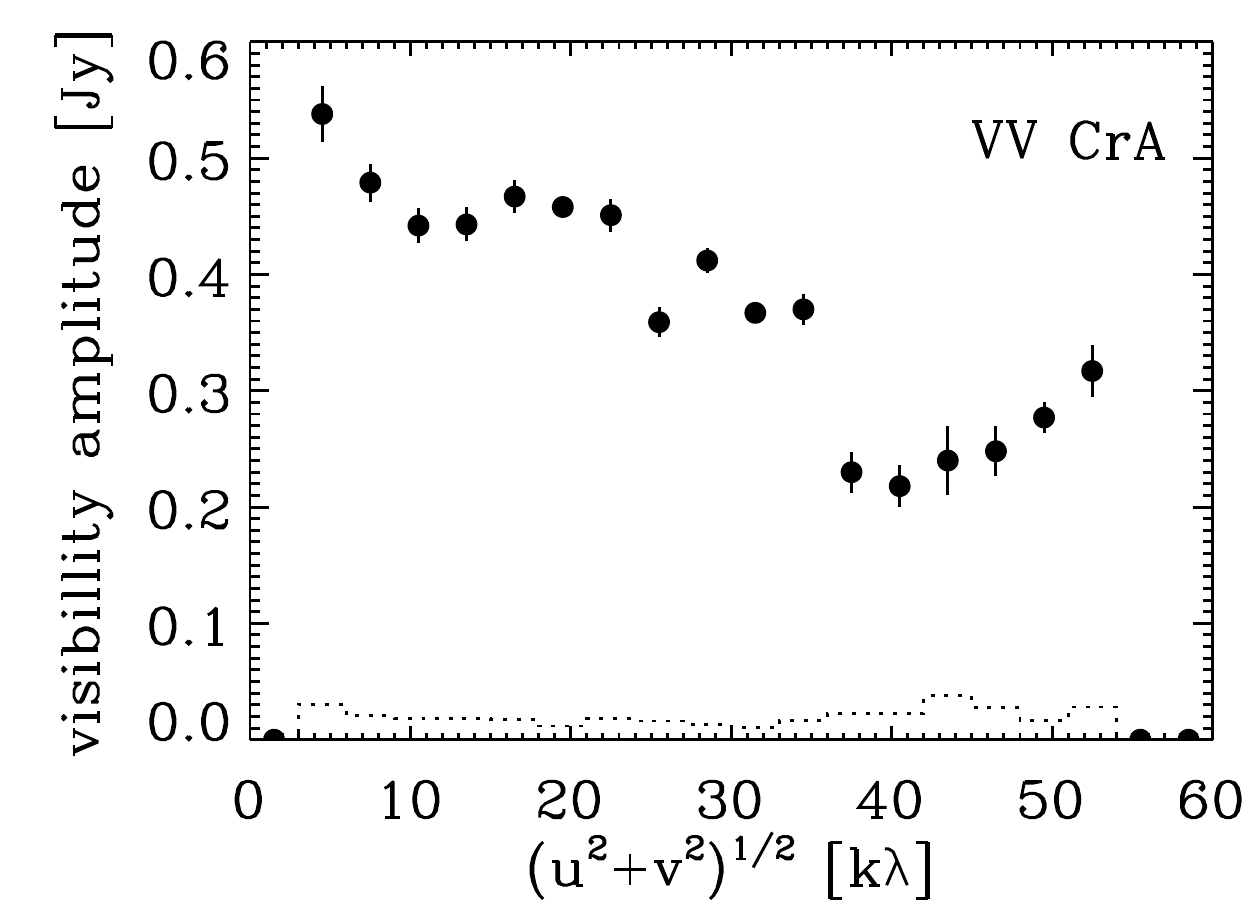}
			\includegraphics[width=5cm]{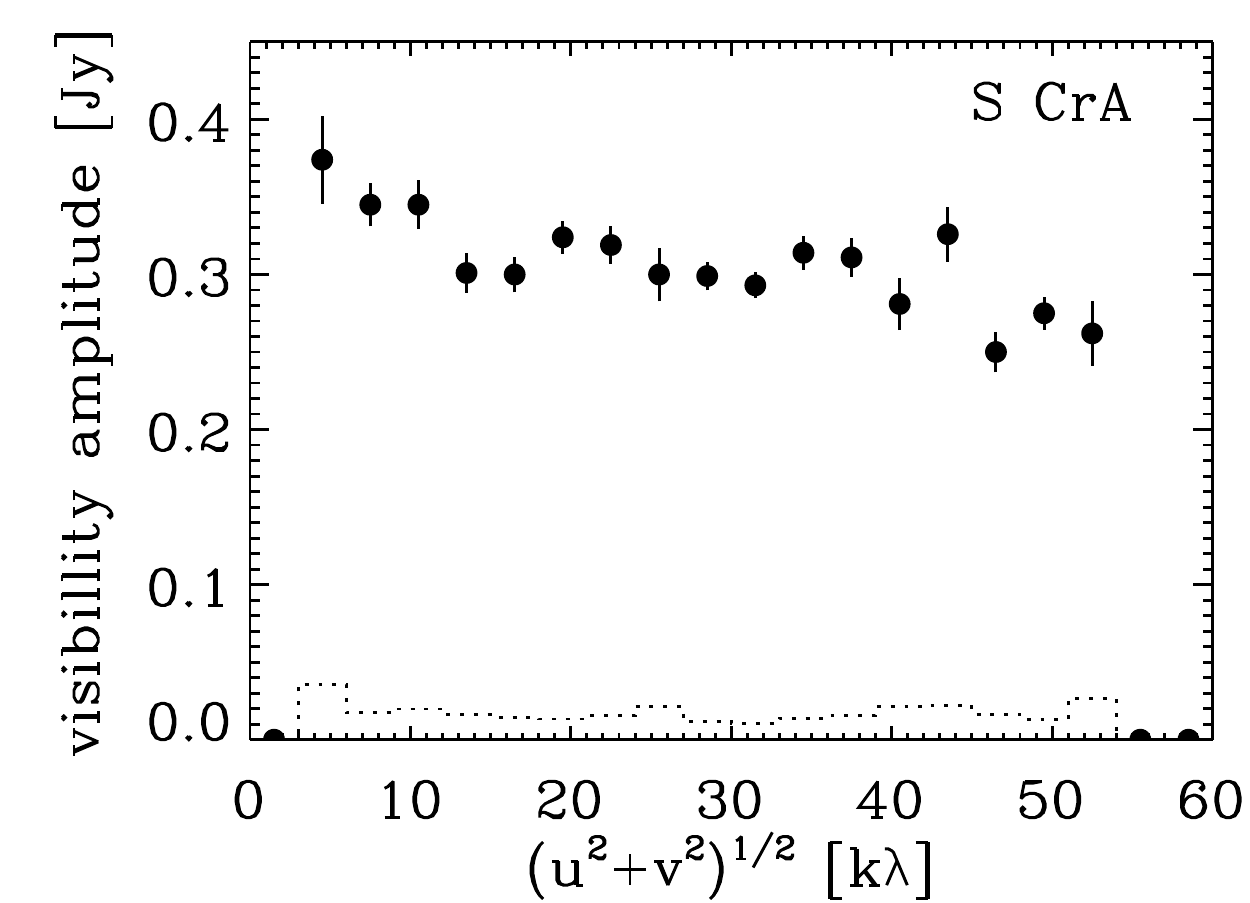}\\
			\includegraphics[width=5cm]{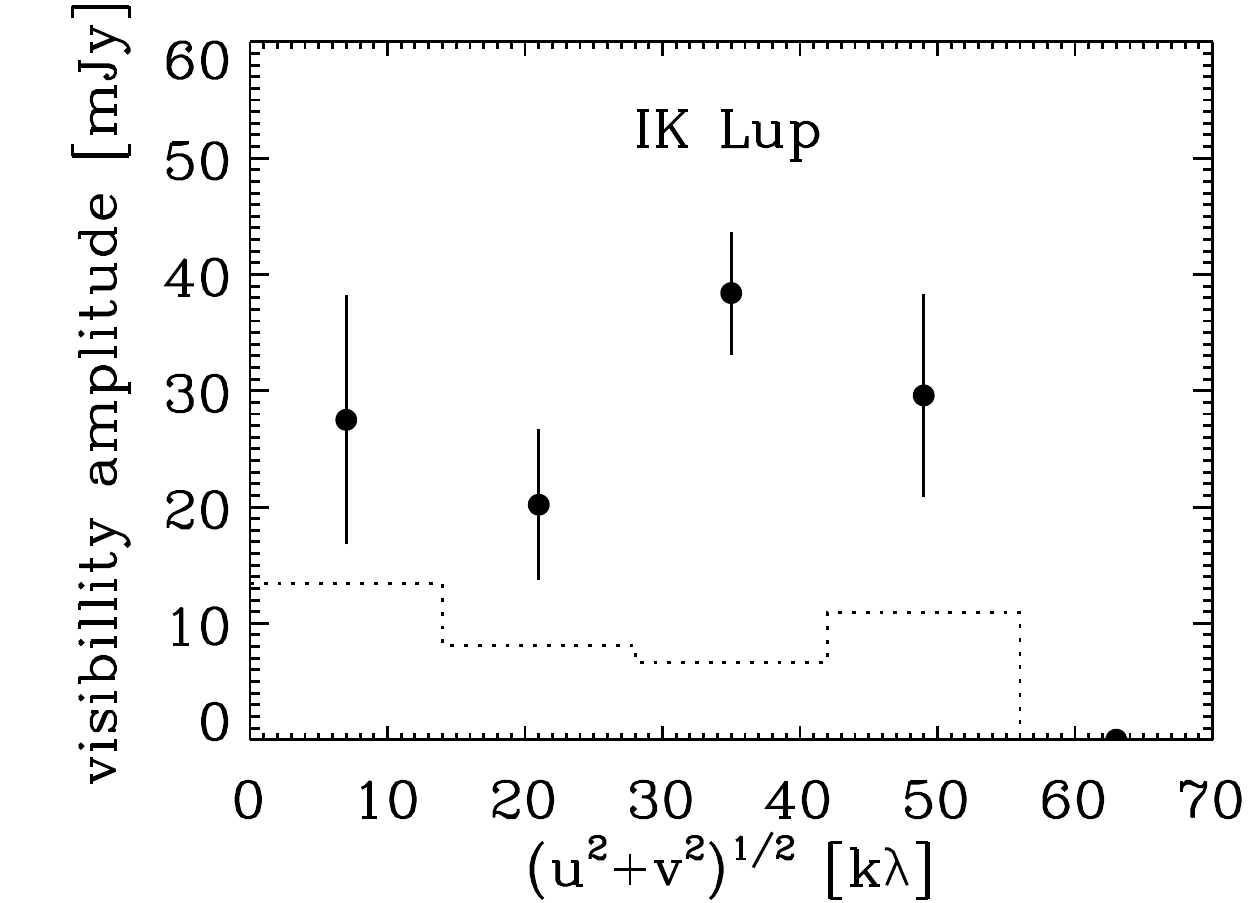}
			\includegraphics[width=5cm]{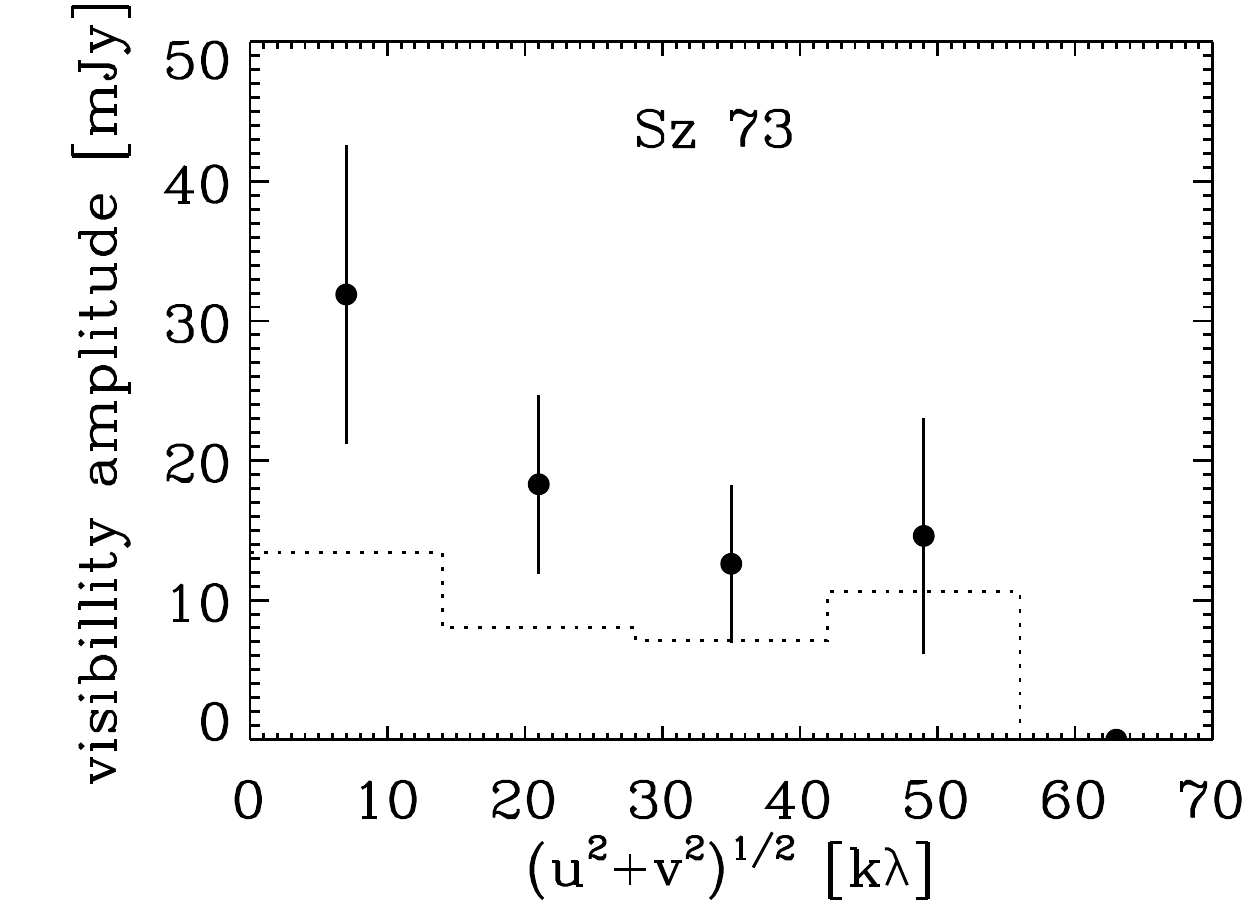}
			\includegraphics[width=5cm]{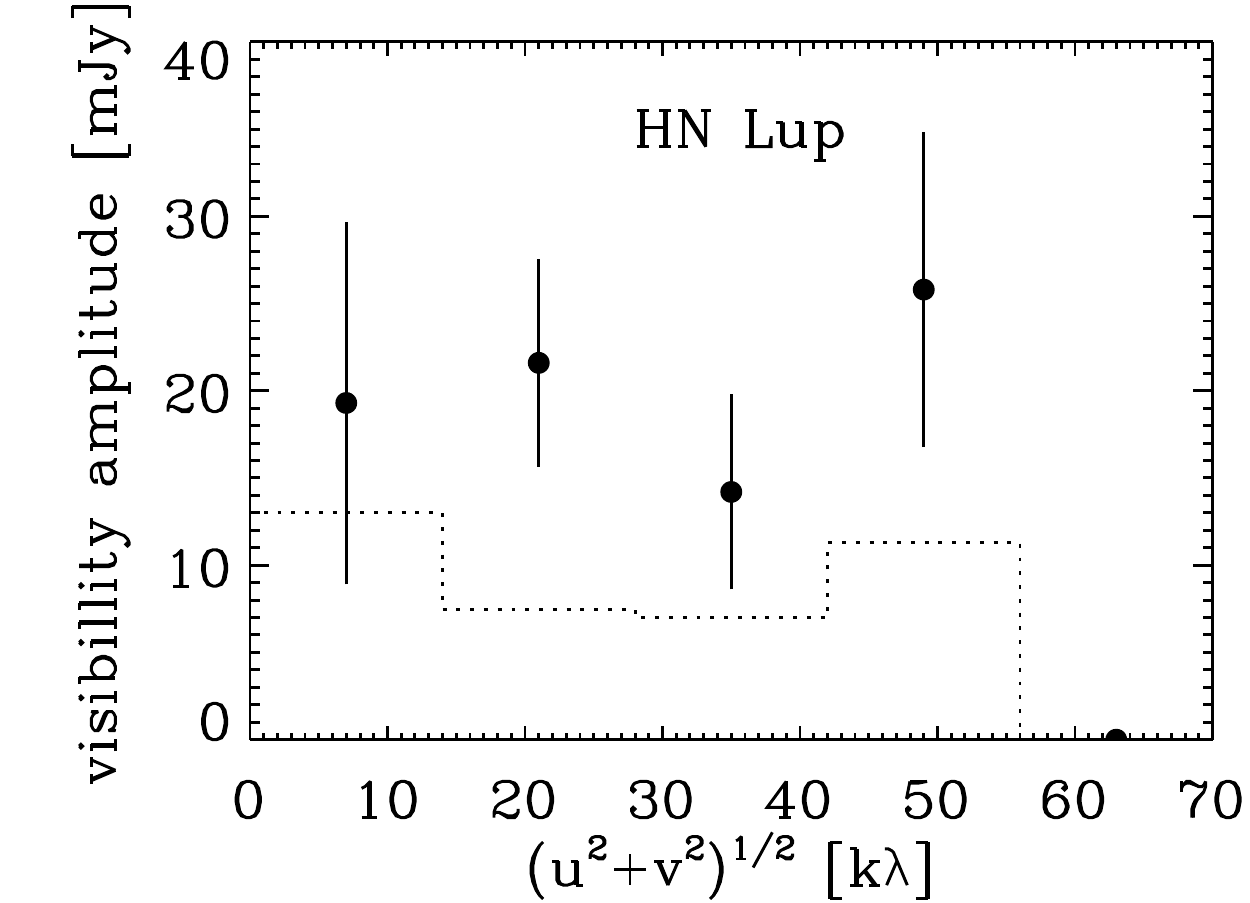}\\
			\includegraphics[width=5cm]{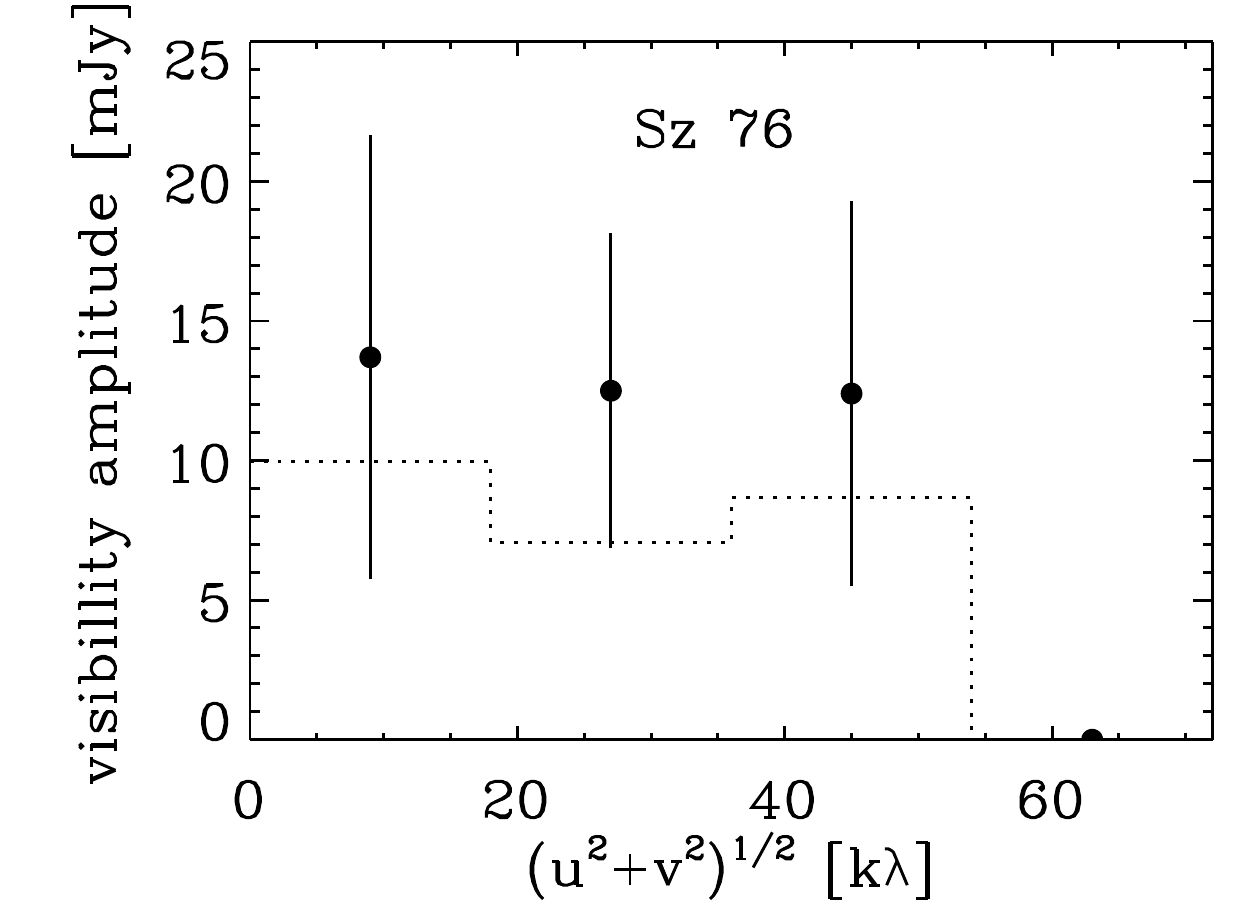}
			\includegraphics[width=5cm]{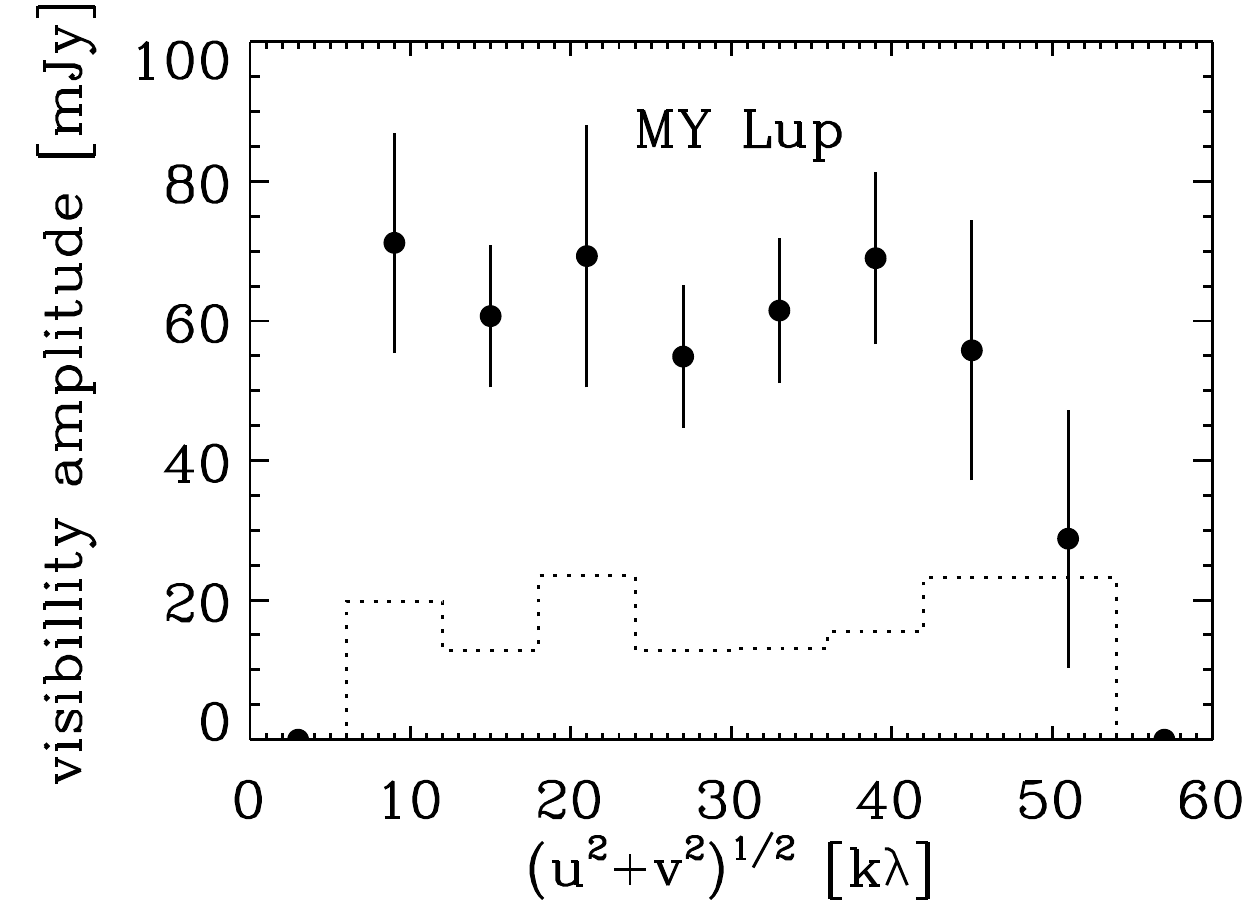}
		\end{center}
		\caption{Amplitude as a function of ($u$, $v$) distance for sources detected with the SMA at 
			1~mm.}\label{fig: SMA UVdist}
	\end{figure*}

	The complete results of the SMA observations are shown in Table~\ref{tab: results log SMA}.
	\begin{table*}
		\caption[]{Complete results of SMA observations at 1.3~mm.}
		\centering
		\begin{tabular}{lcccccccc}
			\hline
			\hline
			Obs. date	& Effective	& Target source$^\mathrm{a}$	& \multicolumn{2}{c}{Continuum flux$^\mathrm{b}$}	& rms$^\mathrm{c}$	& Gaussian size		& RA$^\mathrm{b}$      	& Dec$^\mathrm{b}$	\\
					& wavelength	&				& (P)			& (G)				&			& (arcsec)		& (J2000)		& (J2000)		\\
					& (mm)		&				& \multicolumn{2}{c}{(mJy)}		     		& (mJy/bm)		&			&			&			\\
			\hline
			20080419	& 1.34		& EX Lup			& 19.3		 	& 21.3		     		& 4.0			& $1.00 \pm 1.34$	& 16 03 05.0		& -40 18 20.1		\\
					&		& RX J1615.3-3255		& 131.8			& 169.1				& 3.9			& $1.53 \pm 0.13$	& 16 15 20.2		& -32 55 05.3		\\
					&		& RY Lup			& 78.3		     	& 89.0 		     		& 5.0			& $1.14 \pm 0.30$	& 15 59 28.4 		& -40 21 51.4		\\
					&		& 161029.57-392214.7		& \multicolumn{2}{c}{$<13.1^\mathrm{d}$}		& 4.4			& --			& 16 10 29.6		& -39 22 14.4		\\
					&		& 161159.81-382338.5		& \multicolumn{2}{c}{$<12.7^\mathrm{d}$}  		& 4.2			& --			& 16 11 59.8		& -38 23 38.0		\\
					&		& Sz 111			& 49.3			& 52.5				& 4.2			& $0.78 \pm 0.67$	& 16 08 54.7		& -39 37 43.6		\\
					&		& Sz 91				& \multicolumn{2}{c}{$<13.0^\mathrm{d}$}  		& 4.3			& --			& 16 07 11.6		& -39 03 47.1		\\
					&		& Sz 96				& \multicolumn{2}{c}{$<12.6^\mathrm{d}$}		& 4.2			& --			& 16 08 12.6		& -39 08 33.3		\\	
			20081001	& 1.35		& VV CrA			& 349.6			& 367.0				& 5.1			& $0.96 \pm 0.06$	& 19 03 06.8		& -37 12 49.3		\\
					&		& S CrA				& 301.4			& 322.2				& 3.5			& $0.91 \pm 0.07$	& 19 01 08.6		& -36 57 20.6		\\
					&		& DG CrA			& \multicolumn{2}{c}{$<6.6^\mathrm{d}$}			& 2.2			& --			& 19 01 55.2		& -37 23 40.5	     	\\
			20090507	& 1.34		& IK Lup			& 29.4			& 29.4				& 2.8			& (unresolved)		& 15 39 27.8		& -34 46 17.8		\\
					&		& Sz 66				& \multicolumn{2}{c}{$<8.3^\mathrm{d}$}			& 2.8			& --			& 15 39 28.3		& -34 46 18.0		\\
					&		& HM Lup			& \multicolumn{2}{c}{$<10.2^\mathrm{d}$}		& 3.4			& --			& 15 47 50.6		& -35 28 35.3		\\
					&		& Sz 73	a			& 16.2			& --$^\mathrm{e}$		& 2.9			& --			& 15 47 57.0		& -35 24 35.9		\\
					&		& Sz 73 b			& 15.8			& --$^\mathrm{e}$		& 2.9			& --			& 15 47 57.1		& -35 14 40.0		\\
					&		& Sz 74				& 15.1			& 15.1				& 3.0			& (unresolved)		& 15 48 05.3		& -35 15 53.8		\\
					&		& Sz 76				& 12.4			& 12.4				& 3.3			& (unresolved)		& 15 49 30.8		& -35 49 51.2		\\
					&		& Sz 77				& \multicolumn{2}{c}{$<9.5^\mathrm{d}$}			& 3.2			& --			& 15 51 47.0		& -35 56 44.1		\\
					&		& MY Lup			& 56.4			& 66.1				& 3.4			& $1.43 \pm 0.51$	& 16 00 44.5		& -41 55 31.5		\\
					&		& Sz 102			& \multicolumn{2}{c}{$<11.4^\mathrm{d}$}		& 3.8			& --			& 16 08 29.7		& -39 03 11.0		\\
			\hline
			\hline
		\end{tabular}
		\label{tab: results log SMA}
	 \begin{list}{}{}
	  \item[$^\mathrm{a}$]	In the case of SSTc2d names, only the coordinates (in J2000) are shown.
	  \item[$^\mathrm{b}$]	Continuum flux and position are from fits in the ($u$, $v$) plane. For sources that were detected at 
	  	3$\sigma$, both the point-source flux (P) and the integrated flux for a Gaussian (G) are shown. For sources that were not
		detected, the coordinates of the phase centre are quoted.
	  \item[$^\mathrm{c}$]	Calculated from the cleaned image.
	  \item[$^\mathrm{d}$]	Quoted value is 3$\sigma$ upper limit.
	  \item[$^\mathrm{e}$]	No circular Gaussian could be fit to the source in the ($u$, $v$) plane.
	 \end{list} 
	\end{table*}
	The amplitude as a function of ($u$, $v$) distance is plotted in Fig.~\ref{fig: SMA UVdist}.

	The complete results of the ATCA observations are shown in Table~\ref{tab: results log ATCA}. Several sources were observed at 
	the same wavelength more than once. The data for those sources were co-added in the ($u$, $v$) plane to improve the 
	signal-to-noise ratio. The resulting fluxes or 3$\sigma$ upper limits are presented in Table~\ref{tab: results} in the main text.
	\begin{table*}
		\caption[]{Complete results of ATCA observations at 3 and 7~mm.}
		\centering
		\begin{tabular}{lcccccccc}
			\hline
			\hline
			Obs. date	& Effective	& Target source		& \multicolumn{2}{c}{Continuum flux$^\mathrm{a}$}	& rms$^\mathrm{b}$	& Gaussian size		& RA$^\mathrm{a}$      	& Dec$^\mathrm{a}$	\\
					& wavelength	&			& (P)			& (G)				&			& (arcsec)		& (J2000)		& (J2000)		\\
					& (mm)		&			& \multicolumn{2}{c}{(mJy)}		     		& (mJy/bm)		&			&			&			\\
			\hline
			20080712	& 3.18		& HBC 556		& \multicolumn{2}{c}{$<3.7^\mathrm{c}$}			& 1.2			& --			& 8 10 31.3		& -36 01 46.5		\\
					&		& HBC 557		& \multicolumn{2}{c}{$<3.2^\mathrm{c}$}			& 1.1			& --			& 8 12 47.4		& -36 19 18.0		\\
					&		& HBC 559		& \multicolumn{2}{c}{$<3.3^\mathrm{c}$}			& 1.1		     	& --			& 8 13 56.4		& -36 08 02.1		\\
			20080713	& 6.83		& HBC 556		& \multicolumn{2}{c}{$<0.7^\mathrm{c}$}			& 0.2			& --			& 8 10 31.6		& -36 01 46.5		\\
					&		& HBC 557		& \multicolumn{2}{c}{$<0.4^\mathrm{c}$}			& 0.1			& --			& 8 12 47.7		& -36 19 18.0		\\
					&		& HBC 559		& \multicolumn{2}{c}{$<0.6^\mathrm{c}$}			& 0.2		     	& --			& 8 13 56.8		& -36 08 02.1		\\
			20080728	& 6.85		& HBC 559		& \multicolumn{2}{c}{$<0.3^\mathrm{c}$}		     	& 0.1			& --			& 8 13 56.8		& -36 08 02.1		\\
			20080729	& 3.18		& HBC 559		& \multicolumn{2}{c}{$<4.3^\mathrm{c}$}			& 1.4		     	& --			& 8 13 51.0		& -36 08 02.1		\\
					&		& SZ Cha		& \multicolumn{2}{c}{$<3.0^\mathrm{c}$}			& 1.0			& --			& 10 58 10.0		& -77 17 17.6		\\
					&		& Sz 32			& \multicolumn{2}{c}{$<2.9^\mathrm{c}$}			& 1.0	     		& --			& 11 09 48.0		& -76 34 26.0		\\
			20080730	& 3.18		& SZ Cha		& 3.4		     	& 5.8				& 0.5			& $1.95 \pm 0.63$	& 10 58 16.6		& -77 17 17.0		\\
			20080801	& 6.85		& Sz 111		& \multicolumn{2}{c}{$<0.6^\mathrm{c}$}			& 0.2			& --			& 16 08 53.8		& -39 37 43.1		\\
					&		& RY Lup		& \multicolumn{2}{c}{$<0.6^\mathrm{c}$}			& 0.2			& --		     	& 15 59 27.5		& -40 21 51.2		\\
					&		& RX J1615.3-3255	& \multicolumn{2}{c}{$<0.5^\mathrm{c}$}			& 0.2		     	& --			& 16 15 19.4		& -32 55 05.0 		\\
					&		& VV CrA		& 3.4			& 3.4				& 0.2			& (unresolved)		& 19 03 06.8		& -37 12 49.3		\\
					&		& S CrA			& 3.7			& 5.4				& 0.2			& $3.12 \pm 0.81$	& 19 01 08.6		& -36 57 20.3		\\
					&		& DG CrA		& \multicolumn{2}{c}{$<0.6^\mathrm{c}$}			& 0.2		     	& --	     		& 19 01 54.4		& -37 23 40.5		\\
			20080802	& 3.18		& SZ Cha		& \multicolumn{2}{c}{$<2.9^\mathrm{c}$}			& 1.0		     	& --			& 10 58 15.5		& -77 17 17.6		\\
					&		& Sz 111		& 5.7			& --$^\mathrm{e}$		& 0.7			& --			& 16 08 54.6		& -39 37 53.3		\\
					&		& RY Lup		& \multicolumn{2}{c}{$<2.3^\mathrm{c}$}		     	& 0.8	     		& --			& 15 59 28.0		& -40 21 51.2 		\\
					&		& RX J1615.3-3255	& 6.8		     	& --$^\mathrm{d}$     		& 0.6			& --			& 16 15 20.2 		& -32 55 05.6		\\
					&		& VV CrA a		& 31.0			& --$^\mathrm{e}$		& 1.2			& --$^\mathrm{e}$	& 19 03 06.8		& -37 12 49.8		\\
					&		& VV CrA b		& 25.1		uniform	& --$^\mathrm{e}$		&			& --$^\mathrm{e}$	& 19 03 06.9		& -37 12 48.3		\\
					&		& S CrA			& 22.0			& --$^\mathrm{d}$		& 1.1			& --			& 19 01 08.6		& -36 57 20.2		\\
					&		& DG CrA		& \multicolumn{2}{c}{$<3.0^\mathrm{c}$}		     	& 1.0	     		& --			& 19 01 54.9		& -37 23 40.5		\\
			20080803	& 3.18		& MY Lup		& 8.7		     	& --$^\mathrm{d}$		& 0.4			& --			& 16 00 44.5		& -41 55 31.2		\\
					&		& VV CrA a		& 23.0			& --$^\mathrm{e}$		& 1.9			& --$^\mathrm{e}$	& 19 03 06.8		& -37 12 49.9		\\
					&		& VV CrA b		& 21.9			& --$^\mathrm{e}$		&			& --$^\mathrm{e}$	& 19 03 06.9		& -37 12 48.4		\\
					&		& S CrA			& 24.9			& --$^\mathrm{d}$		& 1.9			& --			& 19 01 08.6		& -36 57 20.6		\\
					&		& DG CrA		& \multicolumn{2}{c}{$<4.3^\mathrm{c}$}		     	& 1.4	     		& --			& 19 01 54.9		& -37 23 40.5 		\\
			20080804	& 3.18		& Sz 65			& 3.4			& --$^\mathrm{d}$		& 0.4			& --			& 15 39 27.7		& -34 46 17.6		\\
					&		& Sz 66			& 2.2		     	& --$^\mathrm{d}$		&			& --			& 15 39 28.2		& -34 46 17.9		\\
			20080805	& 6.65		& MY Lup		& 1.3			& 2.9				& 0.1			& $4.75 \pm 1.31$	& 16 00 44.6		& -41 55 31.5		\\
					&		& IM Lup		& 2.2		     	& 2.2				& 0.2			& (unresolved)		& 15 56 09.2		& -37 56 06.0		\\
			\hline
		\end{tabular}
		\label{tab: results log ATCA}
	 \begin{list}{}{}
	  \item[$^\mathrm{a}$]	Continuum flux and position are from fits in the ($u$, $v$) plane. For sources that were detected at 
	  	3$\sigma$, both the point-source flux (P) and the integrated flux for a Gaussian (G) are shown. For sources that were not
		detected, the coordinates of the phase centre are quoted.
	  \item[$^\mathrm{b}$]	Calculated from the cleaned image.
	  \item[$^\mathrm{c}$]	Quoted value is 3$\sigma$ upper limit.
	  \item[$^\mathrm{d}$]	No circular Gaussian could be fit to the source in the ($u$, $v$) plane.
	  \item[$^\mathrm{e}$]	The two components could not be separated with circular Gaussian fits in the ($u$, $v$) plane. One 
	  	circular Gaussian was fitted to the binary, yielding a flux of 69.5~mJy and a size of $2.56 \pm 0.21$~arcsec on 2 
		August 2008 and a flux of 44.2~mJy and a size of $2.28 \pm 0.38$~arcsec on 3 August 2008.
	 \end{list} 
	\end{table*}
	
\begin{figure}
		\centering
		\includegraphics[width=8.3cm]{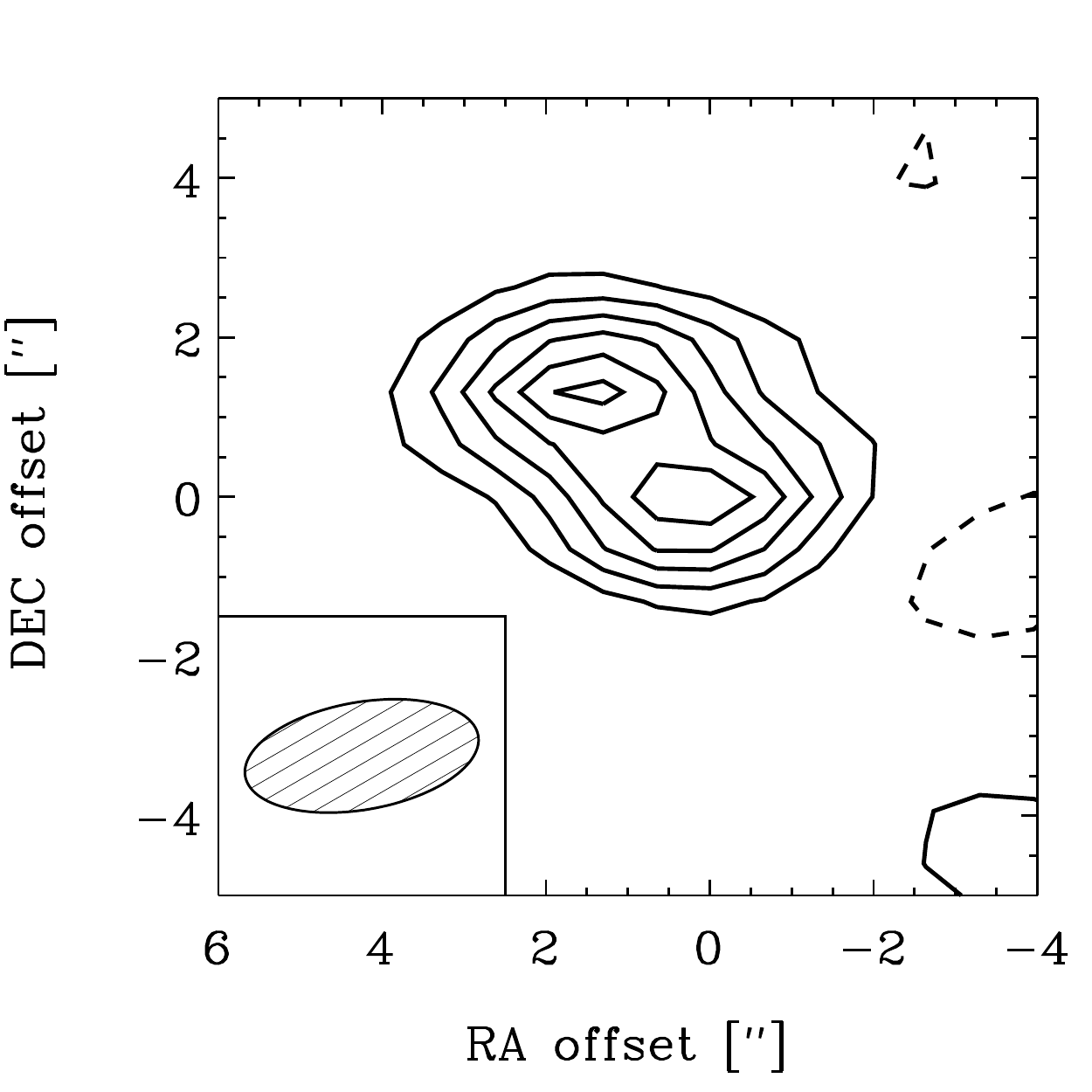}
		\caption[]{Image of VV~CrA, observed at 3.2~mm on 2 and 3 August 2008. The offsets are 
			with respect to the phase centre, which is located at 19:03:06.7, -37:12:49.7. The contours are at 2, 4, 6, ... 
			times the rms of 2.0~mJy/bm; negative contours are dashed.}\label{fig: VV CrA resolved}
	\end{figure}

\begin{figure*}
		\begin{center}
			\includegraphics[width=5cm]{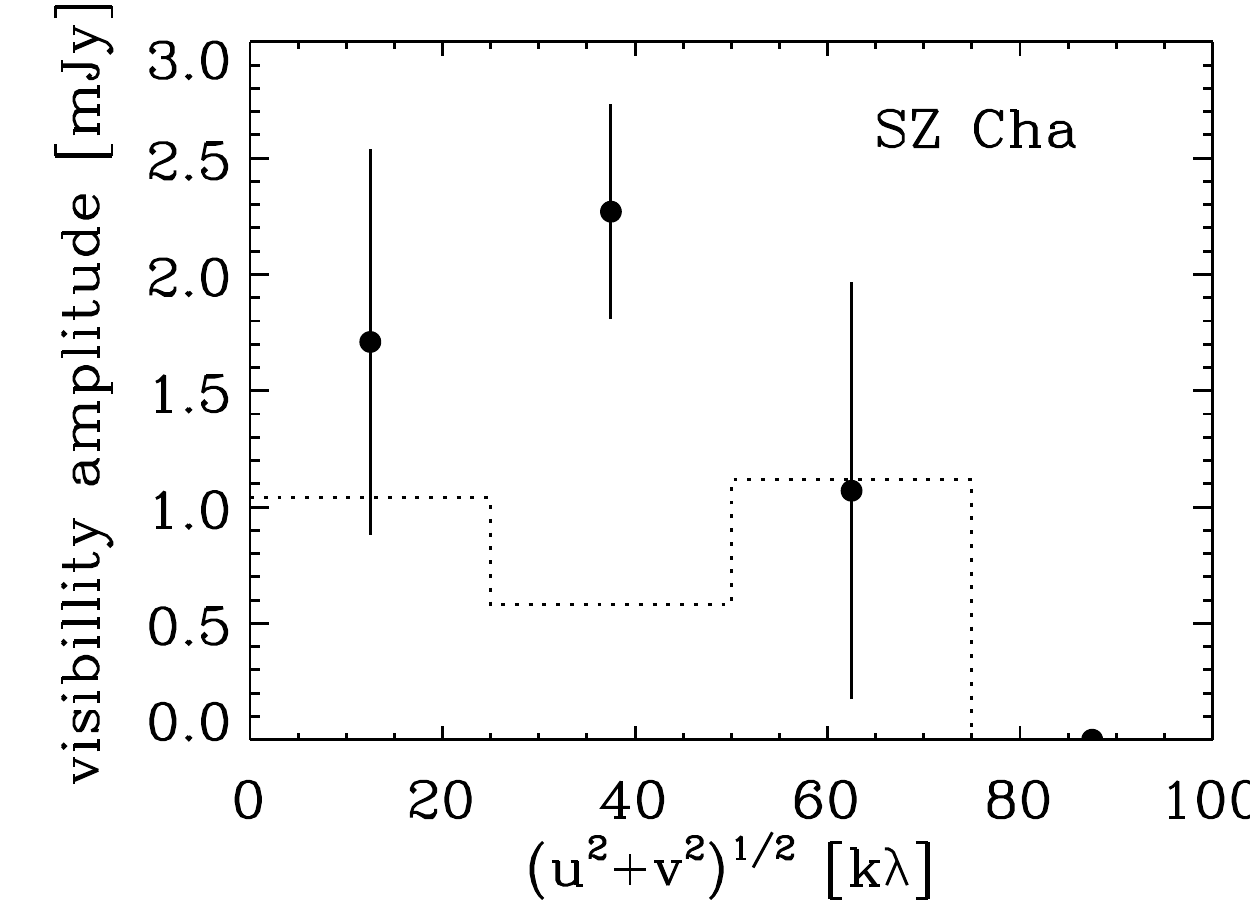}
			\includegraphics[width=5cm]{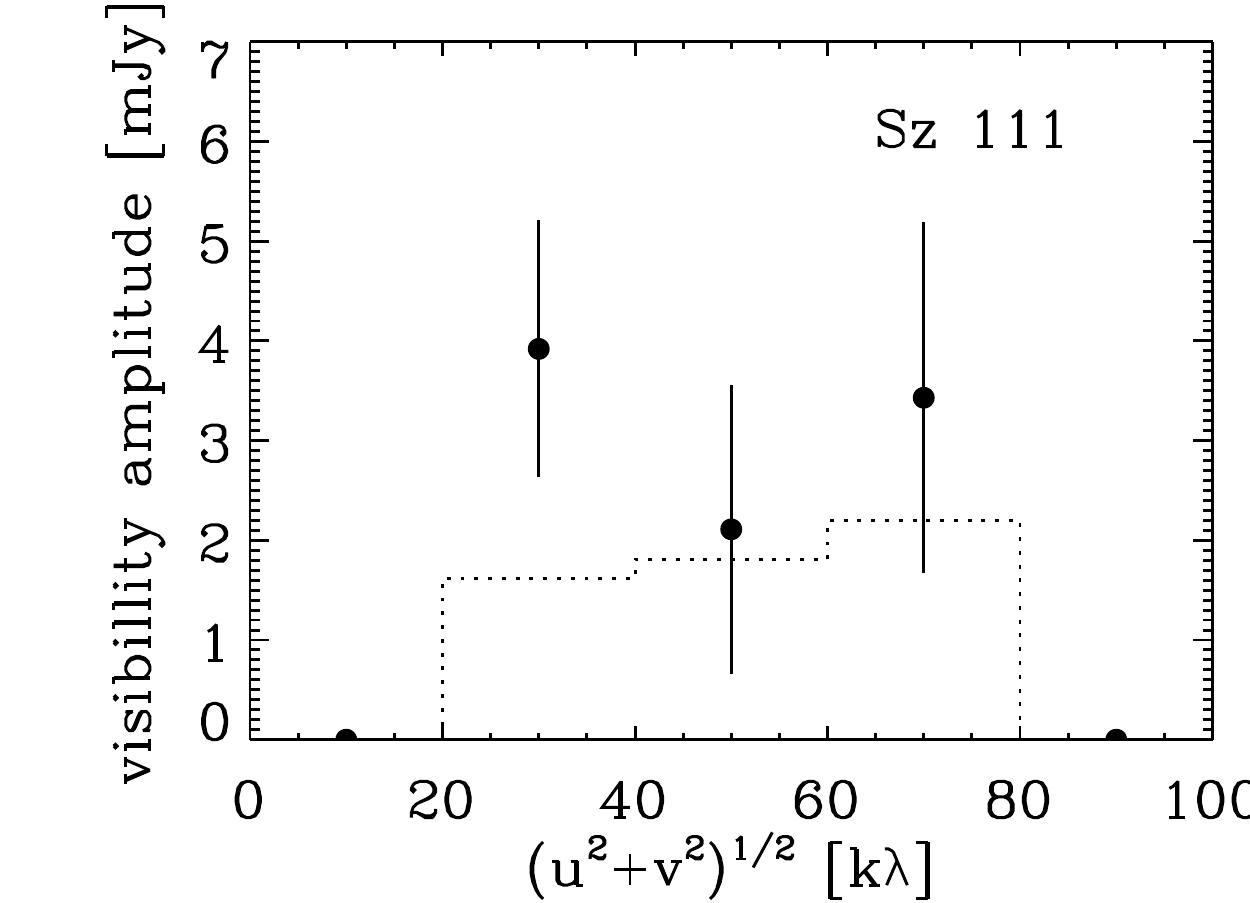}
			\includegraphics[width=5cm]{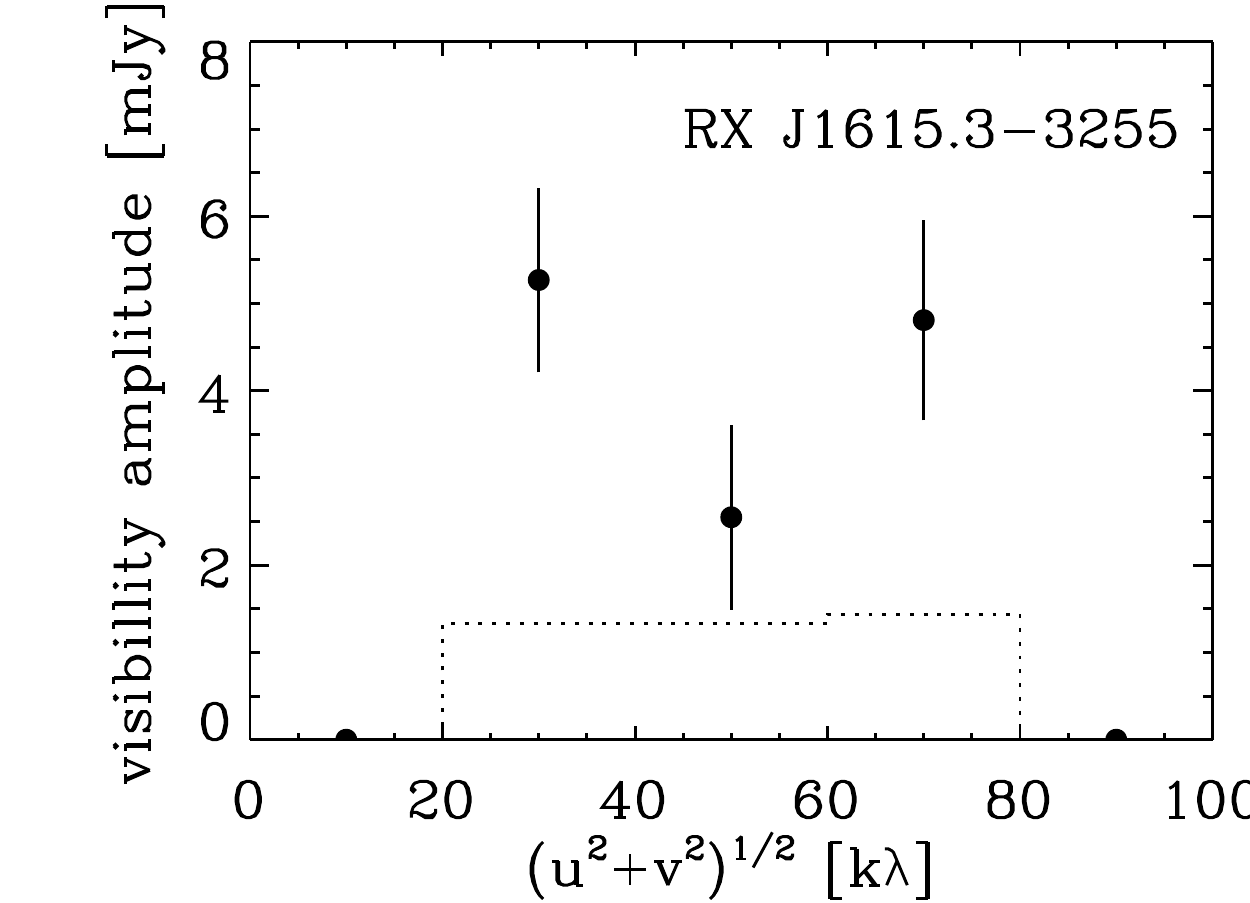}\\
			\includegraphics[width=5cm]{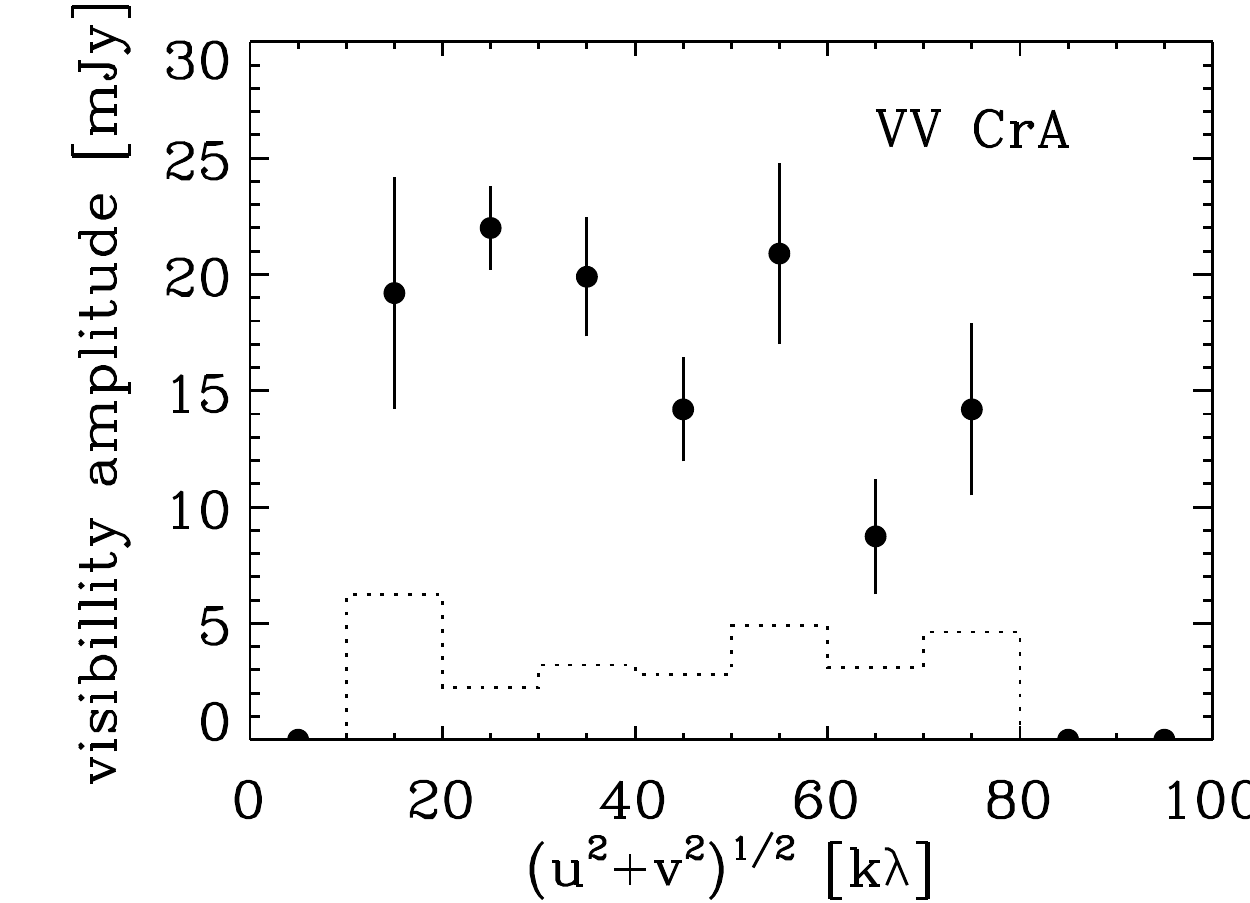}
			\includegraphics[width=5cm]{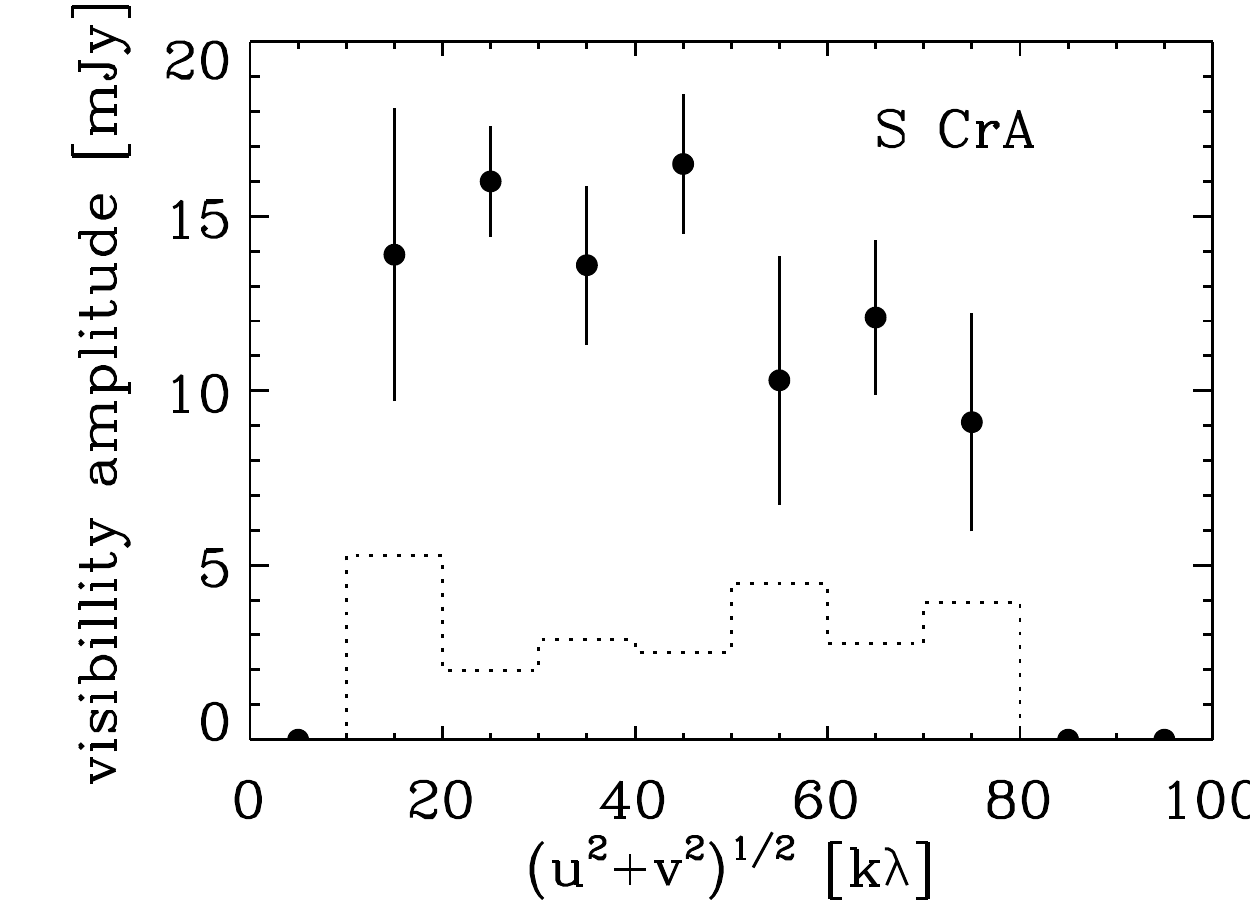}
			\includegraphics[width=5cm]{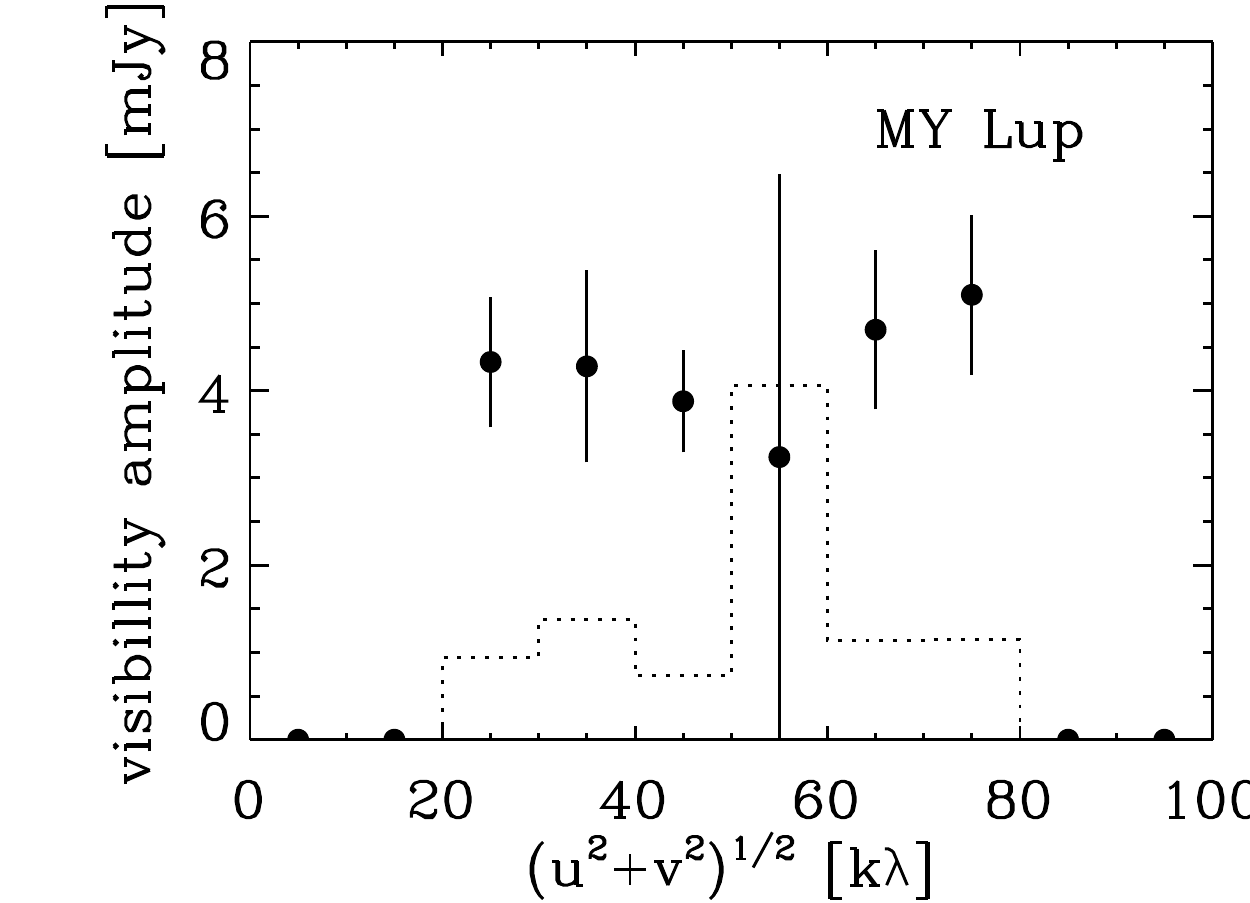}\\
			\includegraphics[width=5cm]{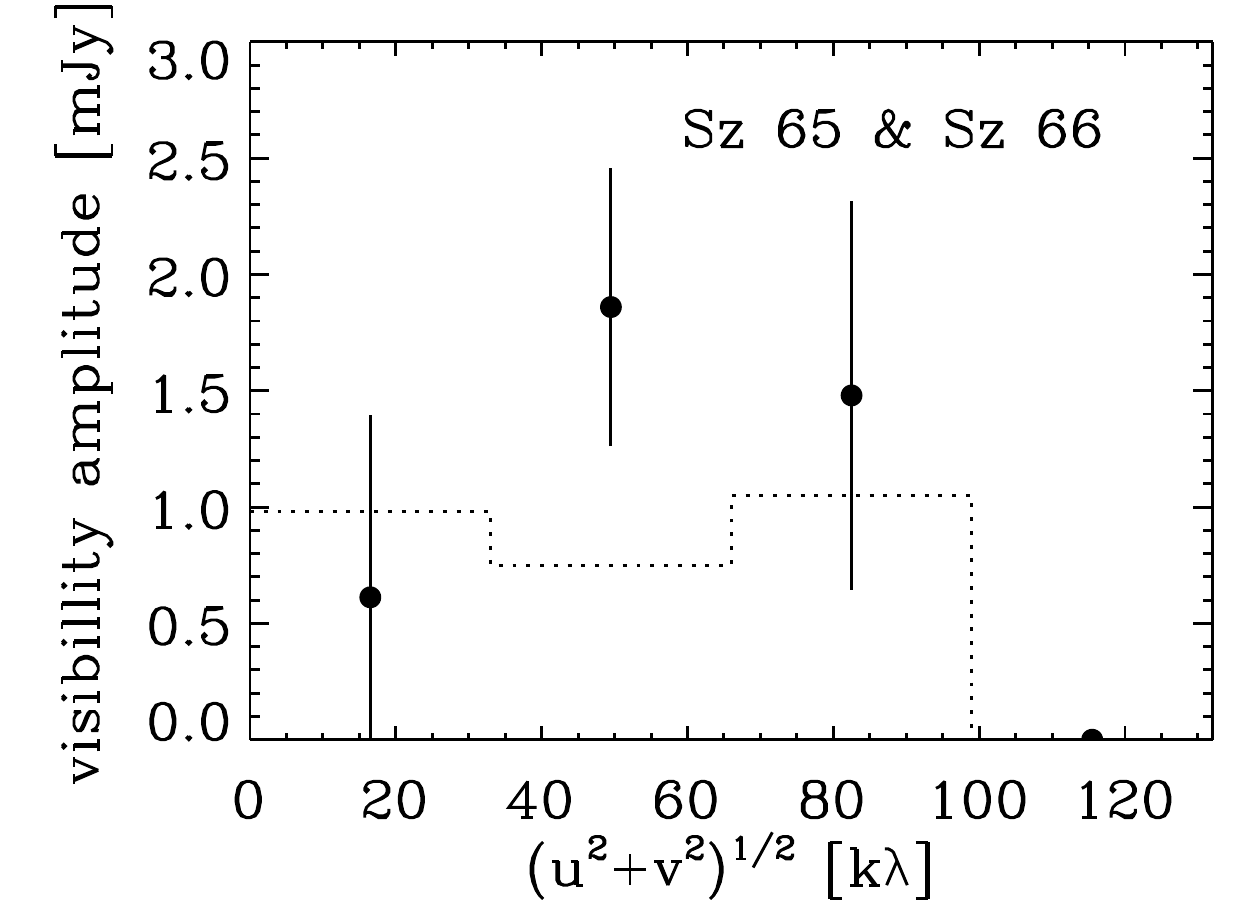}
		\end{center}
		\caption{Amplitude as a function of ($u$, $v$) distance for sources detected with the ATCA at
			3~mm.}\label{fig: ATCA UVdist 3mm}
	\end{figure*}

\begin{figure*}
		\begin{center}
			\includegraphics[width=5cm]{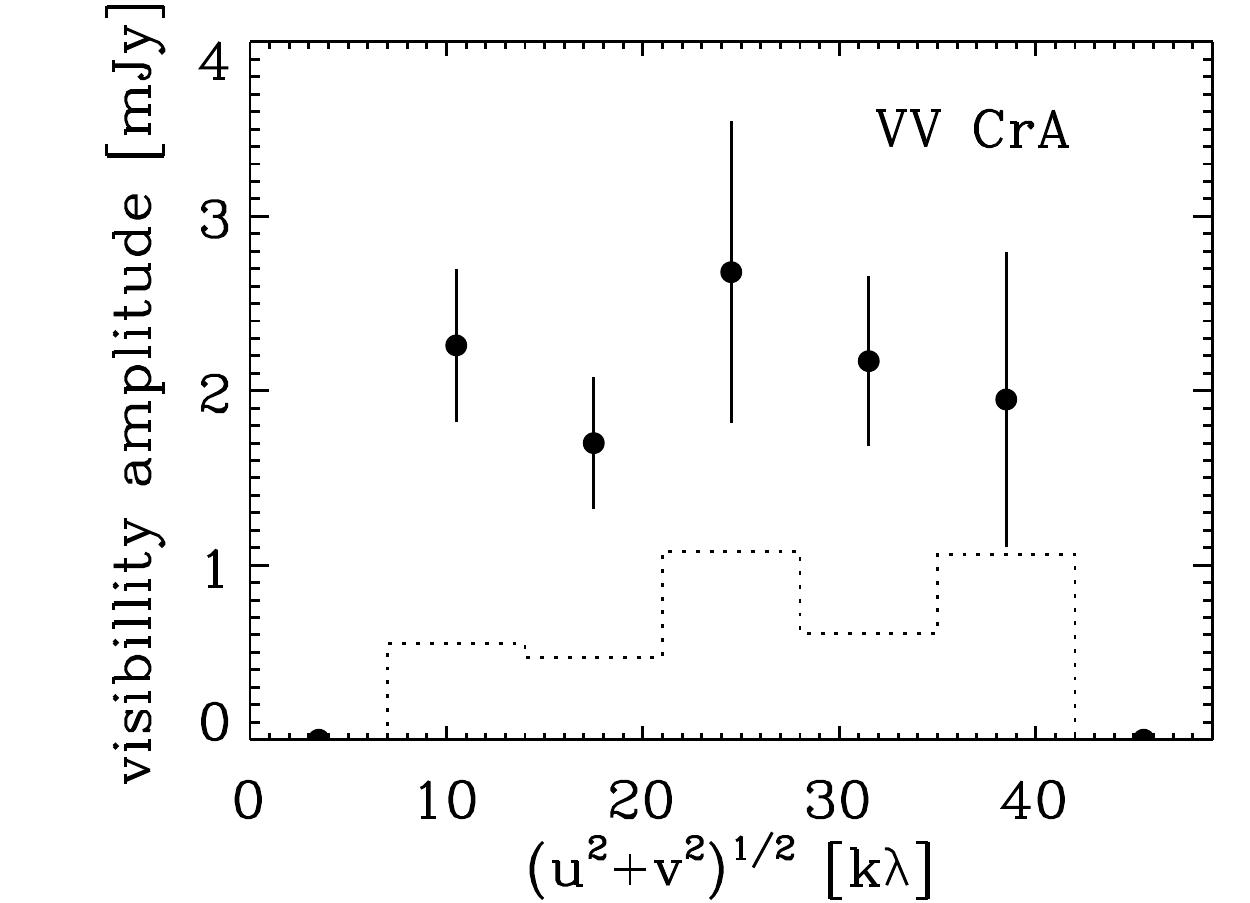}
			\includegraphics[width=5cm]{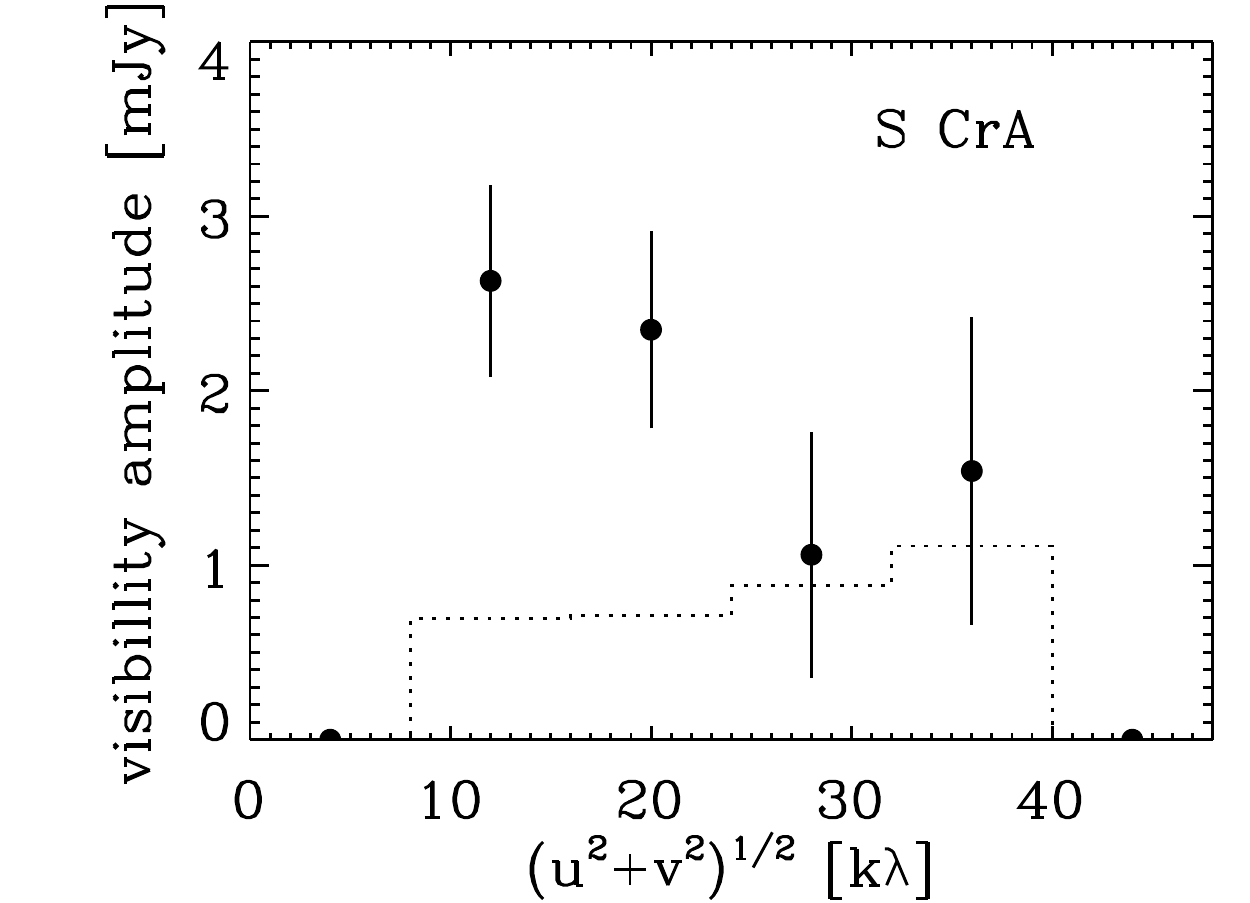}
			\includegraphics[width=5cm]{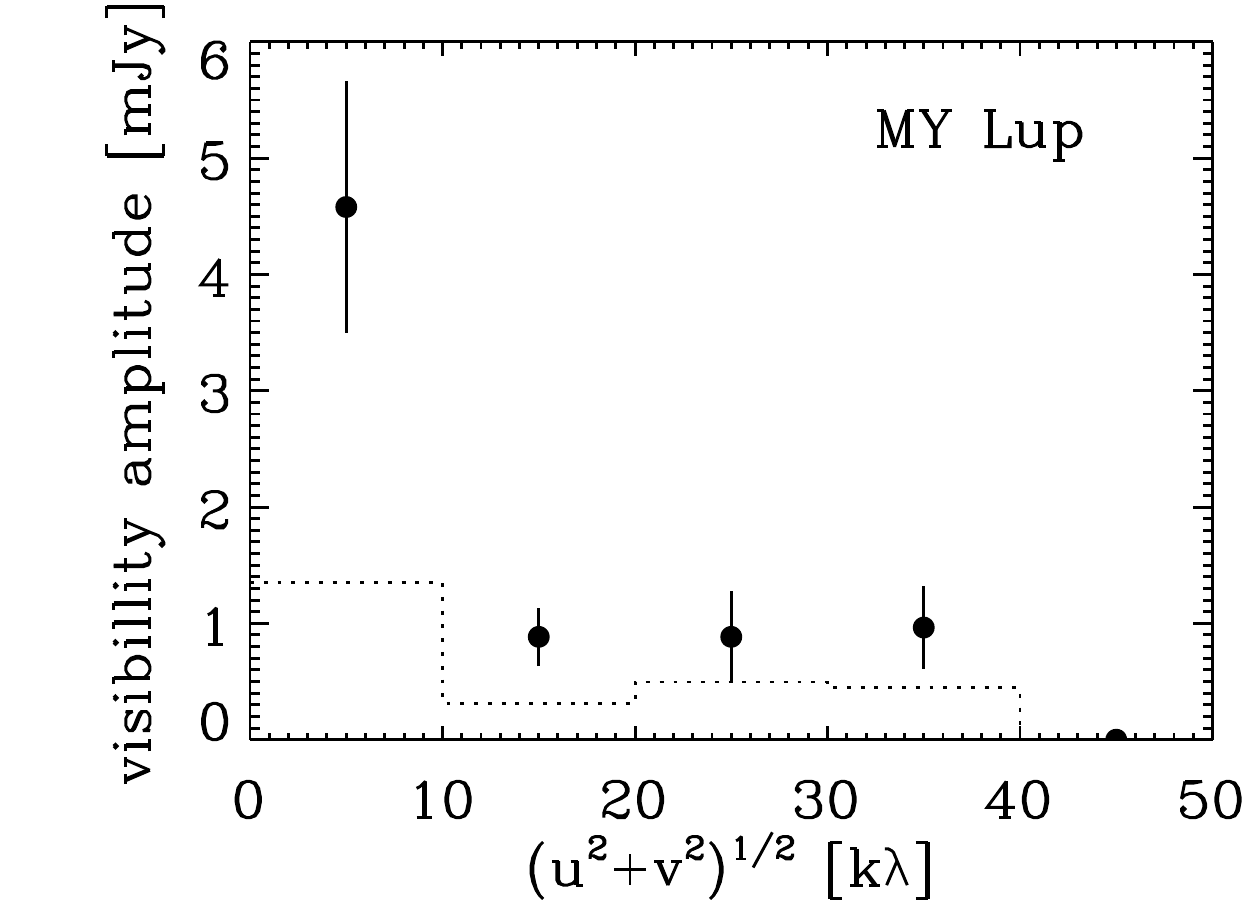}\\
			\includegraphics[width=5cm]{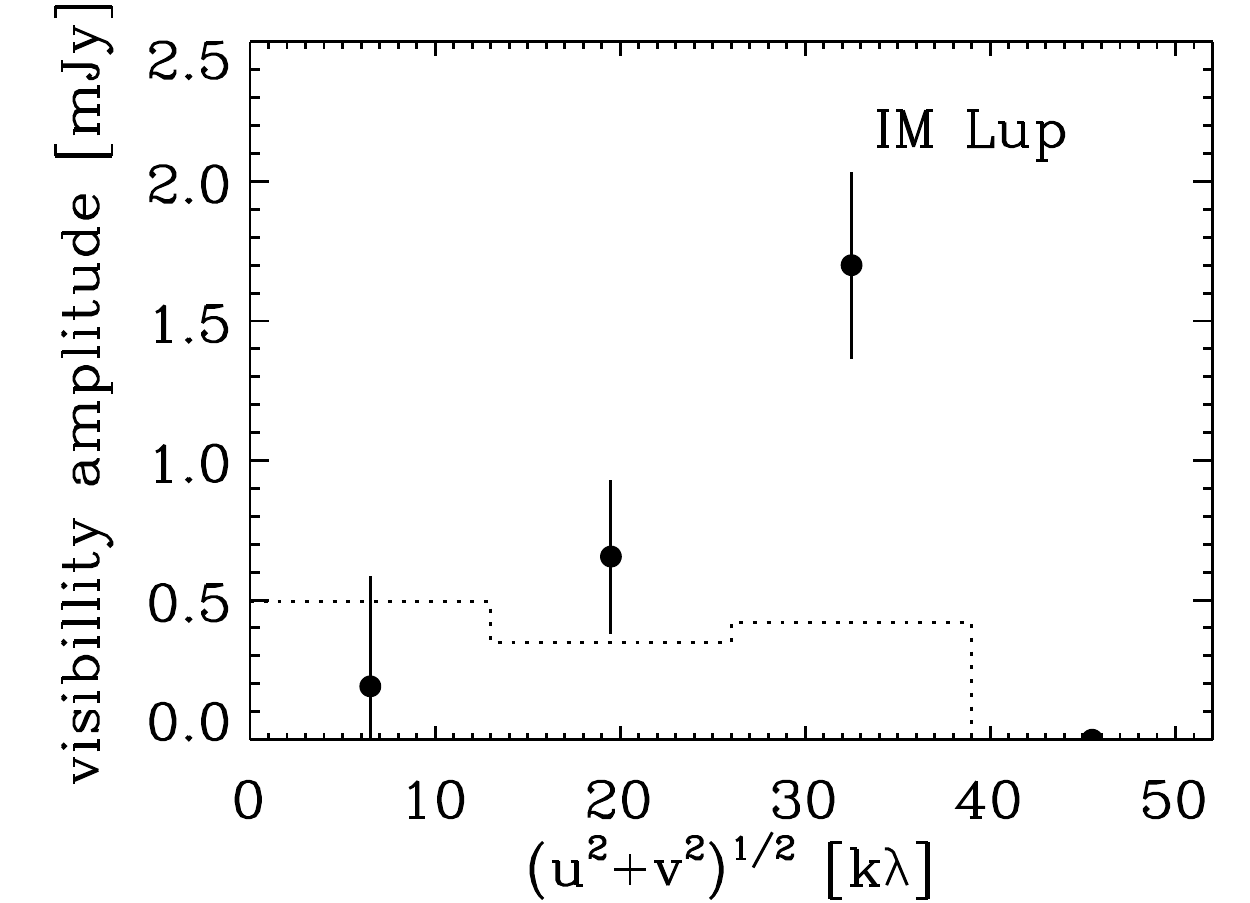}
		\end{center}
		\caption{Amplitude as a function of ($u$, $v$) distance for sources detected with the ATCA at
			7~mm.}\label{fig: ATCA UVdist 7mm}
	\end{figure*}

	The binary VV CrA was not resolved with the ATCA at 3 or 7 mm using natural weighting, which is optimised for sensitivity.
	However, using uniform weighting, which is optimised for resolution, the binary could be resolved at 3~mm. The map is shown in 
	Fig.~\ref{fig: VV CrA resolved}.
	The amplitude as a function of ($u$, $v$) distance of the sources detected with the ATCA at 3~mm is plotted in 
	Fig.~\ref{fig: ATCA UVdist 3mm}, that of the sources detected with the ATCA at 7~mm in
	Fig.~\ref{fig: ATCA UVdist 7mm}.
	
\begin{figure*}
		\begin{center}
			\includegraphics[width=5cm]{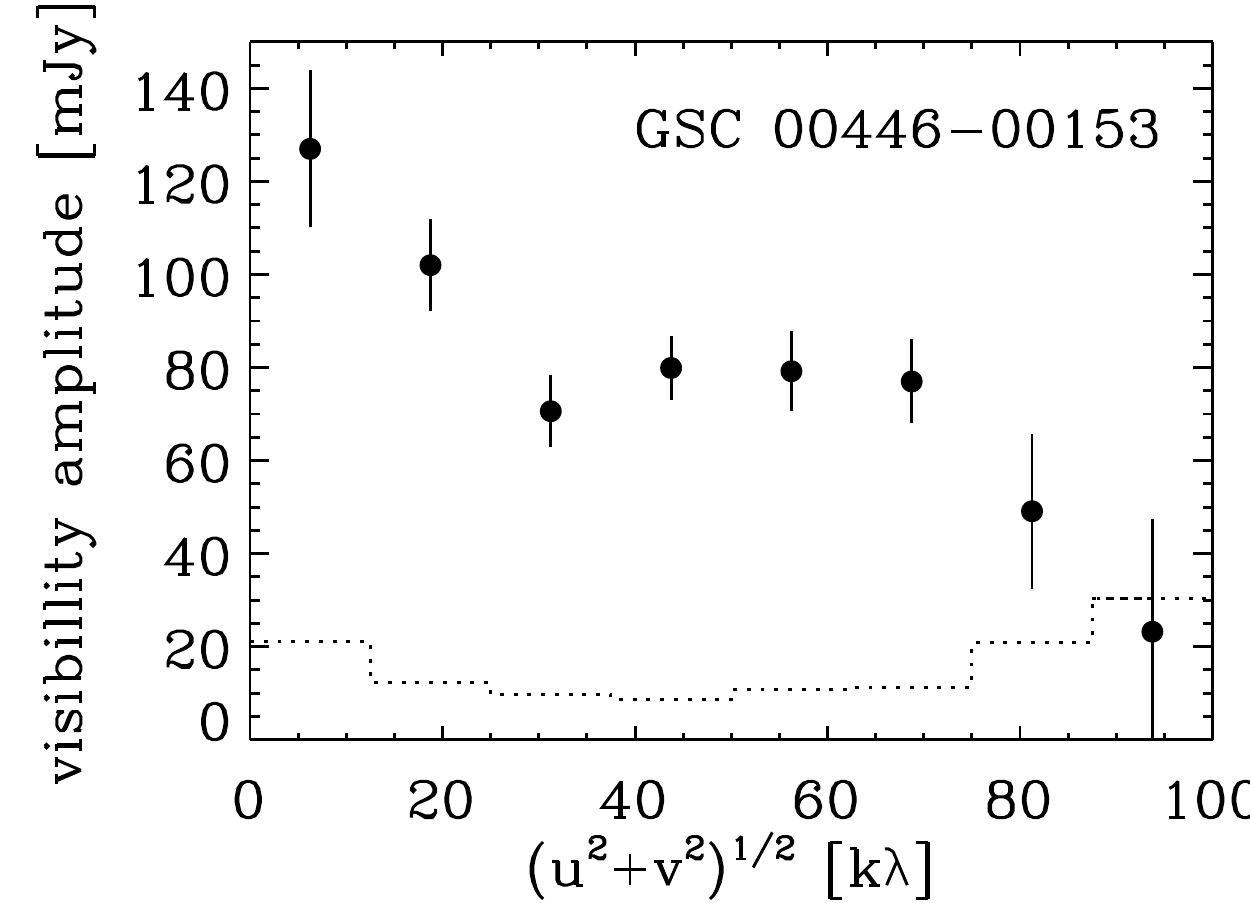}
			\includegraphics[width=5cm]{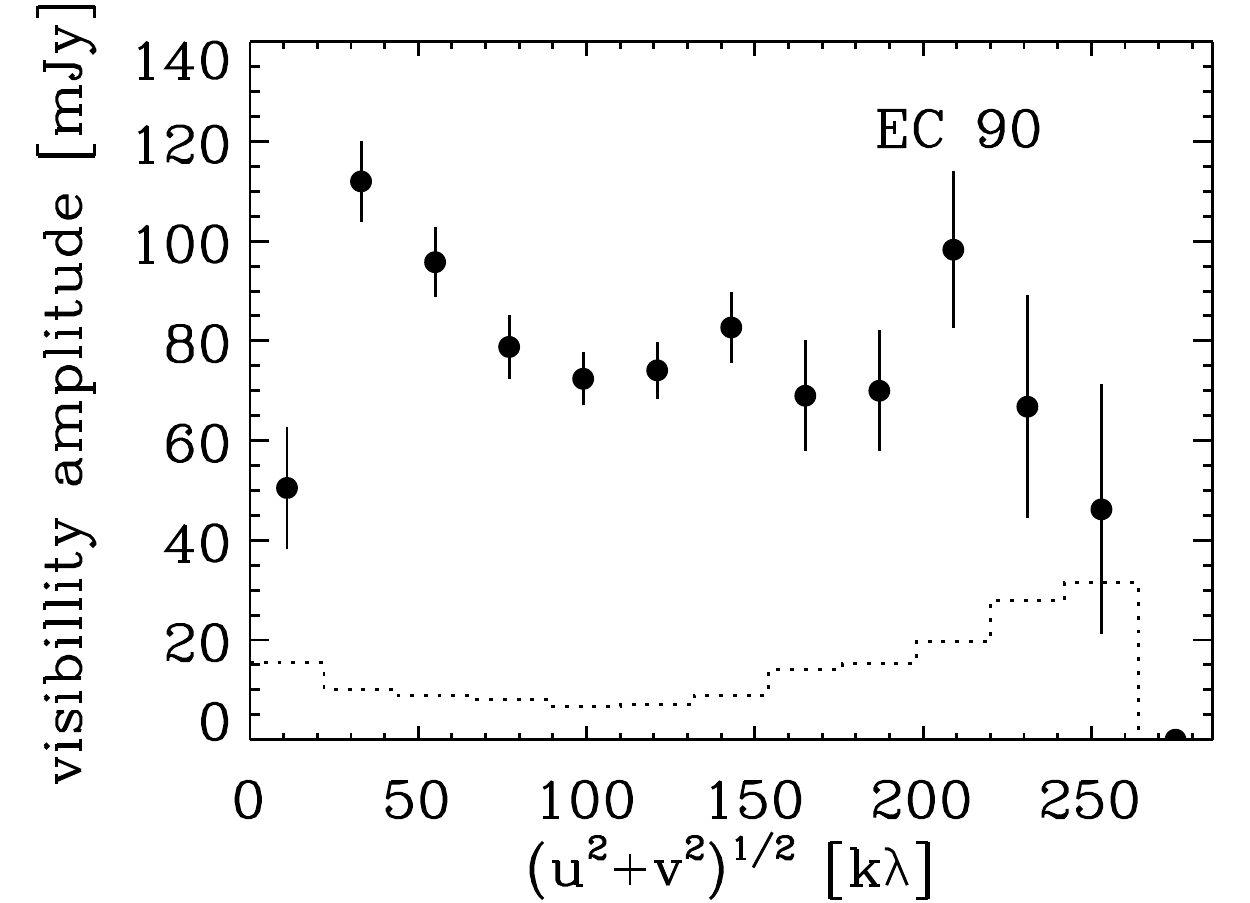}
		\end{center}
		\caption{Amplitude as a function of ($u$, $v$) distance for sources detected with CARMA at
			1~mm.}\label{fig: CARMA UVdist 1mm}
	\end{figure*}

\begin{figure*}
		\begin{center}
			\includegraphics[width=5cm]{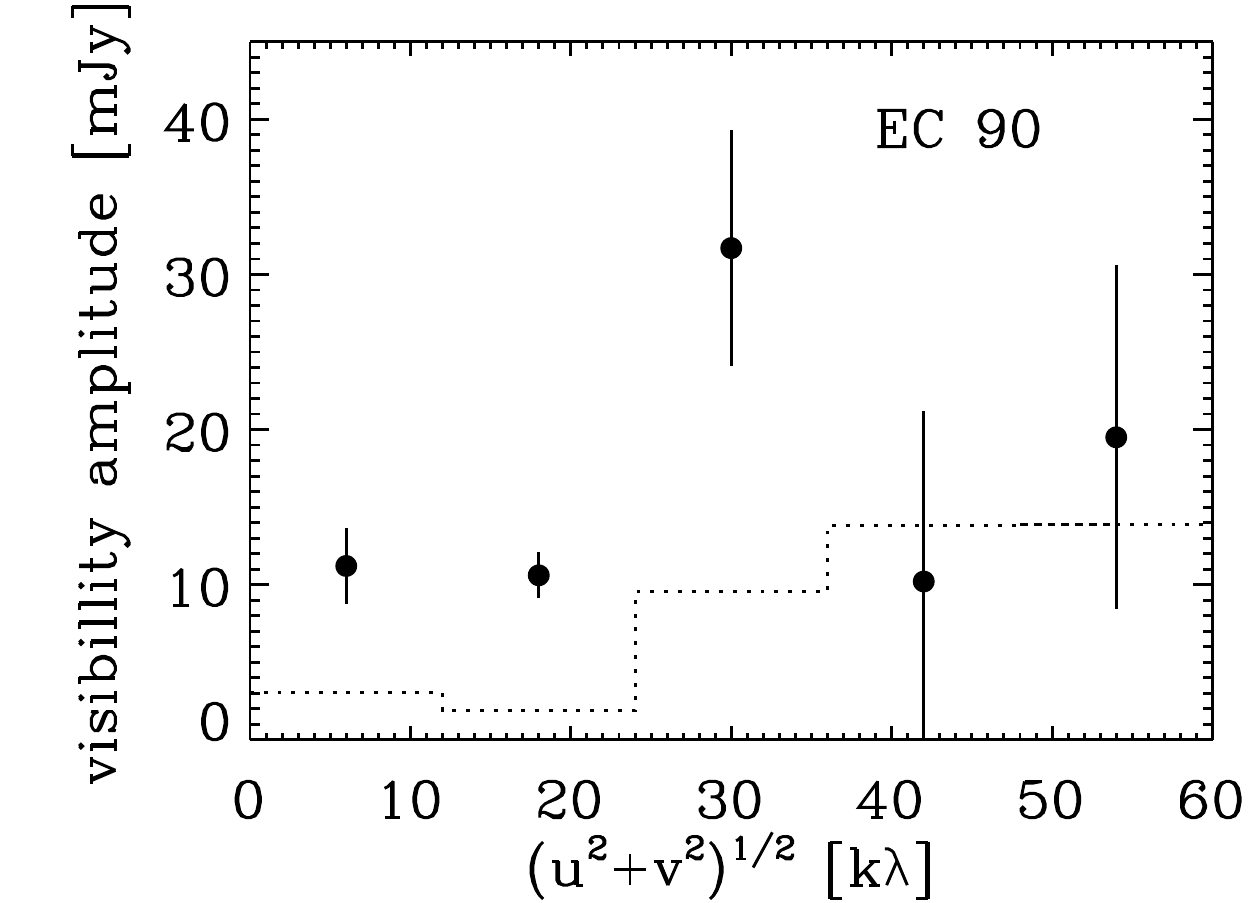}
			\includegraphics[width=5cm]{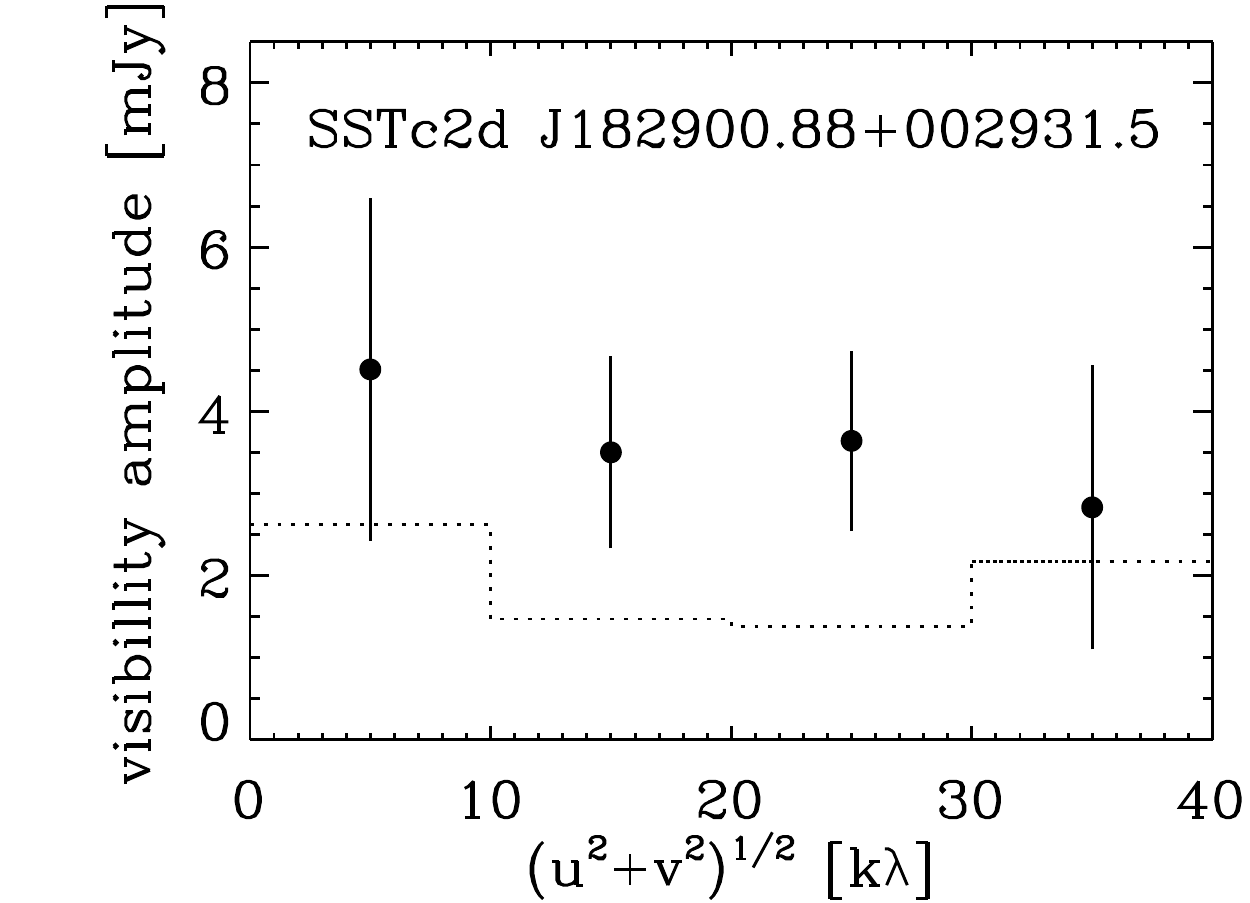}
			\includegraphics[width=5cm]{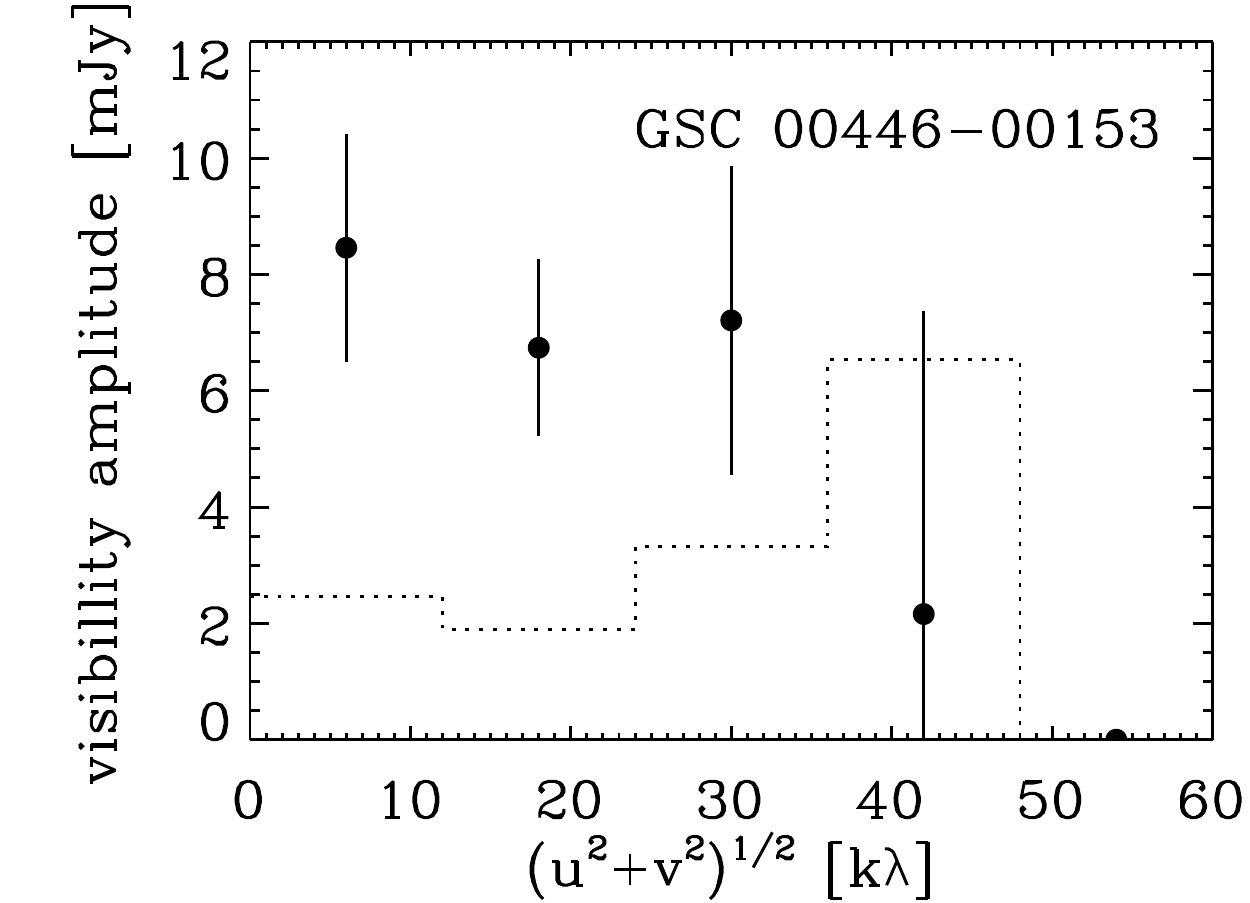}
		\end{center}
		\caption{Amplitude as a function of ($u$, $v$) distance for sources detected with CARMA at
			3~mm.}\label{fig: CARMA UVdist 3mm}
	\end{figure*}

	The complete results of the CARMA observations are shown in Table~\ref{tab: results log CARMA}. Data from tracks that could 
	not be properly calibrated due to a weak gain calibrator are not included. Some sources were observed 
	twice, once in the C and once in the D configuration. If good data were obtained in both occasions, they were co-added in the
	($u$, $v$) plane to improve the signal-to-noise ratio. The resulting fluxes or 3$\sigma$ upper limits are presented in 
	Table~\ref{tab: results} in the main text.
	\begin{table*}
		\caption[]{Complete results of CARMA observations at 1 and 3~mm.}
		\centering
		\begin{tabular}{lcccccccc}
			\hline
			\hline
			Obs. date	& Effective	& Target source$^\mathrm{a}$	& \multicolumn{2}{c}{Continuum flux$^\mathrm{b}$}	& rms$^\mathrm{c}$	& Gaussian size		& RA$^\mathrm{b}$      	& Dec$^\mathrm{b}$	\\
					& wavelength	&				& (P)			& (G)				&			& (arcsec)		& (J2000)		& (J2000)		\\
					& (mm)		&				& \multicolumn{2}{c}{(mJy)}		     		& (mJy/bm)		&			&			&			\\
			\hline
			20080518	& 1.33		& GSC 00446-00153		& 43.6		     	& 50.5 	     			& 6.0			& $0.46 \pm 0.32$	& 18 30 06.2		& +00 42 33.6		\\
			20080618	& 3.15		& EC 82				& \multicolumn{2}{c}{$<2.9^\mathrm{d}$}			& 1.0			& --			& 18 29 56.8		& +01 14 46.0		\\
					&		& EC 90				& 11.3			& 11.6				& 1.0			& $1.69 \pm 0.97$	& 18 29 57.8		& +01 14 06.9		\\
					&		& 182900.88+002931.5		& 3.4			& 3.9				& 0.6			& (unresolved)		& 18 29 00.9		& +00 29 31.7		\\
					&		& IRAS 18268-0025		& \multicolumn{2}{c}{$<1.9^\mathrm{d}$}			& 0.6			& --			& 18 29 28.2		& -00 22 57.1		\\
					&		& CoKu Ser G3			& \multicolumn{2}{c}{$<1.6^\mathrm{d}$}			& 0.5			& --			& 18 29 01.8		& +00 29 54.6		\\
					&		& VV Ser			& \multicolumn{2}{c}{$<1.8^\mathrm{d}$}		     	& 0.6	     		& --			& 18 28 47.9		& +00 08 40.0 		\\
			20080619	& 3.15		& 182858.08+001724.4		& \multicolumn{2}{c}{$<1.9^\mathrm{d}$}			& 0.6			& --			& 18 28 58.1		& +00 17 24.4		\\
					&		& 182850.20+000949.7		& \multicolumn{2}{c}{$<1.9^\mathrm{d}$}		     	& 0.6  			& --			& 18 28 50.2		& +00 09 49.7 		\\
					&		& 182944.10+003356.1		& \multicolumn{2}{c}{$<1.7^\mathrm{d}$}			& 0.6			& --			& 18 29 44.1		& +00 33 56.1		\\
					&		& 182936.19+004216.7		& \multicolumn{2}{c}{$<2.7^\mathrm{d}$}		     	& 0.9	     		& --			& 18 29 36.2		& +00 42 16.7 		\\
			20080620	& 1.33		& EC 82				& \multicolumn{2}{c}{$<15.7^\mathrm{d}$}		& 5.2			& --			& 18 29 56.8		& +01 14 46.0		\\
					&		& EC 90				& 91.8			& 91.7				& 10.4			& (unresolved)		& 18 29 57.7		& +01 14 07.0		\\
					&		& 182900.88+002931.5		& \multicolumn{2}{c}{$<16.4^\mathrm{d}$}		& 5.5		     	& --			& 18 29 00.9		& +00 29 31.6		\\
					&		& IRAS 18268-0025		& \multicolumn{2}{c}{$<15.6^\mathrm{d}$}		& 5.2			& --			& 18 29 28.2		& -00 22 57.1		\\
					&		& CoKu Ser G3			& \multicolumn{2}{c}{$<17.1^\mathrm{d}$}		& 5.7			& --			& 18 29 01.8		& +00 29 54.6		\\
					&		& VV Ser 			& \multicolumn{2}{c}{$<14.8^\mathrm{d}$}	     	& 4.9 			& --			& 18 29 47.9		& +00 08 40.0  		\\
					&		& 182858.08+001724.4		& \multicolumn{2}{c}{$<24.6^\mathrm{d}$}		& 8.2			& --			& 18 28 58.1		& +00 17 24.4		\\
					&		& 182850.20+000949.7		& \multicolumn{2}{c}{$<98.2^\mathrm{d}$}	     	& 32.7	     		& --			& 18 28 50.2		& +00 09 49.7 		\\
			20080622	& 3.15		& 182936.19+004216.7		& \multicolumn{2}{c}{$<5.3^\mathrm{d}$}			& 1.8			& --			& 18 29 36.2		& +00 42 16.7		\\
					&		& GSC 00446-00153		& 6.8		     	& 8.1	 	       		& 1.0			& $2.71 \pm 1.58$	& 18 30 06.3		& +00 42 34.2 		\\
			20080704	& 1.33		& EC 97				& \multicolumn{2}{c}{$<23.3^\mathrm{d}$}		& 7.8			& --			& 18 29 58.2		& +01 15 22.0		\\
					&		& 182850.20+000949.7		& \multicolumn{2}{c}{$<22.8^\mathrm{d}$}	     	& 7.6	     		& --			& 18 28 50.2		& +00 09 49.7		\\
					&		& 182944.10+003356.1 		& \multicolumn{2}{c}{$<15.0^\mathrm{d}$}		& 5.0			& --			& 18 29 44.1		& +00 35 56.1	     	\\
					&		& 182936.19+004216.7		& \multicolumn{2}{c}{$<8.7^\mathrm{d}$}			& 2.9			& --			& 18 29 36.2		& +00 42 16.7		\\
					&		& GSC 00446-00153		& 90.8		     	& 97.6 		     		& 2.8			& $0.67 \pm 0.23$	& 18 30 06.2  		& +00 42 33.6		\\
			\hline
		\end{tabular}
		\label{tab: results log CARMA}
	 \begin{list}{}{}
	  \item[$^\mathrm{a}$]	In the case of SSTc2d names, only the coordinates (in J2000) are shown.
	  \item[$^\mathrm{b}$]	Continuum flux and position are from fits in the ($u$, $v$) plane. For sources that were detected at 
	  	3$\sigma$, both the point-source flux (P) and the integrated flux for a Gaussian (G) are shown. For sources that were not
		detected, the coordinates of the phase centre are quoted.
	  \item[$^\mathrm{c}$]	Calculated from the cleaned image.
	  \item[$^\mathrm{d}$]	Quoted value is 3$\sigma$ upper limit.
	 \end{list} 
	\end{table*}
	The amplitude as a function of ($u$, $v$) distance of the sources detected with CARMA at 1~mm is plotted in 
	Fig.~\ref{fig: CARMA UVdist 1mm}, that of the sources detected with CARMA at 3~mm in
	Fig.~\ref{fig: CARMA UVdist 3mm}.
	
	The complete results of the VLA observations are shown in Table~\ref{tab: results log VLA}.
	\begin{table*}
		\caption[]{Complete results of VLA observations at 6.9~mm and 1.3, 3.5, and 6.2~cm.}
		\centering
		\begin{tabular}{lcccccccc}
			\hline
			\hline
			Obs. date	& Effective	& Target source$^\mathrm{a}$	& \multicolumn{2}{c}{Continuum flux$^\mathrm{b}$}	& rms$^\mathrm{c}$	& Gaussian size		& RA$^\mathrm{b}$      	& Dec$^\mathrm{b}$	\\
					& wavelength	&				& (Peak)		& (Integ.)			&			& (arcsec)		& (J2000)		& (J2000)		\\
					& (mm)		&				& \multicolumn{2}{c}{(mJy)}		     		& (mJy/bm)		&			&			&			\\
			\hline
			20080310	& 6.93		& CoKu Ser G3			& \multicolumn{2}{c}{$<1.2^\mathrm{d}$}			& 0.4			& --			& 18 29 01.8		& +00 29 54.5		\\
					&		& 182900.88+002931.5		& \multicolumn{2}{c}{$<0.7^\mathrm{d}$}			& 0.2			& --			& 18 29 00.9		& +00 29 31.5		\\
					&		& EC 90				& \multicolumn{2}{c}{$<1.0^\mathrm{d}$}			& 0.3			& --			& 18 29 57.7		& +01 14 05.7		\\
					&		& VV Ser			& \multicolumn{2}{c}{$<0.7^\mathrm{d}$}			& 0.2			& --			& 18 28 47.9		& +00 08 39.8		\\
					&		& EC 97				& \multicolumn{2}{c}{$<0.6^\mathrm{d}$}			& 0.2			& --			& 18 29 58.2		& +01 15 21.7		\\
					&		& 182850.20+000949.7		& \multicolumn{2}{c}{$<0.6^\mathrm{d}$}			& 0.2			& --			& 18 28 50.2		& +00 09 49.6		\\
					&		& EC 82				& \multicolumn{2}{c}{$<0.5^\mathrm{d}$}			& 0.2			& --			& 18 29 56.9		& +01 14 46.4		\\
			20080311	& 13.4		& CoKu Ser G3			& 8.66			& 9.81				& 0.06			& $1.17 \times 0.91$	& 18 29 01.8		& +00 29 54.8		\\
					&		& 182900.88+002931.5		& \multicolumn{2}{c}{$<0.17^\mathrm{d}$}		& 0.06			& --			& 18 29 00.9		& +00 29 31.5		\\
					&		& EC 90				& \multicolumn{2}{c}{$<0.24^\mathrm{d}$}		& 0.08			& --			& 18 29 57.7		& +01 14 05.7		\\
					&		& VV Ser			& \multicolumn{2}{c}{$<0.17^\mathrm{d}$}		& 0.06			& --			& 18 28 47.9		& +00 08 39.8		\\
					&		& EC 97				& \multicolumn{2}{c}{$<0.15^\mathrm{d}$}		& 0.05			& --			& 18 29 58.2		& +01 15 21.7		\\
			20080313	& 13.4		& 182850.20+000949.7		& \multicolumn{2}{c}{$<0.20^\mathrm{d}$}		& 0.07			& --			& 18 28 50.2		& +00 09 49.6		\\
					&		& 182909.80+003445.9		& \multicolumn{2}{c}{$<0.23^\mathrm{d}$}		& 0.08			& --			& 18 29 09.8		& +00 34 45.8		\\
					&		& EC 82				& \multicolumn{2}{c}{$<0.28^\mathrm{d}$}		& 0.09			& --			& 18 29 56.9		& +01 14 46.4		\\
			20080314	& 35.5		& EC 82				& \multicolumn{2}{c}{$<0.09^\mathrm{d}$}		& 0.03			& --			& 18 29 56.0		& +01 14 49.9		\\
					&		& EC 90				& \multicolumn{2}{c}{$<0.09^\mathrm{d}$}		& 0.03			& --			& 18 29 56.0		& +01 14 49.0		\\
					&		& EC 97				& \multicolumn{2}{c}{$<0.09^\mathrm{d}$}		& 0.03			& --			& 18 29 56.0		& +01 14 49.0		\\
					&		& CoKu Ser G3			& 1.11			& 1.11				& 0.03			& $2.49 \times 2.35$	& 18 29 01.8		& +00 29 54.7		\\
					&		& 182900.88+002931.5		& \multicolumn{2}{c}{$<0.10^\mathrm{d}$}		& 0.03			& --			& 18 29 05.0		& +00 29 44.0		\\
					&		& VV Ser			& 0.14			& 0.17				& 0.03			& $3.25 \times 2.31$	& 18 28 47.9		& +00 08 40.1		\\
					& 		& 182850.20+000949.7		& \multicolumn{2}{c}{$<0.08^\mathrm{d}$}		& 0.03			& --			& 18 28 40.0		& +00 09 13.0		\\
			20080315	& 61.7		& EC 82				& \multicolumn{2}{c}{$<0.09^\mathrm{d}$}		& 0.03			& --			& 18 29 57.0		& +01 14 40.0		\\
					&		& EC 90				& \multicolumn{2}{c}{$<0.09^\mathrm{d}$}		& 0.03			& --			& 18 29 57.0		& +01 14 40.0		\\
					&		& EC 97				& \multicolumn{2}{c}{$<0.09^\mathrm{d}$}		& 0.03			& --			& 18 29 57.0		& +01 14 40.0		\\
					&		& CoKu Ser G3			& 0.91			& 1.00				& 0.04			& (unresolved)		& 18 29 01.9		& +00 29 54.9		\\
					&		& 182900.88+002931.5		& \multicolumn{2}{c}{$<0.12^\mathrm{d}$}		& 0.04			& --			& 18 29 07.0		& +00 32 09.0		\\
					&		& VV Ser			& \multicolumn{2}{c}{$<0.14^\mathrm{d}$}		& 0.05			& --			& 18 29 47.0		& +00 09 11.0		\\
					&		& 182850.20+000949.7		& \multicolumn{2}{c}{$<0.14^\mathrm{d}$}		& 0.05			& --			& 18 29 47.0		& +00 09 11.0		\\
			\hline
		\end{tabular}
		\label{tab: results log VLA}
	 \begin{list}{}{}
	  \item[$^\mathrm{a}$]	In the case of SSTc2d names, only the coordinates (in J2000) are shown.
	  \item[$^\mathrm{b}$]	Continuum flux and position are obtained using the AIPS task JMFIT, which fits elliptical Gaussians to
	  	the cleaned image. For sources that were detected at 3$\sigma$, both the peak (Peak) and the integrated (Integ.) flux are
		shown. For sources that were not detected, the coordinates of the phase centre are quoted.
	  \item[$^\mathrm{c}$]	Calculated from the cleaned image.
	  \item[$^\mathrm{d}$]	Quoted value is 3$\sigma$ upper limit.
	 \end{list} 
	\end{table*}
\bibliographystyle{aa}
\bibliography{references}

\end{document}